\documentclass[
aps,
physrev,
superscriptaddress,
amsmath,
amssymb,
nofootinbib,
floatfix,
twocolumn,
reprint
]
{revtex4-2}

\usepackage[utf8]{inputenc}
\usepackage{graphicx,color}
\usepackage{url}
\usepackage{bm}
\usepackage{siunitx}
\usepackage[italicdiff]{physics}
\usepackage[all]{xy}
\usepackage{qcircuit}
\usepackage{amsthm}
\usepackage{enumitem}
\usepackage{tikz}

\setlist[description]{leftmargin=\parindent,labelindent=\parindent}

\theoremstyle{plain}

\newtheorem*{thm*}{Theorem}
\newtheorem*{prop*}{Proposition}

\usepackage{hyperref}
\hypersetup{
colorlinks=true,
linkcolor=blue,
citecolor=blue
}

\begin{document}
 
\title{Quantum Algorithm for Direct Estimation of Energy-Level Differences\texorpdfstring{\\}{} via Quantum Dynamics Simulation} 

\author{Norifumi Matsumoto}
\affiliation{Quantum Laboratory, Fujitsu Research, Fujitsu Limited., 4-1-1 Kamikodanaka, Nakahara, Kawasaki, Kanagawa 211-8588, Japan}
\email{m.norifumi@fujitsu.com}

\author{Yuya O. Nakagawa}
\affiliation{QunaSys Inc., Aqua Hakusan Building 9F, 1-13-7 Hakusan, Bunkyo, Tokyo 113-0001, Japan}

\author{Shota Kanasugi}
\affiliation{Quantum Laboratory, Fujitsu Research, Fujitsu Limited., 4-1-1 Kamikodanaka, Nakahara, Kawasaki, Kanagawa 211-8588, Japan}

\author{Yuichiro Hidaka}
\affiliation{QunaSys Inc., Aqua Hakusan Building 9F, 1-13-7 Hakusan, Bunkyo, Tokyo 113-0001, Japan}

\author{Kazunori Maruyama}
\affiliation{Quantum Laboratory, Fujitsu Research, Fujitsu Limited., 4-1-1 Kamikodanaka, Nakahara, Kawasaki, Kanagawa 211-8588, Japan}

\author{Hirotaka Oshima}
\affiliation{Quantum Laboratory, Fujitsu Research, Fujitsu Limited., 4-1-1 Kamikodanaka, Nakahara, Kawasaki, Kanagawa 211-8588, Japan}

\date{\today}

\begin{abstract}
We present a quantum algorithm for the early fault-tolerant quantum computing (early-FTQC) era that estimates minute energy-level differences, such as energy gradients and molecular forces, at a lower sampling cost than existing phase-estimation methods. Such energy derivatives, accessed as the small energy difference induced by an infinitesimal parameter shift, are central to quantum simulation in chemistry and materials science. Although Statistical Phase Estimation (SPE) is well-suited to energy estimation in this era, applying it to an energy difference demands a sampling complexity that scales with the operator norm of the full Hamiltonian, which grows with system size. 
Our algorithm overcomes this by incorporating two time-evolution operators, corresponding to the Hamiltonians before and after the parameter shift, into a single quantum circuit, which is combined with a non-adaptive, statistically sampled reconstruction in the energy-difference domain.
This reduces the relevant scale from the operator norm of the full Hamiltonian to that of the operator norm of the Hamiltonian derivative and the parameter shift, thereby lowering the sampling complexity. We further propose the joint estimation of energy eigenvalues and their differences, giving a holistic view of the energy landscape.
Numerical simulations and resource estimations on prototypical systems, such as the Fermi-Hubbard model, demonstrate the resulting reductions in required resources without compromising precision. 
This positions energy-difference estimation as a practical task for early-FTQC devices, opening a route to molecular forces, response properties, and excited-state energy differences.
\end{abstract}

\maketitle

\begin{figure*}[htbp]
    \centering
        \includegraphics[width=1.\linewidth]{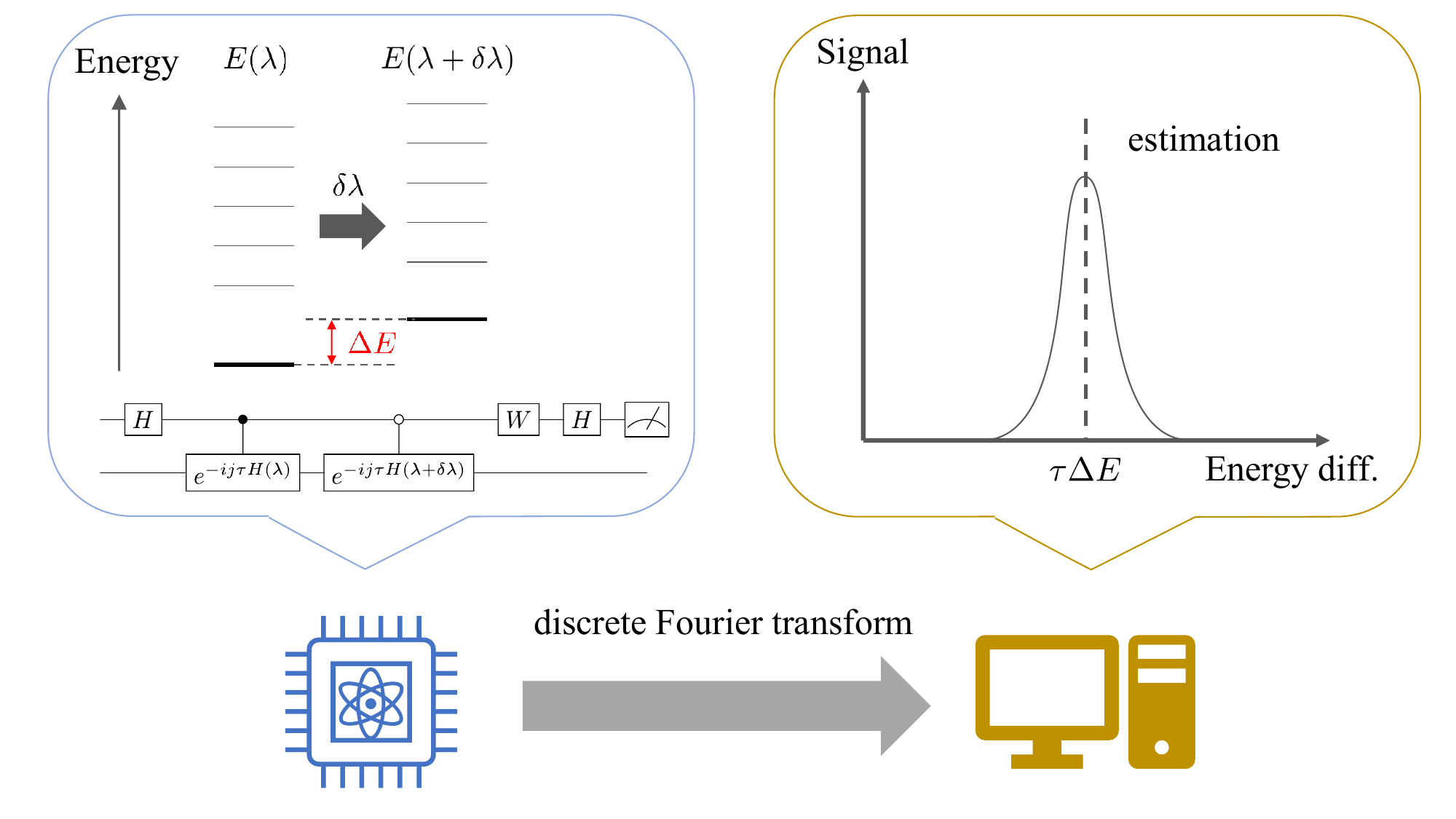}
    \caption{
        Conceptual overview of the proposed quantum-classical hybrid algorithm for efficient energy level difference estimation.
        The method leverages a synergy between a quantum computer (left) and a classical computer (right).
        The quantum computer executes a specially designed Hadamard test circuit, which incorporates time-evolution operators associated with the Hamiltonians $H(\lambda)$ and $H(\lambda+\delta\lambda)$, where $\lambda$ and $\delta \lambda$ denote a parameter and its shift in the Hamiltonian, respectively (see Fig.~\ref{fig:circuit-proposed-simple} for details of the quantum circuit).
        Measurement results from the quantum computer are then sent to the classical computer.
        The classical computer performs statistical processing on these results to directly estimate the energy level difference $\Delta E$.
    }
    \label{fig:proposal_overview}
\end{figure*}

\section{Introduction}\label{sec:Intro}

Quantum computing is a rapidly advancing technology that is expected to impact the simulation of quantum many-body systems~\cite{Feynman1982, Lloyd1996, Aspuru2005, Cao2019}. Simulating these systems is notoriously challenging for classical computers due to the exponential scaling of Hilbert space. In this context, the accurate estimation of energy eigenvalues~\cite{Kitaev1995, Cleve1998, Abrams1999} is a central problem across materials science, chemistry, and condensed matter physics~\cite{Reiher2017, Babbush2018lowdepth, Mcclean2016, Campbell2022}.

Despite its fundamental importance, the energy itself often provides only an incomplete picture for practical applications. Knowing the energy of a molecular system primarily indicates its overall stability; it does not, on its own, reveal how that energy changes in response to perturbations. Yet it is precisely these changes, captured by energy gradients and finite differences, that drive the central tasks of materials science and chemistry~\cite{Jordan2005, Kassal2009, Mitarai2018, Mitarai2020, Schuld2019, OBrien2019, OBrien2022}.

Foremost among these are molecular forces, the energy derivatives with respect to nuclear positions, which are central to geometry optimization~\cite{Pulay1969, Kassal2009, Schlegel2011, Pulay2014, OBrien2019, Delgado2021, OBrien2022}, transition-state identification and reaction-pathway analysis~\cite{Hratchian2005, Henkelman2000, Reiher2017}, and \textit{ab initio} molecular dynamics~\cite{Car1985, Marx2009}. Derivatives with respect to external fields further give response properties such as polarizability and magnetizability~\cite{Helgaker1984, Cai2020, Huang2022}, while energy gradients evaluated on excited-state surfaces yield excited-state forces, extending geometry optimization and ab initio dynamics to excited states~\cite{Higgott2019, Nakanishi2019, Ollitrault2020, Asthana2023}. 
Together, these energy-derivative quantities underpin applications ranging from drug discovery to materials design~\cite{Cournia2017, SaloAhen2021, Malone2022, Kirsopp2022}.

The Quantum Phase Estimation (QPE) algorithm~\cite{Kitaev1995, Cleve1998, Abrams1999} and its variants, such as Iterative Phase Estimation (IPE)~\cite{Kitaev2002, Dobsicek2007, Svore2014}, Bayesian Phase Estimation (BPE)~\cite{Svore2014, Wiebe2016, Paesani2017} and Statistical Phase Estimation (SPE)~\cite{Lin2022, Wan2022, Zhang2022, Dong2022, Ding2023, Wang2023}, are promising algorithms for energy estimation.
BPE, for instance, is a quantum-classical hybrid approach that iteratively refines an estimate of a unitary operator's phase parameter using Bayesian inference, typically by combining iterative executions of controlled-unitary operations with a parametrised phase rotation gate in its quantum circuit (see Appendix~\ref{app:bpe_overview} for details).
However, their practical application on early-FTQC devices~\cite{Campbell2022, Katabarwa2024}, which offer a critical opportunity to leverage quantum parallelism for classically intractable problems, is often hindered by the challenges associated with precisely determining energy differences.
Both IPE and BPE require iterative, adaptively controlled circuit executions, i.e., feedback-conditioned rotations in IPE and Bayesian experiment design in BPE, posing demands for early-FTQC devices.
For estimating small energy differences, SPE also suffers from an unfavorable scaling of the sampling complexity. 
The details and implications of this scaling bottleneck will be discussed in further detail in Sec.~\ref{subsec:limitations-existing-methods}. 
Since the cost of resolving a minute difference grows with the operator norm of the Hamiltonian, which grows with system size and is therefore large for the sizable, strongly correlated systems targeted for quantum advantage, this overhead becomes a serious obstacle in practice. We quantify it for the Fermi-Hubbard model and for molecular force estimation~\cite{OBrien2022} in Sec.~\ref{subsec:advantages-over-existing}. 

This work addresses these challenges by proposing a quantum algorithm for efficiently and precisely estimating energy level differences. Our approach represents a fundamental change: instead of indirectly inferring differences from separate energy-eigenvalue estimations, we directly target the energy difference itself. 
In particular, the small energy difference induced by an infinitesimal parameter shift yields the energy gradient through a finite difference, which is the central quantity for forces and response properties.
This is achieved by the combination of a specially designed quantum circuit, built upon the SPE framework, and a refined statistical processing technique (see Fig.~\ref{fig:proposal_overview}).  
The proposed quantum circuits' outputs inherently encode information about the energy difference, rather than individual energy values. 
This decouples the total computational cost from the overall magnitude of the energy eigenvalues: rather than scaling with the operator norm of the full Hamiltonian $\|H\|$, as in conventional approaches, the cost of our method is governed by the much smaller scale set by the operator norm of the Hamiltonian derivative and the parameter shift [see Eq.~\eqref{eq:diff-d-lower-bound}].  
This substantially suppresses the sampling complexity (Sec.~\ref{sec:numerical-demonstration}), making the precise characterization of minute energy changes practically feasible on early-FTQC quantum hardware.

The remainder of this paper is organized as follows. In Sec.~\ref{sec:review-SPE}, we provide a theoretical background on quantum algorithms for energy estimation, discussing existing methods and their limitations for energy difference estimation. Section~\ref{sec:proposed-method} details our proposed method, elucidating its core principle, algorithmic steps, and theoretical advantages. In Sec.~\ref{sec:numerical-demonstration}, we present numerical simulation results demonstrating the reduction in the sampling complexity of quantum circuits while maintaining high precision. Section~\ref{sec:resource-estimation} provides a detailed analysis of the resource requirements for implementing our algorithm, assuming a specific and viable early-FTQC architecture. Finally, Sec.~\ref{sec:discussion} discusses the implications of our findings, potential applications, and future research directions.

\section{Preliminary}\label{sec:review-SPE}

This section lays the groundwork for our proposed quantum algorithm by defining key concepts and reviewing relevant existing methods for energy estimation. We first explicitly define the problem of energy-level difference estimation and the metric of accuracy. 
Following this, we provide an overview of established QPE algorithms, highlighting their general principles and limitations.
We then delve into the SPE~\cite{Lin2022, Wan2022, Zhang2022, Dong2022, Ding2023, Wang2023} in detail, which serves as a foundational component for our method, discussing its core mechanisms and bottlenecks when used for energy difference calculations. Finally, we review other existing direct energy difference estimation methods~\cite{Sugisaki2021, Sugisaki2022} and their inherent challenges, thereby setting the context for the necessity and novelty of our proposed approach.

\subsection{Definition of Problem}\label{subsec:problem_definition}

The energy eigenvalues $E_k$ of the target Hamiltonian, associated with the corresponding eigenstates $|E_k\rangle$, are ordered such that $E_0 \le E_1 \le E_2 \le \dots$, where $E_0$ denotes the ground state energy. A central task in quantum chemistry and materials science is to accurately determine these energy levels.

Our focus in this work---and the primary object of estimation for our proposed algorithm---is the energy-level difference, most commonly the difference involving the ground state energy. This is defined as the difference between two energy eigenvalues. Specifically, when considering a Hamiltonian $H(\lambda)$ and its perturbed counterpart $H(\lambda+\delta\lambda)$, where $\lambda$ and $\delta\lambda$ represent a parameter and its small shift, respectively, an energy-level difference is typically expressed as
\begin{align}
\Delta E_0 = E_0(\lambda+\delta\lambda) - E_0(\lambda),
\end{align}
where $E_0(\lambda)$ and $E_0(\lambda+\delta\lambda)$ denote the ground state energies of $H(\lambda)$ and $H(\lambda+\delta\lambda)$ respectively. More generally, energy-level differences can also refer to differences between arbitrary energy eigenvalues $\Delta E_k = E_k(\lambda+\delta\lambda) - E_k(\lambda)$. 
When the subscript is omitted and the context does not indicate otherwise, $\Delta E$ (and $E$) denotes the difference for the targeted level, which is the ground state by default, corresponding to the standard situation in which the initial state $|\psi_0\rangle$ has dominant overlap with the ground state.
Accurate estimation of such differences is paramount for uncovering various physical and chemical properties, as detailed in Sec.~\ref{sec:Intro}.

To quantify the performance of quantum algorithms designed for energy-difference estimation, we define the precision $\epsilon$ of an estimation. This refers to the desired absolute error for the estimated energy-level difference. If $\Delta E_{\rm estimated}$ is the estimated value and $\Delta E_{\rm true}$ is the true value, the precision $\epsilon$ represents the maximum allowable deviation:
\begin{align}\label{eq:precision-definition}
\epsilon \ge |\Delta E_{\rm estimated} - \Delta E_{\rm true}|.
\end{align}
Our proposed algorithm aims to estimate $\Delta E$ with a desired precision $\epsilon$ efficiently on early fault-tolerant quantum computing devices.

\subsection{Overview of Energy Estimation Algorithms}

The QPE algorithm is a foundational algorithm for estimating the eigenvalues of a unitary operator. When applied to a Hamiltonian $H$ via the unitary $U=e^{-iHt}$ for some time interval $t$, QPE can determine energy eigenvalues with high precision. While powerful, standard QPE typically requires a large number of ancilla qubits and deep quantum circuits, which are challenging for early-FTQC quantum hardware. Variants like IPE~\cite{Kitaev2002, Dobsicek2007, Svore2014} and BPE~\cite{Svore2014, Wiebe2016, Paesani2017} reduce ancilla-qubit requirements, often to a single one, by iteratively refining the phase estimate through classical post-processing. 
However, IPE uses feedback rotations conditioned on previously measured bits, and BPE utilizes an adaptive, iteratively optimized choice of experiment parameters (see Appendix~\ref{app:bpe_overview}); both therefore require coherently controlled, adaptively scheduled circuit executions.

\subsection{SPE and its Limitations for Energy Difference Estimation}

While IPE and BPE also adopt a quantum-classical hybrid approach, SPE is specifically designed, primarily for estimating the ground-state energy, to be well-suited to early-FTQC devices. It uses a single ancilla qubit and, unlike IPE and BPE, a non-adaptive set of circuits followed by purely classical post-processing, i.e., requiring neither feedback rotations nor iterative parameter optimization, while avoiding the overlap-dependent depth overhead, i.e., the coherent evolution time $\mathcal{O}(p_0^{-1}\epsilon_{\rm E}^{-1})$ with $p_0$ the ground-state overlap for precision $\epsilon_{\rm E}$ of the energy estimation~\cite{Lin2022}, that a naive single-ancilla QPE would need to reach high confidence from a general imperfect initial state.
Our proposed method is inspired by the principles of SPE, adapting and extending them to efficiently address the specific challenge of estimating energy-level differences. Therefore, we first detail the SPE algorithm before describing our proposal.
The SPE's strength lies in its ability to estimate energy information by performing measurements of the expectation value of the time evolution operators $e^{-iHt}$ for various evolution times $t$. Unlike the standard QPE that directly measures phases, SPE employs sophisticated classical post-processing to reconstruct the cumulative distribution function (CDF) of the Hamiltonian's spectral measure from these measurement results.

\subsubsection{The Quantum Part of SPE}

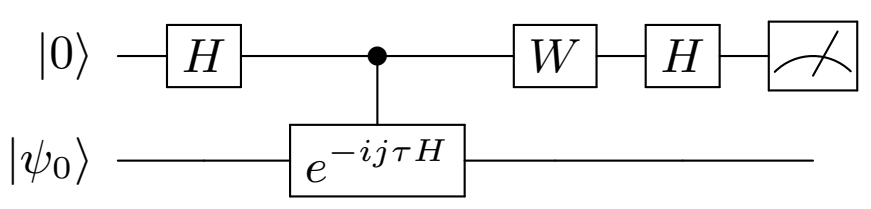
\begin{figure}[bp]
\centering
\[
    	\Qcircuit @C=1em @R=.7em {\\
  	\lstick{\ket{0}} & \gate{H} & \ctrl{1} 
    &\gate{W}
    & \gate{H}      & \meter \\
  	\lstick{\ket{\psi_0}} & \qw      & \gate{ e^{-i  j\tau H}} &\qw  & \qw &\qw 
	}
    \]
    \caption{
    Quantum circuit for SPE. The gate $W$ is set to $W=I$ ($W=S^\dag$) for estimating the real (imaginary) part of the expectation value of the time-evolution operator. The controlled time evolution is implemented by the $j$-times iterations of the time evolution with a fixed time step $\tau$.
    }
    \label{fig:circuit-SPE}
\end{figure}

The SPE employs the single-ancilla Hadamard-test quantum circuit~\cite{Knill2007} (see Fig.~\ref{fig:circuit-SPE}). This circuit allows for the estimation of the real and imaginary parts of the expectation value $\langle\psi_0|U|\psi_0\rangle$, where $U$ is a unitary operator and $|\psi_0\rangle$ is the initial state of the system. In the context of SPE, $U$ is the time evolution operator $e^{-iHt}$.

To estimate the real part, $\mathrm{Re}[\langle\psi_0|e^{-iHt}|\psi_0\rangle]$, the gate $W$ in Fig.~\ref{fig:circuit-SPE} is set to $W=I$. As shown in Fig.~\ref{fig:circuit-SPE}, the Hadamard test performs a controlled time-evolution operation between two Hadamard gates on the ancilla qubit, which is then measured. 
The probability of measuring the ancilla in the $|0\rangle$ state is $P_0 = \frac{1}{2}(1 + \mathrm{Re}[\langle\psi_0|U|\psi_0\rangle])$. From this, $\mathrm{Re}[\langle\psi_0|U|\psi_0\rangle]$ can be estimated as $2P_0 - 1$.
To estimate the imaginary part, $\mathrm{Im}[\langle\psi_0|e^{-iHt}|\psi_0\rangle]$, a slight modification is made: a phase gate $S^\dagger$ is inserted on the ancilla qubit just before the second Hadamard gate (i.e., $W=S^\dag$). This effectively rotates the phase of the $|1\rangle$ component of the ancilla, allowing the imaginary part to be projected onto the measurement basis.

In SPE, these Hadamard tests are performed for a series of discrete evolution times $t_j = j\tau$, where $j$ is an integer and $\tau$ is a chosen time step. The results from these measurements, collected over many repetitions, provide unbiased estimators for $\mathrm{Re}[\langle\psi_0|e^{-ij\tau H}|\psi_0\rangle]$ and $\mathrm{Im}[\langle\psi_0|e^{-ij\tau H}|\psi_0\rangle]$. These expectation values serve as the Fourier components of the Hamiltonian's spectral measure, as detailed in the following.

\subsubsection{The Classical Part of SPE}
\label{subsubsec:SPE-classical}

A key advantage of SPE emerges from its sophisticated classical post-processing, which extracts energy information from the quantum measurements. This process involves constructing and analyzing the CDF of the Hamiltonian's spectral measure.

Consider a Hamiltonian $H$ with eigenvalues $E_k$ and corresponding projectors $\Pi_k$. If the initial state, denoted as $|\psi_0\rangle$ in previous section, is now generalized to a density operator $\rho$, we define $p_k = \mathrm{Tr}[\rho\Pi_k]$. This $p_k$ represents the probability that the system is in the eigenstate corresponding to $E_k$ (or rather, the total weight for the eigenvalue $E_k$).
The spectral measure $p(x)$ is defined as a sum of Dirac delta functions at each scaled eigenvalue:
\begin{align}
p(x) = \sum_k \mathrm{Tr}[\rho\Pi_k] \, \delta(x - \tau E_k),
\end{align}
where $\tau = \mathcal{O}\qty(\|H\|^{-1})$ is a scaling factor chosen to map the eigenvalues into a convenient range, typically within $[-\pi, \pi]$, and corresponds to the time step appearing in the quantum circuit in Fig.~\ref{fig:circuit-SPE}. The expectation value measured by the Hadamard test, $\mathrm{Tr}[\rho e^{-ij\tau H}]$, is essentially the Fourier transform of this spectral measure.

The CDF $C(x)$ is then defined as the convolution of a $2\pi$-periodic Heaviside function
\begin{equation}
\tilde{H}(x) = 
\begin{cases} 
1 & \text{for } x \in [2k\pi, (2k+1)\pi) \text{ with }k \in \mathbb{Z}; \\
0 & \text{otherwise}
\end{cases}
\end{equation}
with the spectral measure $p(x)$:
\begin{align}
C(x) = (\tilde{H} * p)(x) = \int_{-\pi}^{\pi} \tilde{H}(x-y)p(y)dy.
\end{align}
This $C(x)$ is a step function that exhibits a staircase-like structure, starting near zero at the left end of its domain and rising sharply at values corresponding to scaled eigenvalues $\tau E_k$,  before leveling off, i.e., remaining largely horizontal, until the next rise.
The ground-state energy $E_0$ is finally estimated by finding the smallest $x$ for which the estimated CDF exceeds a certain threshold, which corresponds to identifying the first, i.e., left-most, upward step in this staircase-like function.

Since the true Heaviside function is difficult to implement in a finite Fourier series, SPE approximates it using a Fourier series $F_{\beta, d} (x)$ truncated at order $d$, with highest mode $2d+1$~\cite{Wan2022}. This $F_{\beta, d} (x)$ is carefully constructed to approximate $\tilde{H}(x)$ within a desired error tolerance.
Here, $\beta$ is a parameter that dictates the steepness of the approximation to the step function, with larger $\beta$ values leading to a sharper transition. This construction involves approximating the Heaviside function with a rescaled error function, which is then further approximated using Chebyshev and Fourier series (see Appendix~\ref{app:Fourier-Heaviside} for details). The approximate CDF (ACDF), denoted $\tilde{C}(x)$, is then obtained by convolving $F_{\beta, d} (x)$ with $p(x)$:
\begin{align}
\label{eq:SPE-ACDF-Fourier}
\tilde{C}(x) = \qty(F_{\beta, d} * p)(x) = \sum_{j\in\mathcal{S}} \hat{F}_j e^{ijx} \mathrm{Tr}[\rho e^{-ij\tau H}].
\end{align}
This ACDF exhibits a smoothed staircase-like structure, where sharp jumps of the ideal CDF are approximated by rapid increases.
Here, $\{\hat{F}_j \}$ is the set of Fourier coefficients of $F_{\beta, d} (x)$ (see Eq.~\eqref{eq:Fouriercoefficients_appendix} for details), nonzero only for $j\in\mathcal{S}$ with the set of the relevant indices $\mathcal{S} = \{0\}\cup\{\pm(2k+1)\}_{k=0}^{d}$, whose largest element $2d+1$ is the maximum Fourier mode and $d$ is the truncation order.
The maximum evolution time required for the quantum circuit is proportional to this maximum Fourier mode, i.e., $T_{\text{max}} = (2d+1)\tau$.
A sophisticated construction of $F_{\beta, d}(x)$ leads to the following scaling~\cite{Wan2022} (see also Appendix~\ref{app:Fourier-Heaviside}):
\begin{align}
\label{eq:scaling-truncation-order}
d = \mathcal{O} \qty( \frac{1}{\tau \epsilon_{\rm E}} ),
\end{align}
up to logarithmic factors, for an additive error $\epsilon_{\rm E}$ of the estimated energy level, where the subscript distinguishes the precision on a single eigenvalue from the precision $\epsilon$ on the difference defined in Eq.~\eqref{eq:precision-definition}.

Instead of computing the right-hand side of Eq.~\eqref{eq:SPE-ACDF-Fourier} term by term, employing a statistical sampling approach leads to the remarkable Heisenberg-limited scaling~\cite{Giovannetti2006, Higgins2007, Giovannetti2011, Zwierz2010, Zwierz2012}. 
Specifically, an integer random variable $J$ is drawn from $\mathcal{S}$ with probability proportional to $|\hat{F}_J|$. Then, an unbiased estimator 
\begin{align}\label{eq:G_estimator}
G(x; J, Z)= \qty(\sum_{j\in\mathcal{S}} |\hat{F}_j|) Z e^{i (\arg[\hat{F}_J]+Jx)}
\end{align}
for $\tilde{C}(x)$ is constructed using the complex random variable $Z=X_J + i Y_J$ obtained from the Hadamard test with evolution time $t_J = J\tau$. 
We write $\mathcal{F}\equiv\sum_{j\in\mathcal{S}}|\hat{F}_j|$ for the rescaling factor appearing in Eq.~\eqref{eq:G_estimator}; it satisfies $\mathcal{F}=\mathcal{O}(\log d)$ (Appendix~\ref{app:Fourier-Heaviside}).
By averaging $N_{\rm sample}$ independent samples of $(J, Z)$, an estimate for $\tilde{C}(x)$ is obtained as 
\begin{align}
\tilde{C}(x) = \frac{1}{N_{\rm sample}}\sum_{k=1}^{N_{\rm sample}}G(x; J_k, Z_k).
\end{align}
The number of samples $N_{\rm sample}$ scales as $\mathcal{O}(\log^2 d / \eta^2)$ with $\eta$, a known lower bound on the squared overlap $|\braket{E_0(\lambda)}{\psi_0}|^2$ between the initial state $\ket{\psi_0}$ and the ground state, and the total evolution time, i.e., sum of all $t_J$, scales as $\mathcal{O}(\tau d\log d / \eta^2)$~\cite{Lin2022, Wan2022, Zhang2022}. The total evolution time, as well as the maximum evolution time, is proportional to $\epsilon_{\rm E}^{-1}$ [see Eq.~\eqref{eq:scaling-truncation-order}], which achieves the Heisenberg-limited scaling.

The ground state energy $E_0$ is finally estimated by finding the smallest $x$ for which the estimated ACDF $\tilde{C}(x)$ exceeds a certain threshold. This is typically achieved through a classical binary search algorithm, which iteratively refines an interval containing $E_0$, aiming to identify the left-most steep rise in the ACDF corresponding to the ground state.

While SPE primarily focuses on estimating the energy eigenvalue, there have been efforts to extend its applicability. For instance, Ref.~\cite{Zhang2022} proposed an extension of SPE, which generalizes the method to estimate expectation values of general physical observables for energy eigenstates.

\subsubsection{Limitations of SPE for Estimating Energy-Level Differences}
\label{subsec:limitations-existing-methods}

While SPE offers advantages in terms of quantum-resource efficiency compared to the conventional QPE, it faces an unfavorable scaling when applied to the task of estimating energy level differences with high precision. This problem stems primarily from the inherent nature of SPE as a method for estimating individual energy eigenvalues, rather than their differences, and the resulting sampling complexity of quantum circuits.

Estimating an energy level difference 
\begin{align}
\Delta E_0 = E_0 (\lambda + \delta \lambda) - E_0 (\lambda)
\end{align}
via SPE requires two separate SPE runs, one for each eigenvalue. 
Each of the two runs estimates its eigenvalue to an additive precision $\epsilon_{\rm E}$, and since the two are estimated independently, resolving their difference to a precision $\epsilon$ requires $\epsilon_{\rm E}\simeq\epsilon/2$.
This estimation is subject to a constraint on the time step $\tau$ and the maximum absolute value of energy eigenvalue, equivalent to the operator norm $\|H\|$ of the Hamiltonian ~\footnote{We note that the quantity entering Eq.~\eqref{eq:SPE-tau-constraint} is, strictly speaking, the spectral width of $H$ rather than its operator norm: shifting $H \to H - cI$ by a constant, which leaves every energy difference invariant, replaces $\|H\|$ by the spectral half-width $\tfrac{1}{2}[E_{\max}-E_{\min}] \le \|H\|$. This refinement does not alter our conclusions, since the spectral width is itself extensive; moreover, the same constant shift cancels identically in the differential constraint of our method [Eq.~\eqref{eq:eq:difference-tau-constraint-Hderivative}], so it cannot close the gap between the two approaches.}~\cite{Lin2022, Wan2022, Zhang2022}:
\begin{align}
\label{eq:SPE-tau-constraint} 
 \tau \|H\|
 \le \frac{\pi}{3}.
 \end{align}
This constraint ensures that the phase evolution remains within a range where the statistical estimation method is well-behaved and avoids issues like phase wrapping or aliasing.
Combining Eq.~\eqref{eq:SPE-tau-constraint} with the scaling in Eq.~\eqref{eq:scaling-truncation-order} yields a fundamental constraint on the truncation order:
\begin{align}
\label{eq:SPE-d-lower-bound}
d \gtrsim \frac{3}{\pi} \frac{\|H\|}{\epsilon_{\rm E}}.
\end{align}
This derived scaling relation highlights the bottleneck: the sampling complexity of quantum circuits, given by 
\begin{align} \label{eq:N_sample-SPE}
N_{\rm sample} = \mathcal{O}\qty(\frac{\log^2 d }{\eta^2}) 
= \mathcal{O}\qty(\frac{\log^2 (\|H\|/\epsilon_{\rm E}) }{\eta^2}),
\end{align}
exhibits an unfavorable logarithmic growth with the large ratio between $\|H\|$ and the single-energy precision $\epsilon_{\rm E}$, that is, with $\|H\|/\epsilon_{\rm E}$ up to a factor of order unity.
Although the logarithmic dependence partially moderates the growth, the overhead induced by the large truncation order $d \gtrsim (3/\pi)\|H\|/\epsilon_{\rm E}$ remains a nonnegligible burden in practice, particularly for large-scale systems where $\|H\|/\epsilon_{\rm E} \gg 1$.

\subsection{Other Existing Methods for Direct Energy Difference Estimation}

Beyond estimating individual energy eigenvalues, another class of quantum algorithms directly aims to determine energy-level differences. Among these, Bayesian Phase Difference Estimation (BPDE)~\cite{Sugisaki2021, Sugisaki2022} leverages Bayesian inference and quantum circuits that encode an energy difference directly in a relative phase (a detailed overview is provided in Appendix~\ref{app:bpde_overview}). 

While potentially valuable for its direct approach to differences, BPDE faces two limitations when targeting early-FTQC devices.
First, the BPDE literature~\cite{Sugisaki2021, Sugisaki2022} does not rigorously guarantee or demonstrate Heisenberg-limited scaling for general Hamiltonian systems or initial states.
Discussions often remain at a qualitative level, noting trade-offs between precision and cost or requiring ``sufficiently large overlap" in initial states without explicitly defining these bounds. Second, BPDE requires iterative Bayesian optimization of the phase shift parameter $\Delta\epsilon$ in the gate $P(\Delta\epsilon \cdot t)$ (see Appendix~\ref{app:bpde_overview} for details), incurring an additional overhead from the number of optimization iterations required for convergence.

These limitations of BPDE highlight the need for a more robust and hardware-compatible approach for high-precision energy-level difference estimation, especially when targeting early-FTQC devices. Our proposed method, while inspired by the resource efficiency of SPE, rethinks the estimation strategy to overcome both the scaling bottleneck of SPE for energy differences and the practical implementation hurdles of methods like BPDE, providing a path towards more feasible quantum advantage in chemistry and condensed matter physics.

\section{Proposed Method: Direct Estimation of Energy-Level Differences}\label{sec:proposed-method}

In this section, we introduce our quantum algorithm for the efficient estimation of energy level differences in quantum systems. Unlike conventional approaches that either estimate individual energy levels and then compute their differences or directly estimate differences while relying on adaptive, iteratively optimized circuits, e.g., BPDE, we directly target their differences and lower the sampling complexity of quantum circuits, which is the source of its efficiency on early-FTQC devices. 
This addresses the unfavorable scaling of the required truncation order, and consequently the sampling complexity, with the magnitude of the Hamiltonian's operator norm (or, more precisely, with the large ratio between its operator norm and the target energy difference precision), as discussed in Sec. \ref{subsec:limitations-existing-methods}.

\subsection{Core Principle and Algorithmic Steps}\label{subsec:core-principle}

The central idea of our proposed method is to directly estimate the energy level difference $\Delta E_k = E_k(\lambda+\delta\lambda) - E_k(\lambda)$ for a Hamiltonian $H(\lambda)$ and its perturbed counterpart $H(\lambda+\delta\lambda)$, where $\delta\lambda$ represents a small parameter shift. This direct approach offers advantages over indirect methods.

To efficiently estimate the energy-level difference $\Delta E_k = E_k(\lambda+\delta\lambda) - E_k(\lambda)$, our method constructs a quantum circuit whose output directly encodes information about the energy difference, rather than individual energy values. This is achieved through an approach involving a specifically designed quantum circuit and refined classical post-processing, detailed in the following subsections.

\subsubsection{Quantum Circuit}\label{subsubsec:quantum-circuit}

Our proposed method leverages a modified form of the Hadamard test to directly probe the energy difference. Instead of evolving under a single Hamiltonian $H$, our circuit involves evolutions under both $H(\lambda)$ and $H(\lambda+\delta\lambda)$.

The core of the quantum computation involves estimating the expectation value:
\begin{align} \label{eq:simple-propagator-before-expansion}
\mathcal{G}(j\tau) = \langle\psi_0| e^{iH(\lambda+\delta\lambda)j\tau} e^{-iH(\lambda)j\tau} |\psi_0\rangle,
\end{align}
where $j\in\mathcal{S}$, $\tau$ is a selected time step, and $|\psi_0\rangle$ is an initial state, typically chosen to have significant overlap with the ground states of both $H(\lambda)$ and $H(\lambda+\delta\lambda)$.

Let $E_m(\lambda)$ and $E_n(\lambda+\delta\lambda)$ be the eigenvalues of $H(\lambda)$ and $H(\lambda+\delta\lambda)$ corresponding to their respective eigenstates $\ket{E_m(\lambda)}$ and $\ket{E_n(\lambda+\delta\lambda)}$. The expectation value $\mathcal{G}(j\tau)$ can be expanded as
\begin{align}
\label{eq:simple-propagator-expansion}
\mathcal{G}(j\tau) = \sum_{m,n} &e^{-ij\tau[ E_m(\lambda) -  E_n(\lambda+\delta\lambda)]} 
\langle\psi_0|E_n(\lambda+\delta\lambda)\rangle
\nonumber\\
&\times\langle E_n(\lambda+\delta\lambda)|E_m(\lambda)\rangle\langle E_m(\lambda)|\psi_0\rangle.
\end{align}

\begin{figure}[tbp]
\centering
\[
    	\Qcircuit @C=1em @R=.7em {\\
  	\lstick{\ket{0}} & \gate{H} & \ctrl{1} & \ctrlo{1} 
    &\gate{W}
    & \gate{H}      & \meter \\
  	\lstick{\ket{\psi_0}} & \qw      & \gate{ e^{-i  j\tau H(\lambda)}} & \gate{ e^{-i  j\tau H(\lambda+\delta\lambda)}} &\qw  & \qw &\qw 
	}
    \]
    \caption{
    Quantum circuit for our proposed method for energy-difference estimation. The gate $W$ is set to $W=I$ ($W=S^\dag$) for estimating the real (imaginary) part of the expectation value of the time-evolution operator. The controlled time evolutions are implemented by the $j$-times iterations of the time evolution with a fixed time step $\tau$.
    The two controlled evolutions are drawn separately for clarity; their merged implementation is discussed in Sec.~\ref{sec:resource-estimation} and Appendix~\ref{app:merged-rotation}.
    }
    \label{fig:circuit-proposed-simple}
\end{figure}
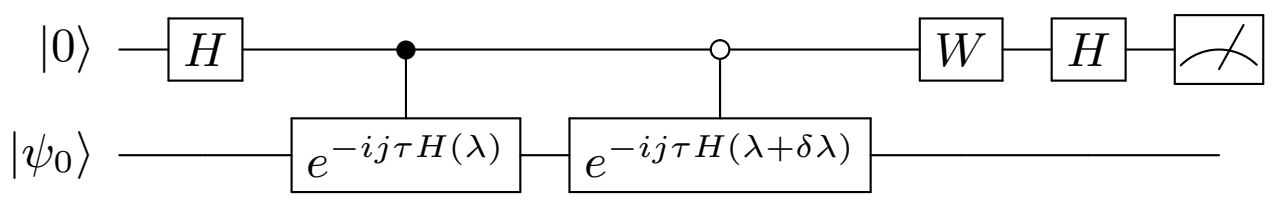

The circuit for estimating $\mathrm{Re}[\mathcal{G}(j\tau)]$ and $\mathrm{Im}[\mathcal{G}(j\tau)]$ is presented in Fig.~\ref{fig:circuit-proposed-simple}, which is constructed as a variant of the Hadamard test. 
It performs an initial Hadamard gate on the ancilla, followed by two distinct controlled time-evolution operations, $e^{-iH(\lambda)j\tau}$ and $e^{-iH(\lambda+\delta\lambda)j\tau}$, conditioned on the ancilla being $\ket{1}$ and $\ket{0}$, respectively, whose interference realizes the composite propagator of Eq.~\eqref{eq:simple-propagator-before-expansion}, and concludes with a final Hadamard gate before the ancilla is measured.
The probability of measuring $|0\rangle$ is 
\begin{align}
P_0 = \frac{1}{2}(1 + \mathrm{Re}[\langle\psi_0| e^{iH(\lambda+\delta\lambda)j\tau} e^{-iH(\lambda)j\tau}|\psi_0\rangle]).
\end{align}
From this, $\mathrm{Re}[\mathcal{G}(j\tau)]$ can be estimated as $2P_0 - 1$. Similarly, by inserting a phase gate $S^\dagger$ on the ancilla just before the second Hadamard gate, i.e., $W=S^\dagger$ in Fig.~\ref{fig:circuit-proposed-simple}, $\mathrm{Im}[\mathcal{G}(j\tau)]$ can be estimated.
In this circuit design, the maximum depth scales directly with the desired precision for the energy difference, which will be discussed in Sec.~\ref{subsec:advantages-over-existing}.
The circuit is logically equivalent to that of Ref.~\cite{Sugisaki2022} apart from the ancilla phase gate $P(\Delta\epsilon\,t)$, whose angle BPDE optimises by Bayesian inference; our circuit needs no such gate, and the evolution time is $j\tau$ with a fixed $\tau$ and an integer $j$ drawn from the fixed distribution $|\hat F_j|$ of Eq.~\eqref{eq:G_estimator}, so the whole set of circuits is fixed before any measurement.
The substance of our proposal therefore lies in the classical post-processing of Sec.~\ref{subsubsec:classical-postprocessing} and in the choice of $\tau$ that it permits [Eq.~\eqref{eq:eq:difference-tau-constraint-Hderivative}].

\subsubsection{Classical Post-processing}\label{subsubsec:classical-postprocessing}

Similar to SPE, our method relies on sophisticated classical post-processing to extract the energy difference from the measured expectation values $\mathcal{G}(j\tau)$. As discussed above, we measure this quantity for a range of discrete time points $t_j = j\tau$, where $j$ takes integer values within $\mathcal{S}$. These values serve as the discrete Fourier components of the energy difference spectrum.

Let the spectral measure for the energy difference be 
\begin{align}
p_{\Delta E}(x) = \sum_k q_k \delta(x + \tau\Delta E_k),
\end{align}
where $\Delta E_k = E_k(\lambda+\delta\lambda) - E_k(\lambda)$ is the exact difference, and $q_k = |\braket{E_k (\lambda)}{\psi_0}|^2$ is the weight defined as the initial state's squared overlap with the target eigenstate. The expectation values $\mathcal{G}(j\tau)$ are related to the Fourier transform of this spectral measure.

We then define a CDF for the energy difference, $C_{\Delta E}(x)$, and approximate it with an ACDF, $\tilde{C}_{\Delta E}(x)$, using an approximated Heaviside function truncated at order $d$, $F_{\beta, d} (x)$, whose Fourier coefficients are $\hat{F}_j$. The ACDF is obtained through a discrete Fourier transform:
\begin{align} \label{eq:acdf-estimator}
\tilde{C}_{\Delta E}(x) = \sum_{j\in\mathcal{S}} \hat{F}_j e^{ijx} \mathcal{G}(j\tau).
\end{align}
This quantity can be written as [see Eq.~\eqref{eq:simple-propagator-expansion}]
\begin{align}
&\hspace{-5mm}\tilde{C}_{\Delta E}(x) \nonumber\\
= &\sum_{m,n} P_{mn} (F_{\beta, d} * \delta)(x - \tau[E_m(\lambda) - E_n(\lambda+\delta\lambda)]),
\end{align}
where $P_{m n} = \langle\psi_0|E_n(\lambda+\delta\lambda)\rangle \nonumber\langle E_n(\lambda+\delta\lambda)|E_m(\lambda)\rangle\langle E_m(\lambda)|\psi_0\rangle$.
For $\delta\lambda \ll 1$, and assuming the relevant eigenstates are nondegenerate, Rayleigh--Schr\"odinger perturbation theory gives the diagonal weights $P_{nn}=|\langle E_n(\lambda)|\psi_0\rangle|^2 + O(\delta\lambda)$ while the off-diagonal weights are suppressed linearly, $P_{mn}=O(\delta\lambda)$ for $m\neq n$. 
This linear suppression originates from the first-order overlap $\langle E_n(\lambda+\delta\lambda)|E_m(\lambda)\rangle \propto \delta\lambda\,\langle E_n(\lambda)|\partial_\lambda H|E_m(\lambda)\rangle/[E_n(\lambda)-E_m(\lambda)]$ $(m\neq n)$, whose energy denominator we take to be non-vanishing; the (near-)degenerate case is discussed below.
Retaining the dominant diagonal contributions, the ACDF reads
\begin{align}
\label{eq:acdf-diff-perturbed}
&\tilde{C}_{\Delta E}(x) \nonumber\\ & = \sum_{k} |\langle E_k(\lambda)|\psi_0\rangle|^2 (F_{\beta, d} * \delta)(x - \tau[E_k(\lambda) - E_k(\lambda+\delta\lambda)]) \nonumber\\
& \hspace{5mm} + \mathcal{O}\qty(\delta \lambda).
\end{align}
The value of $x$ at which $\tilde{C}_{\Delta E}(x)$ shows a sharp increase, commonly referred to as a {\it jump} in the ideal CDF and a {\it steep rise} in the smoothed ACDF, is located at $x_k=\tau[E_k(\lambda) - E_k(\lambda+\delta\lambda)]= -\tau\,\Delta E_k$, from which the energy level difference $\Delta E_k = E_k(\lambda+\delta\lambda) - E_k(\lambda)$ is obtained as $\Delta E_k = -x_k/\tau$.

The neglected off-diagonal terms, although formally of order $\delta\lambda$ rather than $\delta\lambda^2$, do not bias the estimated difference. Each off-diagonal term $(m\neq n)$ in Eq.~\eqref{eq:simple-propagator-expansion} contributes a step at the position $\tau[E_m(\lambda)-E_n(\lambda+\delta\lambda)]$, i.e., near a genuine level gap of $H(\lambda)$ rather than at the target position $-\tau\Delta E_k$, and carries a weight $P_{mn}=\mathcal{O}(\delta\lambda)$ that is smaller than the target weight $P_{kk}=\mathcal{O}(1)$ by a factor $\mathcal{O}(\delta\lambda)$. Generically it therefore adds only a small, well-separated step elsewhere in the ACDF and leaves the target rise, and hence the estimated $\Delta E_k$, unaffected.
The one situation requiring care, where the relaxed time step of our method can bring such a step onto the target strip, and where the same weight suppression still protects the estimate, will be analyzed in Sec.~\ref{subsec:advantages-over-existing}. When $|\psi_0\rangle$ is dominantly supported on a single eigenstate, $|\langle E_k(\lambda)|\psi_0\rangle|^2\approx 1$, the off-diagonal weights are further suppressed by the small overlaps $\langle E_m(\lambda)|\psi_0\rangle$ for $m \ne k$.

The reduction to the diagonal contributions in Eq.~\eqref{eq:acdf-diff-perturbed} presumes that the target level is nondegenerate and separated from the rest of the spectrum by a gap resolved by the chosen time step. The (near-)degenerate case, where this assumption fails, is discussed in Sec.~\ref{subsec:advantages-over-existing} and handled by quasi-degenerate perturbation theory together with the joint estimation of Sec.~\ref{subsec:joint-estimation}. Unless stated otherwise, we assume throughout that the target level is nondegenerate and gapped.

The classical post-processing steps are analogous to SPE, but specifically applied to the difference domain:
\begin{enumerate}
    \item \textbf{Sampling $\mathcal{G}(j\tau)$:} For various $j \in \mathcal{S}$, estimate the real and imaginary parts of $\mathcal{G}(j\tau)$ using the quantum circuit described in Fig.~\ref{fig:circuit-proposed-simple}. Each estimation requires multiple measurement shots.
    Here, employing a statistical sampling of $j$, rather than computing every term in the summation, leads to the Heisenberg-limited scaling (see Sec.~\ref{subsec:advantages-over-existing} for details), as in the case of the SPE.
    \item \textbf{Constructing ACDF for Difference:} Use the sampled $\mathcal{G}(j\tau)$ values as Fourier coefficients to construct the approximate CDF for the energy difference, $\tilde{C}_{\Delta E}(x)$.
    \item \textbf{Estimating $\Delta E$:} A classical binary search, or peak detection on the derivative of $\tilde{C}_{\Delta E}(x)$, locates the notable  steep rises of the staircase-like ACDF; a rise at position $x_k$ yields the energy-level difference $\Delta E_k = -x_k/\tau$, weighted by the overlap $q_k = |\langle E_k(\lambda)|\psi_0\rangle|^2$. This search algorithm iteratively narrows down an interval until the desired precision is achieved.
\end{enumerate}
In contrast to ground-state \emph{energy} estimation, where the left-most jump necessarily marks the ground state because $E_0$ is the smallest eigenvalue, the differences $\{\Delta E_k\}$ are not ordered according to the energy ordering, so the left-most rise does not in general correspond to $\Delta E_0$. The ground-state difference is therefore identified by weight rather than position: preparing $|\psi_0\rangle$ with dominant overlap on the ground state, i.e., $|\langle E_0(\lambda)|\psi_0\rangle|^2 \approx 1$, which is the standard assumption in SPE, makes the single dominant-weight rise encode $\Delta E_0$, and an analogous preparation with dominant overlap on a chosen eigenstate $|E_k(\lambda)\rangle$ allows the corresponding $\Delta E_k$ to be estimated selectively. When $|\psi_0\rangle$ has appreciable overlap with several eigenstates, the rises cannot be assigned to individual levels from the difference spectrum alone; the joint estimation introduced in Sec.~\ref{subsec:joint-estimation}, which pairs each $\Delta E_k$ with its eigenvalue $E_k(\lambda)$, resolves this ambiguity and identifies the ground-state difference as the pair with the smallest $E_k(\lambda)$.

\begin{table*}[htbp]
\renewcommand{\arraystretch}{1.3}
\caption{Comparison between SPE-based energy-difference estimation and our Proposed Method.}
    \label{tab:comparison-proposal-SPE}
    \centering
    \begin{tabular}{|l|c|c|}
        \hline
        \textbf{Feature} & \textbf{SPE-based Difference Estimation} & \textbf{Proposed Method} \\
        \hline
        Quantity Estimated & Individual $E(\lambda)$ and $E(\lambda+\delta\lambda)$ & Direct $\Delta E = E(\lambda+\delta\lambda) - E(\lambda)$ \\
        \hline
        Quantum Circuit & Hadamard Test for single time-evolution  & Hadamard Test for both time-evolution  \\
        & operator (Fig.~\ref{fig:circuit-SPE}) & operators (Fig.~\ref{fig:circuit-proposed-simple})\\
        \hline
        \begin{tabular}{@{}l@{}} Required \\ truncation order $d$ \end{tabular} & $\displaystyle \mathcal{O} \qty(\frac{\|H\|}{\epsilon})$ & $\displaystyle \mathcal{O} \qty(\left\|\dv{H(\lambda)}{\lambda}\right\| \frac{\delta\lambda}{\epsilon} )$ \\
        \hline
        \begin{tabular}{@{}l@{}} Sample complexity \\ of quantum circuits $N_{\rm sample}$
        \end{tabular} & $\displaystyle \mathcal{O}\qty(\frac{\log^2 \qty[ \|H\|/\epsilon] }{ \eta^2})$ & $\displaystyle \mathcal{O}\qty(\frac{1 }{ \eta^2} \log^2 \qty[\left\|\dv{H (\lambda)}{\lambda} \right\| \frac{\delta \lambda}{\epsilon}])$ \\
        \hline
\begin{tabular}{@{}l@{}} Maximum evolution \\ time $T_{\max}=(2d+1)\tau$ \end{tabular} 
& $\displaystyle \mathcal{O}\qty(\frac{1}{\epsilon})$ 
& $\displaystyle \mathcal{O}\qty(\frac{1}{\epsilon})$ \\
        \hline
        Classical Post-processing & Calculating ACDF of energy eigenvalue & Calculating ACDF of energy difference \\
        \hline
    \end{tabular}
\end{table*}

\subsection{Advantages over Existing Methods}
\label{subsec:advantages-over-existing}

The primary advantage of our proposed method lies in its favorable scaling for estimating energy differences.
To clearly demonstrate this, we provide an explicit performance comparison between our approach and existing algorithms, particularly focusing on the SPE framework as applied to the estimation of energy-level differences.

\subsubsection{Reduced truncation order and sampling complexity} \label{subsubsec:reduced-d-Nsample}
As discussed in Sec.~\ref{subsec:limitations-existing-methods}, conventional SPE faces an unfavorable scaling when applied to energy-difference estimation. To achieve a target precision $\epsilon$ on the difference, SPE must estimate the two energy eigenvalues $E(\lambda)$ and $E(\lambda+\delta\lambda)$ separately, each to $\epsilon_{\rm E}\simeq\epsilon/2$. Since the required truncation order for SPE scales as $d = \mathcal{O}(\|H\|/\epsilon_{\rm E})$, the growth of $\|H\|$ with system size renders the ratio $\|H\|/\epsilon_{\rm E}$ unfavorably large, leading to an unfavorable growth in $d$. 
This, in turn, leads to a further increase in the sampling complexity of the quantum circuits, which scales as $\mathcal{O}(\log^2 d / \eta^2)$.

In contrast, our proposed method directly targets $\Delta E_0$, leading to the favorable scaling of the truncation order $d$.
This advantageous scaling arises from a difference in how time evolution parameters are constrained compared to the SPE. The output of our direct estimation is the phase $\tau \Delta E_k (\lambda)$, allowing us to judiciously choose $\tau$ such that
\begin{align} \label{eq:difference-tau-constraint-DeltaE}
\max_k\qty|\tau \Delta E_k (\lambda)| \le \frac{\pi}{3}.
\end{align}
By Weyl's inequality, the left-hand side of this inequality is bounded as
\begin{align}
\max_k\qty|\tau \Delta E_k|
&\le \tau \| H(\lambda+\delta \lambda) - H(\lambda) \| \nonumber\\
&\le \tau \sup_{\lambda'\in[\lambda,\lambda+\delta\lambda]}
\left\|\dv{H (\lambda')}{\lambda'} \right\| \delta\lambda ,
\end{align}
which reduces to $\tau\|\dv*{H}{\lambda}\|\,\delta\lambda$ exactly when $\lambda$ enters $H$ linearly, as for the hopping and interaction strengths $h$ and $U$ or an external field.
Hence a \emph{sufficient} condition for Eq.~\eqref{eq:difference-tau-constraint-DeltaE} is
\begin{align} \label{eq:eq:difference-tau-constraint-Hderivative}
\tau \left\|\dv{H (\lambda)}{\lambda} \right\| \delta\lambda
\lesssim \frac{\pi}{3}.
\end{align}
Because the ACDF constructed in Sec.~\ref{subsubsec:classical-postprocessing} resolves the difference spectrum rather than the energy spectrum, Eq.~\eqref{eq:scaling-truncation-order} applies with the precision $\epsilon$ on $\Delta E$ in place of the single-energy precision $\epsilon_{\rm E}$.
Combining this inequality with Eq.~\eqref{eq:scaling-truncation-order} then yields the following scaling for the truncation order $d$ of the ACDF:
\begin{align}\label{eq:diff-d-lower-bound}
d \gtrsim \frac{3}{\pi} \left\|\dv{H (\lambda)}{\lambda} \right\| \frac{\delta \lambda}{\epsilon}.
\end{align}
Here and in Table~\ref{tab:comparison-proposal-SPE}, we write $\epsilon$ for the precision resolved by the ACDF, which coincides with the target precision when the whole budget is assigned to the algorithmic error; when a time-evolution implementation error is budgeted separately, as in the resource estimates of Sec.~\ref{sec:resource-estimation}, $\epsilon$ is to be read as $\epsilon_{\rm alg}<\epsilon$.
This result is in sharp contrast to the scaling of the conventional SPE, where $d \gtrsim (3/\pi) \|H\|/\epsilon_{\rm E}$, as shown in Eq.~\eqref{eq:SPE-d-lower-bound}.
In the lattice models considered in Secs.~\ref{sec:numerical-demonstration} and \ref{sec:resource-estimation}, where $\lambda$ is a global parameter such as an external field or a coupling constant that acts identically on every site or bond (e.g., the hopping amplitude $h$), both $\|H\|$ and $\left\| \dv*{H(\lambda)}{\lambda} \right\|$ are extensive quantities, scaling linearly with the system size. For molecular Hamiltonians, the growth is not linear in general, and the two norms scale differently; this case is examined below.

The advantage of our method lies in its dependence on the small parameter $\delta\lambda$.
While the SPE requires a massive $d$ due to the large magnitude of $\|H\|$ compared with $\epsilon$, our approach suppresses the scaling of $d$ through the factor $\delta\lambda$, resulting in a manageable ratio $\delta\lambda/\epsilon$.
The reduction in $d$ directly benefits the sampling complexity of the quantum circuits, which scales as [see also Eq.~\eqref{eq:N_sample-SPE}]
\begin{align} \label{eq:N_sample_difference}
N_{\rm sample} = \mathcal{O}\qty(\frac{\log^2 d }{ \eta^2})
= \mathcal{O}\qty(\frac{1 }{ \eta^2} \log^2 \qty[\left\|\dv{H (\lambda)}{\lambda} \right\| \frac{\delta \lambda}{\epsilon}]).
\end{align}
The reduction in $d$ should not be read as a shallower circuit. Because $d=\mathcal{O}(1/\tau\epsilon)$ [Eq.~\eqref{eq:scaling-truncation-order}], the smaller $d$ of our method is offset by a proportionally larger time step $\tau$, so that the maximum evolution time $T_{\max}=(2d+1)\tau=\mathcal{O}(1/\epsilon)$ is fixed by the target precision alone and coincides with the Heisenberg-limited value of SPE. What our method reduces is thus not the depth of any single circuit but the number of repetitions, through $N_{\rm sample}\propto\log^2 d$ and the avoidance of two separate eigenvalue estimations, and hence the total \emph{serial} runtime. This is why the $T_{\max}$ row of Table~\ref{tab:comparison-proposal-SPE} is common to both methods while the sampling complexity, and with it the serial runtime, differs.

This advantage is largest in the small-shift regime $\|dH/d\lambda\|\,\delta\lambda \ll \|H\|$, which is precisely that of \emph{gradient} estimation (see Sec.~\ref{subsubsec:budget-gradient}): molecular forces and other response properties follow from the derivative $\dv*{E}{\lambda}$, accessed through the finite difference $\Delta E = E(\lambda+\delta\lambda)-E(\lambda)\approx(\dv*{E}{\lambda})\,\delta\lambda$. For a large parameter shift, i.e., $\delta\lambda$ comparable to the intrinsic scale of $\lambda$ itself, so that $H(\lambda+\delta\lambda)$ is no longer a perturbation of $H(\lambda)$, the product $\|dH/d\lambda\|\,\delta\lambda$ becomes comparable to $\|H\|$ and the gain over SPE diminishes. This gradient regime is the practically important one, underlying geometry optimization, ab initio molecular dynamics, and response-property calculations.

To give a concrete sense of the scaling advantage discussed above, i.e., the reduction in $d$ set by $\|\partial H/\partial\lambda\|\,\delta\lambda$ rather than $\|H\|$, we consider two representative examples.
In each example the single parameter $\lambda$ of the analysis above is identified with a specific physical parameter on which the Hamiltonian depends, and the total derivative $dH/d\lambda$ is accordingly written as a partial derivative with the remaining parameters held fixed.
Since the operator norm is generally hard to compute for large systems, we use the Pauli $L_1$-norm as a convenient, computable proxy: for an operator $A=\sum_k a_k P_k$ expanded in Pauli strings $\{P_k\}$ with real coefficients $a_k$, its Pauli $L_1$-norm $\Lambda_A \equiv \sum_k |a_k|$ upper-bounds the operator norm, $\|A\| \le \Lambda_A$, and in particular $\|H\| \le \Lambda_H$. 
We stress that the truncation order $d$ is set by the operator norms themselves [Eqs.~\eqref{eq:SPE-d-lower-bound} and \eqref{eq:diff-d-lower-bound}], both of which grow with system size as noted above; the $L_1$-norm ratios below therefore serve as order-of-magnitude indicators rather than exact prefactors. 
First, for the 2D Fermi-Hubbard model $H(h,U)$ studied in Sec.~\ref{sec:resource-estimation} with $h=1$, $U=4$, and an $N_{\rm size}\times N_{\rm size}$ lattice, we take $\lambda=h$, i.e., shifting the hopping at fixed $U$.
The $L_1$-norm scales as $\Lambda_H = 5N_{\rm size}^2$, while the corresponding quantity for the derivative satisfies $\Lambda_{\partial H/\partial h}\,\delta h = 4 N_{\rm size}^2 \,\delta h = 0.08\,N_{\rm size}^2$ for a representative shift $\delta h = 0.02$, adopted in our numerical demonstration (Sec.~\ref{sec:numerical-demonstration}), yielding a ratio of approximately $62.5$ for this fixed shift.
Second, for molecular force estimation in quantum chemistry, the molecular Hamiltonian $H(\{R_i\})$ depends on the nuclear coordinates $\{R_i\}$, and the relevant parameter is a single such coordinate, $\lambda = R_i$.
O'Brien \textit{et al.}~\cite{OBrien2022} numerically evaluated the $L_1$-norms of the Hamiltonian and force operators $\partial H/\partial R_i$, as block-encoding rescaling factors in localized molecular orbitals, for hydrogen chains of up to $N_H = 100$ atoms using STO-6G basis set~\cite{Hehre1969} (Fig.~4 therein).
They found that while $\Lambda_H^{(\rm sparse)} \sim \mathcal{O}(N_H^{1.33})$, the mean $L_1$-norm of the force operators scales as $\Lambda_F^{(\rm sparse)} \sim \mathcal{O}(N_H^{0.28})$, tending toward a constant for large $N_H$, with qualitatively similar results for water clusters.
This stark difference in scaling suggests that in molecular systems the reduction in $d$ of our method over SPE-based approaches is expected to grow with system size.

\subsubsection{Robustness of the direct-difference estimate} \label{subsubsec:robustness}
The relaxed time step fixed by Eq.~\eqref{eq:eq:difference-tau-constraint-Hderivative} warrants a comment on the validity of the perturbative reduction of Sec.~\ref{subsubsec:classical-postprocessing}. An off-diagonal term $(m\neq n)$ in Eq.~\eqref{eq:simple-propagator-expansion} contributes a step to the ACDF at the scaled position $\tau[E_m(\lambda)-E_n(\lambda+\delta\lambda)]=\tau[E_m(\lambda)-E_n(\lambda)]+\mathcal{O}(\tau\delta\lambda)$, i.e., near a genuine level gap of $H(\lambda)$ rather than at the target position $-\tau\Delta E_k=\mathcal{O}(\tau\delta\lambda)$. In the conventional SPE regime $\tau\|H\|\lesssim\pi/3$ the two are positionally well separated whenever the level gap exceeds the spectral resolution, so the off-diagonal term merely adds a small, distinct step elsewhere and leaves the target rise untouched. In our method, by contrast, $\tau$ is set by $\|\dv*{H}{\lambda}\|$ rather than by $\|H\|$ [Eq.~\eqref{eq:eq:difference-tau-constraint-Hderivative}], so that $\tau\|H\|$ can exceed $\pi$ and the gap phase may wrap modulo $2\pi$; when $\tau[E_m(\lambda)-E_n(\lambda)]\equiv-\tau\Delta E_k \pmod{2\pi}$, a rise in the ACDF originating from the off-diagonal term with $m \ne n$ is aliased onto the target strip and can, in principle, displace the located target peak. Only in this aliased case does the estimate of $\Delta E_k$ acquire a bias at all.

What bounds this bias is not positional separation alone but also the weight. The off-diagonal amplitude $P_{mn}=\mathcal{O}(\delta\lambda)$ is smaller than the target weight $P_{kk}=\mathcal{O}(1)$ by a factor $\mathcal{O}(\delta\lambda)$, so an aliased term pulls the located peak toward itself by that fraction of its separation from the target rise. Aliasing matters only when the off-diagonal rise falls inside the target strip, whose extent is set by the target positions themselves, $|x_k|=\tau|\Delta E_k|=\mathcal{O}(\tau\delta\lambda)$, so that separation is at most $\mathcal{O}(\tau\delta\lambda)$. The two factors multiply: the located position is displaced by $\mathcal{O}(\tau\delta\lambda^2)$, i.e., the estimated difference $\Delta E_k$ is shifted by $\mathcal{O}(\delta\lambda^2)$ and the extracted gradient $\Delta E_k/\delta\lambda$ by $\mathcal{O}(\delta\lambda)$.
This bias is therefore one order higher in $\delta\lambda$ than the target signal $\Delta E_k\sim\langle\partial_\lambda H\rangle\,\delta\lambda$ itself, and vanishes as $\delta\lambda\to0$.

When $|\psi_0\rangle$ is dominantly supported on a single eigenstate, $|\langle E_k(\lambda)|\psi_0\rangle|^2\approx1$, the off-diagonal weights are further suppressed by the small overlaps $\langle E_m(\lambda)|\psi_0\rangle$ for $m \ne k$, and their \emph{total} is, by an exact sum rule, second order in $\delta\lambda$ rather than first. Setting $|\psi_0\rangle=|E_k(\lambda)\rangle$ in Eq.~\eqref{eq:simple-propagator-expansion} gives $P_{kn}=|\langle E_n(\lambda+\delta\lambda)|E_k(\lambda)\rangle|^{2}\ge0$ and hence
\begin{align}\label{eq:offdiag-sum-rule}
\sum_{n\neq k}P_{kn}
&= 1-\bigl|\langle E_k(\lambda+\delta\lambda)|E_k(\lambda)\rangle\bigr|^{2} \nonumber\\
&= \chi_F^{(k)}\,\delta\lambda^{2}
+\mathcal{O}(\delta\lambda^{3}),
\end{align}
where $\chi_F^{(k)}=\sum_{n\neq k}|\langle E_n(\lambda)|\partial_\lambda H|E_k(\lambda)\rangle|^{2}/[E_n(\lambda)-E_k(\lambda)]^{2}$ is the fidelity susceptibility of the target level. 
This aggregate statement is what the per-term bound $P_{mn}=\mathcal{O}(\delta\lambda)$ does not by itself supply: Eq.~\eqref{eq:offdiag-sum-rule} shows that it is the single combination $\chi_F^{(k)}\delta\lambda^{2}$, and not the number of terms, that governs the total leakage; how it is bounded, and why the choice of $\delta\lambda$ made below already controls it, is taken up in Sec.~\ref{subsubsec:budget-gradient}.

Both the positional-separation and the weight-suppression arguments presume that the target level is separated from the rest of the spectrum by a gap $\Delta_{\rm gap}$ resolved by the chosen time step, i.e., $\tau\Delta_{\rm gap}$ exceeds the spectral resolution $\sim1/d$. In the presence of a (near-)degeneracy, $E_n(\lambda)\approx E_m(\lambda)$, this picture breaks down in two ways: the first-order overlap, $\langle E_n(\lambda+\delta\lambda)|E_m(\lambda)\rangle \propto \delta\lambda\,\langle E_n(\lambda)|\partial_\lambda H|E_m(\lambda)\rangle/[E_n(\lambda)-E_m(\lambda)]$, acquires a small energy denominator, so the corresponding off-diagonal weight is no longer $\mathcal{O}(\delta\lambda)$, and the associated off-diagonal step, located near $\tau[E_m(\lambda)-E_n(\lambda)]$, approaches the cluster of target peaks at $-\tau\Delta E_k=\mathcal{O}(\tau\delta\lambda)$ and can no longer be resolved from them. Such (near-)degeneracies arise generically from symmetry and at conical intersections in molecules, and near quantum critical points in lattice models.
They are handled by replacing the nondegenerate perturbation theory of Eq.~\eqref{eq:acdf-diff-perturbed} with its quasi-degenerate counterpart, treating the near-degenerate manifold as a block and diagonalizing $\partial_\lambda H$ within it. The resulting cluster of closely spaced differences is then disentangled by the joint estimation of Sec.~\ref{subsec:joint-estimation}, which labels each difference by its energy $E_k(\lambda)$ and so separates cluster members whose energies remain distinct.

\subsubsection{Gradient estimation: error budget and choice of $\delta\lambda$} \label{subsubsec:budget-gradient}
So far the precision $\epsilon$ on the difference $\Delta E$ and the shift $\delta\lambda$ have both been treated as given, and the costs, such as the time step $\tau$, the truncation order $d$, and the sampling complexity, have been derived from them [Eqs.~\eqref{eq:eq:difference-tau-constraint-Hderivative}--\eqref{eq:N_sample_difference}]. 
A gradient calculation is specified differently. What is physically prescribed is the precision $\epsilon_{\rm g}$ on the derivative $\dv*{E}{\lambda}$, e.g.\ the accuracy required of a molecular force, whereas $\delta\lambda$ is not a physical input but a free parameter of the algorithm, and the precision demanded of $\Delta E$ follows from the two. This subsection therefore reverses the logic: we ask how the total gradient error decomposes into contributions with different $\delta\lambda$ dependence, and hence how $\delta\lambda$ should be chosen, and at what cost, to reach a given $\epsilon_{\rm g}$. 

The total error in estimating the gradient may be decomposed as
\begin{align}\label{eq:error-budget-gradient}
\epsilon_{\rm g} = \frac{\epsilon_{\rm alg} + \epsilon_{\rm TE}}{\delta\lambda} + \epsilon_{\rm fd}.
\end{align}
Here $\epsilon_{\rm alg}$ and $\epsilon_{\rm TE}$ are, respectively, the algorithmic error in analyzing the ACDF and the time-evolution implementation error. Both are errors on the measured energy difference $\Delta E$, and are therefore divided by $\delta\lambda$ to convert them into gradient errors, since the gradient is estimated as $\Delta E/\delta\lambda$.
The remaining term $\epsilon_{\rm fd}$, by contrast, is already a bias of the gradient: it arises from identifying $\Delta E/\delta\lambda$ with $\dv*{E}{\lambda}$ and so requires no such conversion. 
The error $\epsilon_{\rm TE}$ denotes the error of implementing the time-evolution operators by whatever method is chosen, e.g., Trotter product formula~\cite{Lloyd1996, Childs2021}, linear combination of unitaries (LCU)~\cite{Berry2015}, or quantum signal processing (QSP)~\cite{Low2017}, and specializes to the Trotter error $\epsilon_{\rm Trotter}$ for the Trotterized implementation adopted in Secs.~\ref{sec:numerical-demonstration} and~\ref{sec:resource-estimation}.
We adopt the linear sum above as a conservative upper bound that guarantees the target gradient precision.
Adopting the lowest-order forward finite difference $dE/d\lambda \approx [E(\lambda+\delta\lambda) - E(\lambda)]/\delta\lambda$ employed in this work, a Taylor expansion of the finite-difference bias $\epsilon_{\rm fd}$ on the \emph{gradient} gives
\begin{align}
\epsilon_{\rm fd} \sim \dfrac{1}{2}\left|\dfrac{\partial^2 E}{\partial\lambda^2}\right|\delta\lambda.
\end{align}
The second derivative can in turn be bounded without explicit knowledge of $E_0$. Standard second-order perturbation theory gives
\begin{align}
\left|\dfrac{\partial^2 E}{\partial\lambda^2}\right|\le 
\left|\langle\partial_\lambda^2 H\rangle_{E_0}\right| +
\frac{2\,\mathrm{Var}_{E_0}(\partial_\lambda H)}{\Delta_{\rm gap}},
\label{eq:fd_var_bound}
\end{align}
where $\mathrm{Var}_{E_0}(\partial_\lambda H) = \langle(\partial_\lambda H)^2\rangle_{E_0} - \langle\partial_\lambda H\rangle_{E_0} ^2$ is the ground-state variance of $\partial_\lambda H$, $\langle\partial_\lambda^2 H\rangle_{E_0}$ is the ground-state expectation of the curvature operator $\partial_\lambda^2 H$, and $\Delta_{\rm gap}$ is the energy gap between the ground state and the lowest excited state coupled to it by $\partial_\lambda H$.
For parameters that enter $H$ linearly, such as the hopping and interaction strengths $h$ and $U$ of the Fermi-Hubbard model, one has $\partial_\lambda^2 H = 0$ and the bound reduces to the variance term alone; for a nonlinear dependence, such as a nuclear coordinate in a molecular force calculation, the curvature term $\langle\partial_\lambda^2 H\rangle_{E_0}$ no longer vanishes and must be kept. In either case the bound is governed by the derivative operators and the gap rather than the operator norm $\|H\|$, so that
\begin{align}\label{eq:fd_deltalambda_var_bound}
\epsilon_{\rm fd}\lesssim \frac{1}{2}\left[\,\left|\langle\partial_\lambda^2 H\rangle_{E_0}\right| + \frac{2\,\mathrm{Var}_{E_0}(\partial_\lambda H)}{\Delta_{\rm gap}}\,\right]\delta\lambda.
\end{align}
Equation~\eqref{eq:fd_var_bound} thus bounds the admissible shift from above in terms of the derivative operators and the gap alone.

On the other hand, the algorithmic error, obtained from the constraint $\tau\|\partial H/\partial\lambda\|\delta\lambda \lesssim \pi/3$ together with the relation between $\tau$ and $d$, scales as 
\begin{align}
\epsilon_{\rm alg} \sim \left\|\pdv{ H}{\lambda}\right\|\, \frac{\delta\lambda}{d}. 
\end{align}
Converted into a gradient error, $\epsilon_{\rm alg}/\delta\lambda \sim \|\partial H/\partial\lambda\|/d$ is therefore \emph{independent of} $\delta\lambda$.
Equivalently, allocating the budget of Eq.~\eqref{eq:error-budget-gradient} in fixed proportions, say $\epsilon_{\rm alg}/\delta\lambda=\epsilon_{\rm TE}/\delta\lambda=\epsilon_{\rm fd}=\epsilon_{\rm g}/3$, gives
\begin{align} \label{eq:d-from-gradient-precision}
d \sim \frac{3\left\|\dv*{H}{\lambda}\right\|}{\epsilon_{\rm g}},
\end{align}
so that the truncation order required for a given gradient precision is fixed by $\epsilon_{\rm g}$ and the derivative norm alone, independently of $\delta\lambda$: a smaller $\epsilon_{\rm g}$ is reached simply by increasing $d$. 
The shift $\delta\lambda$ affects not $d$ but the time step $\tau\sim\pi/(3\|\partial_\lambda H\|\delta\lambda)$, and hence only the length of the circuits, as discussed below.

The same $\delta\lambda$-independence holds for the time-evolution term whenever the two evolutions are implemented by the same deterministic scheme, since the difference structure then suppresses $\epsilon_{\rm TE}$ by a factor $\mathcal{O}(\delta\lambda)$ relative to a single-energy estimate; this is established below [Eq.~\eqref{eq:trotter-cancellation}].
Both $\delta\lambda$-independent contributions, $\epsilon_{\rm alg}/\delta\lambda$ and $\epsilon_{\rm TE}/\delta\lambda$, are then balanced only against the monotonically decreasing finite-difference bias $\epsilon_{\rm fd}\propto\delta\lambda$, so reducing $\delta\lambda$ lowers the total gradient error until the two $\delta\lambda$-independent terms dominate; there is no trade-off, and $\delta\lambda$ can simply be taken sufficiently small. This contrasts sharply with central-difference gradient estimation, as analyzed by O'Brien \textit{et al.}~\cite{OBrien2022}, where each energy is estimated to a fixed precision, so that after division by $\delta\lambda$ the statistical error grows as $1/\delta\lambda$ and must be balanced against the finite-difference bias, necessitating an optimal $\delta\lambda$. In our method the resolvable energy-difference precision $\epsilon_{\rm alg}$ itself scales as $\delta\lambda$, through the larger admissible time step $\tau\sim 1/(\|\partial_\lambda H\|\delta\lambda)$, so the gradient error $\epsilon_{\rm alg}/\delta\lambda$ is $\delta\lambda$-independent and this amplification does not occur.

This also makes the comparison with SPE explicit at the level of the gradient. SPE must resolve each of the two energies separately to an absolute accuracy $\sim\epsilon_{\rm g}\delta\lambda$, requiring $d\sim\|H\|/(\epsilon_{\rm g}\delta\lambda)$, larger than Eq.~\eqref{eq:d-from-gradient-precision} by a factor $\sim(\|H\|/\|\partial_\lambda H\|)/\delta\lambda$. The finite-difference bias enters only by bounding $\delta\lambda$ from above [Eq.~\eqref{eq:fd_deltalambda_var_bound}], and therefore bounds this factor from \emph{below}. Since that bound is governed by the derivative operators and the gap, the operator norm $\|H\|$ does not re-enter through $\delta\lambda$, and the ratio $\|H\|/\|\partial_\lambda H\|$ survives intact.

We stress that the absence of an \emph{error} trade-off does not license taking $\delta\lambda$ arbitrarily small. The $\delta\lambda$-independence of $\epsilon_{\rm alg}/\delta\lambda$ relies on saturating the relaxed constraint $\tau\sim 1/(\|\partial_\lambda H\|\delta\lambda)$, so that the maximum evolution time $T_{\max}=(2d+1)\tau\sim 1/(\epsilon_{\rm g}\delta\lambda)$, and hence the error-mitigation overhead $\gamma^2$ in the partially fault-tolerant setting (Sec.~\ref{sec:resource-estimation}), grows as $\delta\lambda$ shrinks; equivalently, the signal $\Delta E\sim(\partial_\lambda E)\delta\lambda$ must be resolved to a correspondingly finer absolute precision. The trade-off is thus moved from the error budget to the cost, and $\delta\lambda$ is taken as large as the finite-difference bias permits, $\delta\lambda\lesssim 2\epsilon_{\rm g}/|\partial^2 E/\partial\lambda^2|$, i.e.\ just small enough that $\epsilon_{\rm fd}$ is subdominant rather than arbitrarily small.
The off-diagonal aliasing bias discussed earlier in Sec.~\ref{subsubsec:robustness} is of the same order as $\epsilon_{\rm fd}$, i.e., $\mathcal{O}(\delta\lambda)$ on the gradient and $\mathcal{O}(\delta\lambda^2)$ on the difference, and is therefore controlled by this same choice of $\delta\lambda$, imposing no further constraint.

The same is true of the weight that leaves the diagonal, Eq.~\eqref{eq:offdiag-sum-rule}.
For the ground state ($k=0$), bounding the energy denominators by the gap gives $\chi_F^{(k)}\le\mathrm{Var}_{E_0}(\partial_\lambda H)/\Delta_{\rm gap}^{2}$, so that the total off-diagonal weight $\chi_F^{(k)}\delta\lambda^{2}$ differs from the finite-difference bias on the difference, $\epsilon_{\rm fd}\delta\lambda\simeq\mathrm{Var}_{E_0}(\partial_\lambda H)\,\delta\lambda^{2}/\Delta_{\rm gap}$ [Eq.~\eqref{eq:fd_deltalambda_var_bound}], only by the factor $\Delta_{\rm gap}$. Whenever a fraction $f$ of the target precision is allocated to the finite-difference bias, the leaked weight is therefore $f\epsilon/\Delta_{\rm gap}$, independent of $\delta\lambda$ and of the system size: although $\chi_F^{(k)}$ is extensive in a gapped phase, holding $\epsilon_{\rm fd}\delta\lambda$ fixed forces $\delta\lambda=\mathcal{O}(N^{-1/2})$, which cancels that growth. The perturbative reduction of Sec.~\ref{subsubsec:classical-postprocessing} therefore imposes no constraint on $\delta\lambda$ beyond the one already made.

The bound in Eq.~\eqref{eq:fd_deltalambda_var_bound} further determines how the optimal shift scales with system size. 
Holding the finite-difference bias at a fixed fraction of the target precision requires $\delta\lambda \lesssim \epsilon_{\rm fd}/\big(\tfrac12|\langle\partial_\lambda^2 H\rangle_{E_0}| + \mathrm{Var}_{E_0}(\partial_\lambda H)/\Delta_{\rm gap}\big)$, which for the linearly-coupled lattice model, where $\langle\partial_\lambda^2 H\rangle_{E_0}=0$, considered in Secs.~\ref{sec:numerical-demonstration} and \ref{sec:resource-estimation} reduces to $\delta\lambda \lesssim \epsilon_{\rm fd}\,\Delta_{\rm gap}/\mathrm{Var}_{E_0}(\partial_\lambda H)$. 
For a gapped system, $\Delta_{\rm gap}=\mathcal{O}(1)$, and a global parameter coupling to an extensive operator, $\mathrm{Var}_{E_0}(\partial_\lambda H)=\mathcal{O}(N)$, so the admissible shift shrinks as $\delta\lambda=\mathcal{O}(1/N)$, where $N$ is the system size, e.g., the number of sites $N=N_{\rm size}^2$ for the lattice model in Sec.~\ref{subsec:estimation-condition}.
The extensivity of the variance, rather than the naive $\mathcal{O}(N^2)$ of $\langle(\partial_\lambda H)^2\rangle$, follows because the disconnected part cancels in the variance, leaving a sum of connected correlators that is short-ranged in a gapped ground state of a local Hamiltonian~\cite{HastingsKoma2006, NachtergaeleSims2006}.
This shrinkage does not, however, degrade the performance: since $\partial_\lambda H$ is likewise extensive, $\|\partial H/\partial\lambda\|=\mathcal{O}(N)$, the product $\|\partial H/\partial\lambda\|\,\delta\lambda$ entering Eq.~\eqref{eq:eq:difference-tau-constraint-Hderivative}, and with it the admissible time step, remains $\mathcal{O}(1)$, so that the time step need not shrink with system size; equivalently, $\delta\lambda$ is chosen so that the targeted difference $\Delta E\sim\langle\partial_\lambda H\rangle\,\delta\lambda$ stays $\mathcal{O}(1)$ and resolvable at fixed cost. 
By contrast, SPE estimates the two energies separately, and its truncation order is set by $\|H\|=\mathcal{O}(N)$ irrespective of $\delta\lambda$~\footnote{Two specifications must be distinguished. If $\epsilon_{\rm g}$ is prescribed, $\delta\lambda=\mathcal{O}(1/N)$, $d=\mathcal{O}(N/\epsilon_{\rm g})$ against $\mathcal{O}(N^{2}/\epsilon_{\rm g})$ for SPE, so the advantage in $d$ grows as $\mathcal{O}(N)$. If instead $\epsilon$ on $\Delta E$ is prescribed and $\delta\lambda$ is taken as large as the bias permits, $\delta\lambda=\mathcal{O}(N^{-1/2})$, $d=\mathcal{O}(N^{1/2}/\epsilon)$, and the delivered gradient precision $\epsilon/\delta\lambda$ degrades as $\mathcal{O}(N^{1/2})$. Table~\ref{tab:hubbard_resource_estimation_Pphys3} adopts the latter.}. 
We stress that the absolute cost of both methods is ultimately limited by the error-mitigation ceiling, i.e., the overhead $\gamma^2$ growing exponentially in $\Lambda T_{\max}=\mathcal{O}(N/\epsilon_{\rm alg})$, discussed in Sec.~\ref{subsec:resource-STAR}, so the present scheme remains restricted to the moderate-$N$ regime characteristic of early-FTQC devices.

For a local parameter, such as a single nuclear coordinate in a molecular force calculation, we do not attempt a general statement about extensivity, and instead use the empirical norms quoted in Sec.~\ref{subsubsec:reduced-d-Nsample}. Since $\mathrm{Var}_{E_0}(\partial_\lambda H)\le\|\partial_\lambda H\|^2\le (\Lambda^{(\rm sparse)}_F)^2$, the slow growth of the mean $\Lambda^{(\rm sparse)}_F\sim\mathcal{O}(N_H^{0.28})$ found for hydrogen chains~\cite{OBrien2022} suggests, neglecting the curvature term, that a shift $\delta\lambda=\mathcal{O}(N_H^{-0.56})$ suffices, so that the product $\|\partial_\lambda H\|\,\delta\lambda$ controlling the time-step constraint and the truncation order does not grow with system size, and in fact decreases as $\mathcal{O}(N_H^{-0.28})$.
We note that this analysis presumes a non-vanishing gap, which enters both through Eq.~\eqref{eq:fd_deltalambda_var_bound} and through the clustering that makes the variance extensive; gapless or critical systems, for which $\Delta_{\rm gap}\to0$ and correlations become long-ranged, would demand a faster shrinkage of $\delta\lambda$.

We now establish the $\mathcal{O}(\delta\lambda)$ suppression of $\epsilon_{\rm TE}$ asserted above. Any deterministic implementation of the time evolution, e.g., a Trotter product formula of any order~\cite{Lloyd1996, Childs2021}, an LCU expansion~\cite{Berry2015}, or a QSP circuit~\cite{Low2017}, realizes, in place of $e^{-iH(\lambda)j\tau}$, some fixed unitary $V(\lambda)$.
Since every unitary is the exponential of a Hermitian operator, $V(\lambda)$ is the exact propagator of an effective Hamiltonian,
\begin{align} \label{eq:effective-hamiltonian}
V(\lambda)=e^{-iH_{\rm eff}(\lambda)j\tau}, \qquad 
H_{\rm eff}(\lambda)=H(\lambda)+\delta H(\lambda),
\end{align}
where $\delta H(\lambda)$ is the implementation-error operator.
Provided the same scheme is applied to both Hamiltonians with its approximation parameters held fixed, the implemented unitary $V$ is a fixed function of the Hamiltonian coefficients, so that $\delta H(\lambda)$ depends smoothly on $\lambda$. 
Writing $E_k^{\rm eff}(\lambda)$ for the $k$-th eigenvalue of $H_{\rm eff}(\lambda)$, the implementation shifts each level by $\delta E^{(k)}(\lambda)\equiv E_k^{\rm eff}(\lambda)-E_k(\lambda)$, which equals $\langle E_k(\lambda)|\delta H(\lambda)|E_k(\lambda)\rangle$ to leading order in $\delta H$.
Since the algorithm estimates the difference of the resulting effective eigenvalues, the difference it locates is $\Delta E_k+[\delta E^{(k)}(\lambda+\delta\lambda)-\delta E^{(k)}(\lambda)]$, so that the per-evolution shift cancels to leading order and the residual bias is
\begin{align} \label{eq:trotter-cancellation}
\delta E^{(k)}(\lambda+\delta\lambda)-\delta E^{(k)}(\lambda) = \delta\lambda\,\partial_\lambda \delta E^{(k)}(\lambda)+\mathcal{O}(\delta\lambda^2),
\end{align}
i.e., smaller by a factor $\mathcal{O}(\delta\lambda)$ than the time-evolution implementation error incurred in estimating either energy separately. The argument does not apply in this form to randomized compilations, e.g., qDRIFT~\cite{Campbell2019}, where the implemented channel is not generated by a fixed effective Hamiltonian.

For the Trotterized implementation adopted in Secs.~\ref{sec:numerical-demonstration} and \ref{sec:resource-estimation}, $\epsilon_{\rm TE}$ reduces to the Trotter error $\epsilon_{\rm Trotter}$ and $\delta H=\delta H_{\rm Trot}$, whose dependence on $\lambda$ enters through the Hamiltonian coefficients.
Consequently the time-evolution implementation error admissible on each evolution is larger, by a factor $\sim\lambda/\delta\lambda$, than that required to resolve a single energy to the same absolute accuracy, so that the Trotter step needed to reach a target accuracy on $\Delta E$ is correspondingly coarser, and the $\epsilon_{\rm Trotter}$ allocated in the resource estimates of Sec.~\ref{sec:resource-estimation} is a conservative upper bound. A precise characterization of $\partial_\lambda \delta E_{\rm Trot}$ and the resulting savings is left to future work.

\subsubsection{Comparison with BPDE}
Our proposed method offers distinct advantages rooted in its robust SPE framework, surpassing the limitations seen in BPDE (as discussed in Appendix \ref{app:bpde_overview}).
First, regarding theoretical performance guarantees, our approach delivers Heisenberg-limited scaling, i.e., $\mathcal{O}(\epsilon^{-1})$ for total evolution time, through statistical sampling. 
The scaling is an explicit property of the estimator, whereas the BPDE literature~\cite{Sugisaki2021, Sugisaki2022} does not analyse it for a general Hamiltonian or initial state.
Second, concerning practical implementation on early-FTQC devices, our method avoids the iterative parameter search inherent in BPDE. In BPDE, the phase shift parameter $\Delta\epsilon$ in the gate $P(\Delta\epsilon \cdot t)$ must be iteratively optimized through Bayesian inference, as detailed in Appendix~\ref{app:bpde_overview}, incurring an additional overhead from the number of Bayesian optimization iterations required for convergence. 
In contrast, our method requires no such iterative parameter tuning: the energy difference is extracted directly from statistical sampling of the circuit outputs, without any adaptive optimization of circuit parameters, which also enables parallel execution on multiple quantum devices without coherent interconnection. 
This structural simplicity is a clear advantage for practical implementation on early-FTQC devices.

\subsection{Joint Estimation of Energy Levels and Differences}
\label{subsec:joint-estimation}

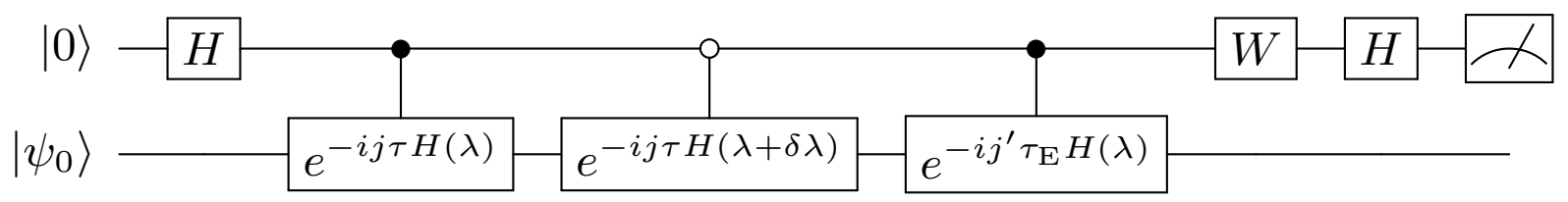
\begin{figure*}[htbp]
\centering
\[
    	\Qcircuit @C=1em @R=.7em {\\
  	\lstick{\ket{0}} & \gate{H} & \ctrl{1} & \ctrlo{1} & \ctrl{1}
    &\gate{W}
    & \gate{H}      & \meter \\
  	\lstick{\ket{\psi_0}} & \qw      & \gate{ e^{-i  j\tau H(\lambda)}} & \gate{ e^{-i  j\tau H(\lambda+\delta\lambda)}} & \gate{ e^{-i  j'\tau_{\rm E} H(\lambda)}} &\qw  & \qw &\qw 
	}
    \]
    \caption{
    Quantum circuit for our proposed method for the joint estimation of the energy-levels and their differences. The gate $W$ is set to $W=I$ ($W=S^\dag$) for estimating the real (imaginary) part of the expectation value. The controlled time evolutions with time width $j\tau$ ($j'\tau_{\rm E}$) are implemented by the $j$- ($j'$-) times iterations of the time evolution with a fixed time step $\tau$ ($\tau_{\rm E}$).
    }
    \label{fig:circuit-proposed-joint}
\end{figure*}

While the primary focus of our method is the direct estimation of energy level differences, the framework can be extended to jointly estimate both energy eigenvalues and their differences. This capability is particularly valuable for a comprehensive characterization of the system's energy landscape, providing both the overall energy scale and the fine-grained differences.

The significance of joint estimation lies in its ability to clarify the origin of each estimated energy difference. 
This situation arises naturally when the initial state $|\psi_0\rangle$ has non-negligible overlap with multiple eigenstates~\cite{Ding2023simultaneous}, e.g., when an approximate ground state preparation also carries weight on low-lying excited states, causing the one-dimensional ACDF (1D-ACDF) of Sec.~\ref{subsec:core-principle} to exhibit multiple peaks.
This is not a marginal case: for the model studied in Sec.~\ref{sec:numerical-demonstration}, the free-fermion initial state loses its dominant weight on the ground state already at $U\simeq7.3$.
In such cases, it can be challenging to determine which difference corresponds to which energy level. By estimating the joint distribution, we can directly associate a specific energy difference with its corresponding energy eigenvalue. For instance, when several difference values are estimated, this makes it possible to identify which of them originates from the ground state, i.e., which corresponds to $\Delta E_0 = E_0(\lambda+\delta\lambda) - E_0(\lambda)$.
The precision required for the energy eigenvalue in this context need only be sufficient to distinguish relevant energy levels, which is typically less demanding than the high precision required for the energy differences themselves, keeping the additional cost of joint estimation moderate.

To enable this joint estimation, we generalize the single-time measurement $\mathcal{G}(j\tau)$ to a composite quantity $\mathcal{G}_{\rm joint}(j\tau, j'\tau_{\rm E})$ carrying a second, independent time argument $j'\tau_{\rm E}$, measured by a generalized quantum circuit shown in Fig.~\ref{fig:circuit-proposed-joint}. This quantum circuit performs an initial Hadamard gate on the ancilla, followed by three distinct controlled time-evolution operations, $e^{-iH(\lambda)j\tau}$, $e^{-iH(\lambda+\delta\lambda)j\tau}$, and $e^{-iH(\lambda)j'\tau_{\rm E}}$, and concludes with a final Hadamard gate before the ancilla is measured. The probability of measuring the ancilla in $|0\rangle$ yields the real part of $\mathcal{G}_{\rm joint}(j\tau, j'\tau_{\rm E})$, while an additional phase gate $S^\dagger$ before the second Hadamard, i.e., $W=S^\dagger$ in Fig.~\ref{fig:circuit-proposed-joint}, enables the estimation of the imaginary part. The choice of two independent time parameters, $\tau$ and $\tau_{\rm E}$, allows for the simultaneous sampling of the spectral information relevant to both the energy eigenvalue and the energy difference.
The measured quantity is given by
\begin{align}
\mathcal{G}_{\rm joint}(j\tau, j'\tau_{\rm E}) =  \langle\psi_0| e^{iH(\lambda+\delta\lambda)j\tau} e^{-iH(\lambda)j'\tau_{\rm E}}  e^{-iH(\lambda)j\tau} |\psi_0\rangle.
\label{eq:joint_expectation_value_def}
\end{align}
This expectation value can be expanded, in terms of the eigenstates and eigenvalues of the Hamiltonians $H(\lambda)$ and $H(\lambda+\delta\lambda)$, as
\begin{align}
\label{eq:joint-propagator-expansion}
&\mathcal{G}_{\rm joint}(j\tau, j'\tau_{\rm E}) \nonumber\\
&= \sum_{m,n}
\Bigl[ \langle\psi_0|E_n(\lambda+\delta\lambda)\rangle \langle E_n(\lambda+\delta\lambda)|E_m(\lambda) \rangle \langle E_m(\lambda)|\psi_0\rangle \nonumber\\
&\hspace{15mm}  \times e^{-i[(j\tau+j'\tau_{\rm E}) E_m(\lambda) - j\tau E_n(\lambda+\delta\lambda)]} \Bigr].
\end{align}
Under the common assumption of a small perturbation $\delta\lambda$ and an initial state $|\psi_0\rangle$ chosen to approximate a specific eigenstate, e.g., the ground state, the sum is dominated by its diagonal terms $m=n$, which pair each unperturbed eigenstate $|E_m(\lambda)\rangle$ with its perturbed counterpart $|E_m(\lambda+\delta\lambda)\rangle$. The off-diagonal terms $m\neq n$ are suppressed, since the overlap $\langle E_n(\lambda+\delta\lambda)|E_m(\lambda)\rangle$ is $\mathcal{O}(1)$ for $m=n$ and $\mathcal{O}(\delta\lambda)$ otherwise.
Retaining these diagonal terms, Eq.~\eqref{eq:joint-propagator-expansion} reduces to
\begin{align} \label{eq:joint-propagator-dominant}
&\mathcal{G}_{\rm joint}(j\tau, j'\tau_{\rm E}) \nonumber\\
&= \sum_{k}
\Bigl[ \langle\psi_0|E_k(\lambda+\delta\lambda)\rangle \langle E_k(\lambda+\delta\lambda)|E_k(\lambda) \rangle \langle E_k(\lambda)|\psi_0\rangle \nonumber\\
&\hspace{15mm}  \times e^{-i[(j\tau+j'\tau_{\rm E}) E_k(\lambda) - j\tau E_k(\lambda+\delta\lambda)]} \Bigr]
+ \mathcal{O}\qty(\delta \lambda),
\end{align}
where the diagonal weight reduces to $|\langle E_k(\lambda)|\psi_0\rangle|^2 + \mathcal{O}\qty(\delta \lambda)$ in the leading order.
The phase part of this quantity encodes information about both the energy eigenvalue $E_k(\lambda)$ and the corresponding energy difference $\Delta E_k = E_k(\lambda+\delta\lambda) - E_k(\lambda)$.

As in the one-dimensional case, the off-diagonal contributions ($m\neq n$) neglected in Eq.~\eqref{eq:joint-propagator-dominant} carry weights $\mathcal{O}(\delta\lambda)$, smaller than the diagonal weight $\mathcal{O}(1)$, and contribute a rise at $(x_1,x_2)=(\tau[E_m(\lambda)-E_n(\lambda+\delta\lambda)],\,\tau_{\rm E} E_m(\lambda))$. The two axes behave differently. On the energy axis, $\tau_{\rm E}$ is chosen conventionally, $\tau_{\rm E}\|H\|\lesssim\pi/3$, since $x_2$ encodes the absolute energies $E_k(\lambda)=\mathcal{O}(\|H\|)$; the levels are therefore mapped faithfully without aliasing, and every term, regardless of whether diagonal or not, sits at the genuine energy of level $m$ because the energy coordinate $x_2=\tau_{\rm E}E_m(\lambda)$ is tied to the same index $m$ that anchors the difference. On the difference axis, by contrast, $\tau$ is set by $\|\dv*{H}{\lambda}\|$ rather than by $\|H\|$ [Eq.~\eqref{eq:eq:difference-tau-constraint-Hderivative}], so $\tau\|H\|$ can exceed $\pi$ and the gap phase $\tau[E_m(\lambda)-E_n(\lambda)]$ may wrap modulo $2\pi$; a rise arising from an off-diagonal term can then be aliased onto the target strip $x_1=-\tau\Delta E_k=\mathcal{O}(\tau\delta\lambda)$. 
Within each energy column $x_2=\tau_{\rm E}E_m(\lambda)$ the situation thus reduces to the one-dimensional case of Sec.~\ref{subsec:advantages-over-existing}: the target rise at $-\tau\Delta E_m$ has weight $\mathcal{O}(1)$, while any aliased off-diagonal rise has weight $\mathcal{O}(\delta\lambda)$, so by the same two-factor argument as in Sec.~\ref{subsubsec:robustness}, the located pair is displaced only by a \emph{relative} $\mathcal{O}(\delta\lambda)$ along the difference axis, i.e., a bias of $\mathcal{O}(\delta\lambda^2)$ on $\Delta E_m$, with the energy coordinate unaffected, which vanishes as $\delta\lambda\to0$ (with the same caveat near (quasi-)degeneracies).
Crucially, because the energy label is faithful and tied to the difference anchor, no off-diagonal term ever places a target-scale difference at the energy of a \emph{different} level; the joint distribution therefore does not generate spurious $(\Delta E,E)$ pairings across levels, which is precisely the ambiguity the joint estimation is designed to remove.

The measured values of $\mathcal{G}_{\rm joint}(j\tau, j'\tau_{\rm E})$ serve as the discrete two-dimensional Fourier coefficients for the joint spectral measure of energy levels and their differences. Similar to the 1D case for difference estimation, these values are fed into a classical post-processing step to construct a {\it two-dimensional ACDF} (2D-ACDF). We define the 2D-ACDF $\tilde{C}_{\text{joint}}(x_1, x_2)$ by:
\begin{align}
&\tilde{C}_{\text{joint}}(x_1, x_2) \nonumber\\
&= \sum_{j\in\mathcal{S}} \sum_{j'\in\mathcal{S}_{\rm E}} \hat{F}_j \hat{F}_{j'}^{\rm (E)} e^{ijx_1} e^{ij'x_2} \mathcal{G}_{\rm joint}(j\tau, j'\tau_{\rm E}),
\label{eq:2d_acdf_def}
\end{align}
where $\hat{F}_j$ and $\hat{F}_{j'}^{\rm (E)}$ are the Fourier coefficients of the Heaviside function, $F_{\beta, d} (x)$ and $F_{\beta_{\rm E}, d_{\rm E}} (x)$, for each dimension (see Appendix~\ref{app:Fourier-Heaviside} for details) and $\mathcal{S}=\{0\}\cup\{\pm(2k+1)\}_{k=0}^{d},\quad \mathcal{S}_{\rm E}=\{0\}\cup\{\pm(2k+1)\}_{k=0}^{d_{\rm E}}$. The truncation orders are set to satisfy $d = \mathcal{O}\qty(\tau^{-1}\epsilon_{\rm alg}^{-1})$ and $d_{\rm E}= \mathcal{O}\qty(\tau_{\rm E}^{-1}\epsilon_{\rm E}^{-1})$, where $\epsilon_{\rm E}$ denotes the additive error for the energy eigenvalue rather than its difference. Under the assumption of $\delta\lambda \ll 1$, this 2D-ACDF approximates the joint cumulative distribution function of the energy level and its difference:
\begin{align}\label{eq:2D-ACDF-final}
&\tilde{C}_{\text{joint}}(x_1, x_2) \nonumber\\
&= \sum_k
|\langle E_k(\lambda)|\psi_0\rangle|^2 \nonumber\\
 &\hspace{5mm} \times(F_{2 \rm D} * \delta_2)(x_1 - \tau[E_k(\lambda)-E_k(\lambda+\delta\lambda)], x_2 - \tau_{\rm E} E_k(\lambda)) \nonumber\\
 & \hspace{5mm} + \mathcal{O}\qty(\delta \lambda)
 \end{align}
to first order in perturbation theory, where $(F_{2 \rm D} * \delta_2)$ represents a 2D convolution between the 2D approximate step function $F_{2 \rm D} (x,y) = F_{\beta, d} (x) F_{\beta_{\rm E}, d_{\rm E}} (y)$ and the 2D delta function $\delta_2(x,y)=\delta(x) \delta(y)$.
The 2D-ACDF, representing the joint spectral measure,  will exhibit distinct steep rises, forming a multi-dimensional staircase-like landscape, at coordinates corresponding to the scaled pairs $(\tau [E_k(\lambda) - E_k(\lambda+\delta\lambda)], \tau_{\rm E} E_k(\lambda))= (-\tau\Delta E_k,\, \tau_{\rm E} E_k(\lambda))$. Just as in the 1D ACDF, these {\it steps} or {\it rises} yield the estimated values for $(\Delta E_k, E_k(\lambda))$. A classical algorithm, analogous to a 2D binary search or peak detection on the derivative of the 2D-ACDF, can then pinpoint these values with high precision.

The ability to perform joint estimation provides several advantages. Foremost, it offers clear identification: by resolving pairs of energy and its difference, any ambiguity in associating a specific difference with its corresponding energy level is removed. This enhanced diagnostic power is critical for detailed spectroscopic analysis and understanding complex energy landscapes.
Here the energy axis serves only to \emph{label} each difference by its level, not to determine $E_k(\lambda)$ accurately: its additive error $\epsilon_{\rm E}$ need only resolve the relevant levels, i.e.\ $\epsilon_{\rm E}\lesssim\min_{k\neq k'}|E_k(\lambda)-E_{k'}(\lambda)|$, the spacing between participating levels. Since these spacings are $\mathcal{O}(1)$ whereas the target differences are $\Delta E_k=\mathcal{O}(\delta\lambda)$, the energy axis may be resolved far more coarsely than the difference axis, $\epsilon_{\rm E}\gg\epsilon_{\rm alg}$. Coarsening $\epsilon_{\rm E}$ does not introduce spurious peaks; it merely broadens each level's rise along $x_2$, which is harmless as long as the broadening stays below the level spacing. 

Because the two axes are governed by independent parameters, i.e., $(\tau,d)$ for the difference and $(\tau_{\rm E},d_{\rm E})$ for the energy, this asymmetry lets the smaller $d_{\rm E}$ keep the additional depth of joint estimation moderate while the high precision on the difference is retained.
The 2D estimator rescales a single Hadamard-test outcome by the product of the two rescaling factors, so its variance carries $(\mathcal{F}\mathcal{F}_{\rm E})^2$ in place of the $\mathcal{F}^2$ of the one-dimensional case, and
\begin{align}\label{eq:N-sample-joint}
  N_{\rm sample} = C\qty(\frac{\mathcal{F}\mathcal{F}_{\rm E}}{\eta})^2 ,
\end{align}
with $\mathcal{F}=\sum_{j\in\mathcal{S}}|\hat{F}_j|$ and $\mathcal{F}_{\rm E}=\sum_{j'\in\mathcal{S}_{\rm E}}|\hat{F}_{j'}^{\rm (E)}|$, and with the same dimensionless prefactor $C$ as in the one-dimensional case, fixed in Sec.~\ref{subsec:simulation-setup}.
Since $\mathcal{F}_{\rm E}=\mathcal{O}(\log d_{\rm E})$, the penalty is the logarithmic factor $\mathcal{F}_{\rm E}^2$ relative to estimating the difference alone: for the parameters of Fig.~\ref{fig:joint-CDF} this is a factor of about six. The joint estimation therefore costs a constant factor in repetitions, not a change of scaling, and leaves the maximum evolution time along the difference axis essentially unchanged.

\section{Numerical Demonstrations}
\label{sec:numerical-demonstration}

This section presents numerical simulations to validate the proposed quantum algorithm. We primarily focus on the one-dimensional Fermi-Hubbard model, a canonical system in condensed matter physics that exhibits rich phenomena~\cite{Hubbard1963, Imada1998}.

\subsection{Model}

The one-dimensional Fermi-Hubbard model is described by the Hamiltonian:
\begin{align} \label{eq:Fermi-Hubbard-Hamiltonian}
H = -h \sum_{\langle i,j \rangle, \sigma} (c_{i\sigma}^\dagger c_{j\sigma} + c_{j\sigma}^\dagger c_{i\sigma}) + U \sum_i n_{i\uparrow} n_{i\downarrow}
-\mu \sum_{i, \sigma} n_{i\sigma},
\end{align}
where $c_{i\sigma}^\dagger$ ($c_{i\sigma}$) creates (annihilates) an electron with spin $\sigma$ at site $i$, $n_{i\sigma} = c_{i\sigma}^\dagger c_{i\sigma}$ is the number operator, $h$ is the hopping parameter between nearest-neighbor sites $\langle i,j \rangle$, $U$ is the on-site Coulomb repulsion, and $\mu$ denotes the chemical potential.

This model serves as an excellent testbed for our algorithm for two reasons. 
First, exact diagonalization is available at the system sizes considered here, so that every estimate can be compared directly with the exact value.
Second, the derivatives of the ground-state energy with respect to $h$ and $U$ are, by the Hellmann--Feynman theorem~\cite{Feynman1939}, ground-state expectation values of local operators, so that the quantity the algorithm estimates has a transparent physical meaning and a non-trivial dependence on the coupling.
The expectation value of the kinetic energy $H_h$ can be obtained through the derivative of the ground state energy with respect to $h$: 
\begin{align}
\langle H_h \rangle_{E_0}
= \langle -h \sum_{\langle i,j \rangle, \sigma} (c_{i\sigma}^\dagger c_{j\sigma} + c_{j\sigma}^\dagger c_{i\sigma}) \rangle_{E_0}
= h \frac{\partial E_0}{\partial h}. 
\end{align}
Similarly, the derivative with respect to $U$ is the double occupancy up to a known constant. 
Throughout we tie the chemical potential to the interaction, $\mu=U/2$, the particle-hole symmetric value, and discard the $U$-dependent c-number that this leaves (Sec.~\ref{subsec:simulation-setup}); differentiating along this line at half filling, Hellmann--Feynman gives
\begin{align}\label{eq:HF-double-occupancy}
      \frac{\partial E_0}{\partial U}
      = \Big\langle \sum_i n_{i\uparrow} n_{i\downarrow} \Big\rangle_{E_0}
 - \frac{N_{\rm site}}{4},
\end{align}
the constant being $-N_{\rm site}/2$ from the $\mu$ term and $+N_{\rm site}/4$ from the discarded c-number.
Our method estimates these derivatives by approximating them as finite differences, $\frac{\Delta E_0}{\delta h}$ and $\frac{\Delta E_0}{\delta U}$, respectively. 

We stress that these particular quantities are chosen because they can be checked exactly, not because a quantum computer would be needed to obtain them; for this model they follow directly from a classical ground-state calculation. The target of the algorithm is the energy difference itself, and the applications that motivate it are ones for which no comparably cheap classical route exists.

\subsection{Simulation Setup}
\label{subsec:simulation-setup}

Our numerical simulations are performed on a one-dimensional Fermi-Hubbard model with $N_{\rm site}=6$ sites and $N_{\rm electron}=6$ electrons, i.e., half-filling with $\mu=U/2$, under the periodic boundary condition.
The chemical potential is taken to follow the shift, $\mu=U/2$, and the c-number it leaves, $-\tfrac{U}{4}N_{\rm site}I$, is dropped. 
Both operations are constants within the $N_{\rm electron}$-electron sector, so the spectrum there is unchanged up to a known shift, which is the constant appearing in Eq.~\eqref{eq:HF-double-occupancy}. The same convention is applied to $H$ itself, so that our method and the SPE reference are compared on a single representation, and the $L_1$ norm is evaluated throughout on the traceless part of the operator: $\|\partial_U H\|_1=N_{\rm site}/4=3/2$ and $\|H\|_1=18$ at $U=4$.
The factor $1/4$ comes from the two number operators, the terms linear in $Z$ being cancelled by the chemical potential.
The model is implemented on quantum circuits via the Jordan-Wigner transformation.
The exact values of the ground-state energies and their finite differences are obtained using exact diagonalization, serving as the benchmark for our algorithm's performance.

The initial state $|\psi_0\rangle$ is prepared in the exact ground state of the noninteracting system, i.e., $U=0$, by using the Givens rotation network~\cite{Jiang2018, Kivlichan2018}.
This state has a large overlap with the target ground state of $H(\lambda)$ for small values of $U$.
Since the difference spectrum is not ordered by energy, the ground-state difference is identified by weight (Sec.~\ref{subsubsec:classical-postprocessing}), which requires $|\langle E_0(\lambda)|\psi_0\rangle|^2$ to be the largest of the overlaps; for this state that holds up to $U\simeq7.3$, and we scan $U\le5$, where it exceeds every other by at least a factor of two ($\eta=0.995$ at $U=0.5$, $0.569$ at $U=5$).

We target a precision $\epsilon = 10^{-3}$ on the energy difference throughout this section.
The time-evolution implementation error is not allocated a share of the budget but measured a posteriori and verified to be subdominant, as described below, so that $\epsilon_{\rm alg}\simeq\epsilon$.
Throughout this work, the truncation order $d$ is chosen to meet the condition in Eq.~\eqref{d_appendix} for $\beta$ satisfying Eq.~\eqref{eq:beta_final_theorem} with $\delta = \tau \epsilon_{\rm alg}$ ($\tau_{\rm E}\epsilon_{\rm E}$ on the energy axis of Sec.~\ref{subsec:joint-estimation}), minimizing $d$ over $\varepsilon_1,\varepsilon_2,\varepsilon_3$ subject to $\varepsilon_1+\varepsilon_2+\varepsilon_3\le\eta/2$ (see Appendix~\ref{app:Fourier-Heaviside} for details). Since this budget scales with $\eta$, $d$ also depends logarithmically on $\eta$.
For the 2-step SPE reference we likewise assign each individual energy estimate the precision $\epsilon_{\rm E}=\epsilon$, rather than the $\epsilon_{\rm E}\simeq\epsilon/2$ that resolving the difference to $\epsilon$ would require [Sec.~\ref{subsec:limitations-existing-methods}], which is conservative in favor of SPE.

The time-evolution operators in the circuit are implemented by Trotterization, using the first-order decomposition with an effective step of $0.025$: each evolution by $\tau$ is split into segments of length $0.1$, each Trotterized into four steps. 
For the SPE reference, each evolution by $\tau$ is Trotterized into eight steps, an effective step of $0.0073$.
The Trotter error is measured rather than budgeted: since the algorithm converges to the eigenvalues of the effective Hamiltonian of Eq.~\eqref{eq:effective-hamiltonian}, we diagonalize the Trotterized propagator and compare with exact diagonalization. The residual shift of the difference is at most $1.1\times10^{-4}$ over the range scanned, suppressed relative to the shift of either eigenvalue separately, which reaches $3.6\times10^{-3}$ at $U=5$, by the $\mathcal{O}(\delta\lambda)$ cancellation of Eq.~\eqref{eq:trotter-cancellation}.
The first-order formula therefore suffices here: the quantities this demonstration is designed to validate, i.e., the truncation order $d$ and the sampling complexity $N_{\rm sample}$, are fixed by $\tau$, $\epsilon_{\rm alg}$ and $\eta$ alone and involve no Trotter parameter, while the Trotter order affects only the gate count needed to reach a given accuracy. That compilation question belongs to the large-scale resource estimates of Sec.~\ref{sec:resource-estimation}, where the more gate-efficient second-order Trotterization is adopted.

For each sampled index $J$ the circuit is executed twice, once for the real and once for the imaginary part, so that the outcome is $Z=X_J+iY_J$ with $X_J,Y_J\in\{\pm1\}$; no averaging is performed at fixed $J$, and the estimator is based on Ref.~\cite{Lin2022} [Eq.~\eqref{eq:G_estimator}]. 
The single-shot randomness is already contained in the bound $\mathrm{var}[G]\le2\mathcal{F}^{2}$ on the estimator of Eq.~\eqref{eq:G_estimator}, with $\mathcal{F}=\sum_{j\in\mathcal{S}}|\hat F_j|$ its rescaling factor. The standard deviation of the ACDF estimator of Eq.~\eqref{eq:acdf-estimator} at fixed $x$ therefore obeys
\begin{align}\label{eq:sigma_C}
\sigma_{\rm ACDF} \;\le\; \mathcal{F}\sqrt{\frac{2}{N_{\rm sample}}}
\;=\; \mathcal{O}\!\left(\frac{\log d}{\sqrt{N_{\rm sample}}}\right),
\end{align}
the last equality using $\mathcal{F}=\mathcal{O}(\log d)$ (Appendix~\ref{app:Fourier-Heaviside}). The estimate $\Delta E_k=-x_k/\tau$ is read off from the position $x_k$ at which $\tilde C_{\Delta E}$ rises. Near this step the ACDF increases by the overlap weight $q_k=|\langle E_k(\lambda)|\psi_0\rangle|^{2}$ over the spectral resolution $\sim1/d$ in the scaled variable $x=\tau\Delta E$, so its slope is $|\mathrm{d}\tilde C_{\Delta E}/\mathrm{d}x|\sim q_k d$; a fluctuation $\sigma_{\rm ACDF}$ displaces the located position by $\delta x\sim\sigma_{\rm ACDF}/(q_k d)$, and hence
\begin{align}\label{eq:sigma_dE_def}
\sigma_{\Delta E} \;=\; \frac{\delta x}{\tau} \;\sim\; \frac{\sigma_{\rm ACDF}}{q_k\,d\,\tau}.
\end{align}

Fixing the number of samples as $N_{\rm sample}=C\,(\mathcal{F}/\eta)^2$, with $C$ a dimensionless prefactor, makes the bound of Eq.~\eqref{eq:sigma_C} equal to $\sigma_{\rm ACDF}=\eta\sqrt{2/C}$.
With $q_k\simeq\eta$ and $1/(d\tau)=\mathcal{O}(\epsilon)$ implied by Eq.~\eqref{eq:scaling-truncation-order}, Eq.~\eqref{eq:sigma_dE_def} then gives
\begin{align}\label{eq:sigma-dE-final}
\sigma_{\Delta E} \simeq \sqrt{2/C}\;\epsilon .
\end{align}
The prefactor $\sqrt{2/C}$ is thus fixed by a single ratio $\sigma_{\rm ACDF}/\eta$, the statistical error of the ACDF as a fraction of the step height $\eta$. 
We adopt $C=32$, corresponding to $\sigma_{\rm ACDF}\le\eta/4$ and $\sigma_{\Delta E}\simeq\epsilon/4$: the statistical error of the reconstruction is at most a quarter of the height of the rise, and the target precision corresponds to about four standard deviations or more of the located difference~\footnote{The rigorous requirement of Ref.~\cite{Lin2022} corresponds to $C=512$ for each of several independent batches combined by a majority vote, together with an approximation error of $\eta/8$.
}. 

Over the settings of Fig.~\ref{fig:compare-estimate-exact} this places all of $2\times10^3$ independent realizations of $\Delta E$, each obtained from a fresh set of $N_{\rm sample}$ samples $(J_k,Z_k)$, within $\epsilon$ of the exact value for $U\le4$.
At $U=5$, where the initial state's dominance $q_0/\max_{k\neq0}q_k$ has fallen to $2.13$, the located maximum occasionally moves to the neighbouring rise; the fraction within $\epsilon$ is then $99.8\%$ for $\Delta E_h$ and $99.5\%$ for $\Delta E_U$. 
      This is the onset of the ambiguity that the joint estimation of Sec.~\ref{subsec:joint-estimation} is designed to remove, and which becomes decisive at $U\simeq7.3$ (Sec.~\ref{subsec:joint-estimation}).

\subsection{Results: Direct Estimation of Energy Level Differences and Performance Comparison with SPE}

We first present the results for the direct estimation of energy level differences, focusing on the ground state energy with respect to shift in $h$ and $U$, and compare the performance of our proposed method with the conventional SPE approach.

\subsubsection{Concrete Examples of ACDF and UDF}

\begin{figure*}[htbp]
\includegraphics[keepaspectratio, scale=0.55]{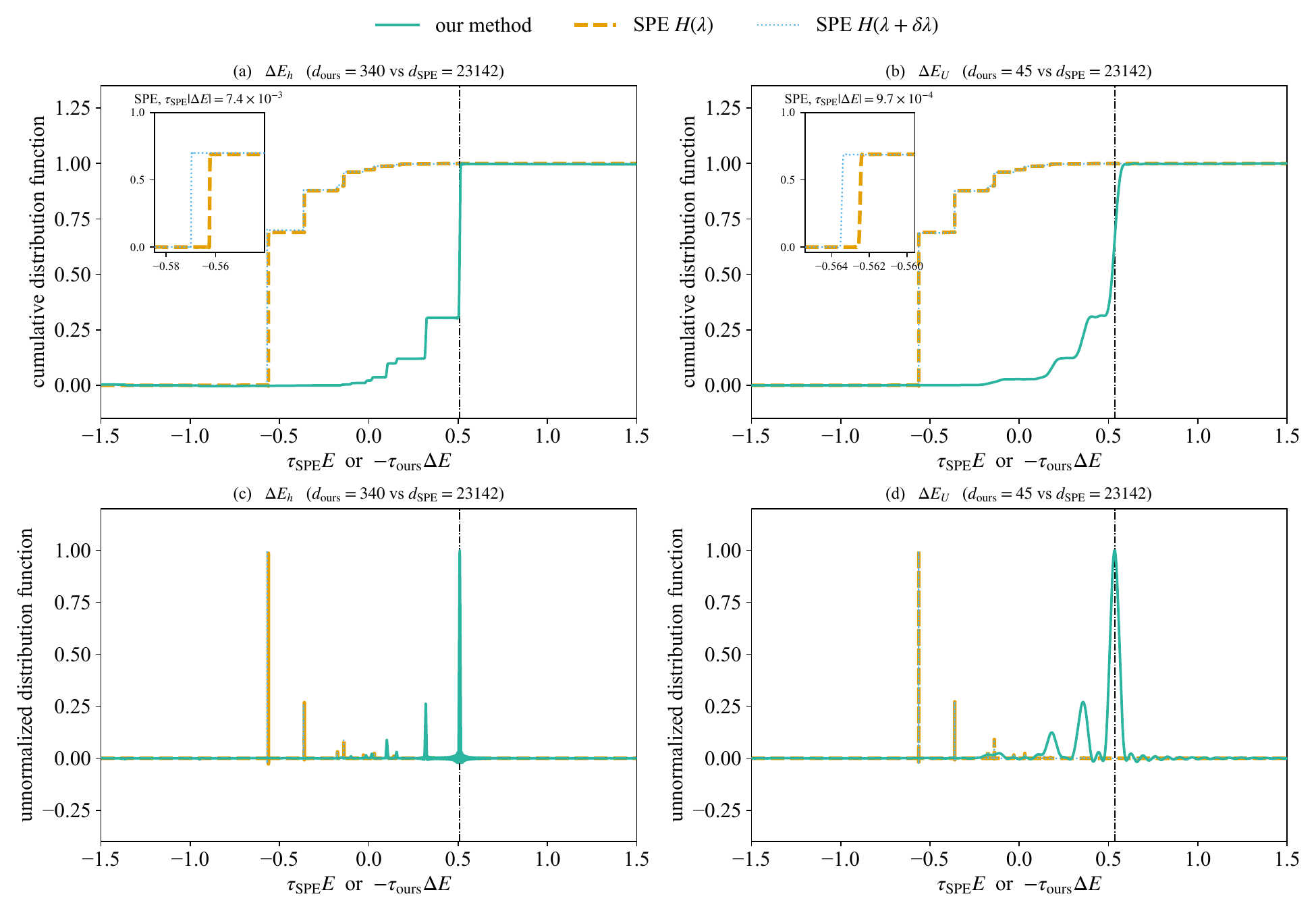}
\caption{
Comparison of energy-level difference estimation between our proposed method and SPE, at the same target precision $\epsilon=10^{-3}$. 
The subscripts ``ours'' and ``SPE'' on the time step $\tau$ and the truncation order $d$ refer to our method and to SPE, respectively; the panel titles quote the truncation orders $d_{\rm ours}$ and $d_{\rm SPE}$ required for each method.
Upper row: ACDF; lower row: UDF, i.e., the analytic derivative of ACDF, scaled to unit maximum.
(a), (c) $\Delta E_h$ with $h=1$, $\delta h=0.02$, $U=4$; our method uses $\tau_{\rm ours}=4$ and SPE $\tau_{\rm SPE}=0.058$, from Eqs.~\eqref{eq:eq:difference-tau-constraint-Hderivative} and \eqref{eq:SPE-tau-constraint} respectively. (b), (d) $\Delta E_U$ with $U=4$, $\delta U=0.02$, $h=1$; our method uses $\tau_{\rm ours}=32$.
Matching the precision leaves the maximum evolution time common to the two methods to within 10\%, so the depth of the longest circuit is comparable on both sides. What differs is the truncation order, given in the panel titles. 
The insets of (a) and (b) magnify the ground-state rise of the SPE reconstruction, whose two curves are indistinguishable on the scale of the main panels: what SPE must resolve is a splitting of $\tau_{\rm SPE}|\Delta E|=7.4\times10^{-3}$ and $9.7\times10^{-4}$ on an $\mathcal{O}(1)$ energy, whereas our method places the same information at $\mathcal{O}(1)$ on the difference axis. The height of each rise is the ground-state overlap $\eta=0.69$.
Reconstructions are shown in the infinite-sample limit; the statistical error is $\sigma_{\Delta E} \lesssim \epsilon\sqrt{2/C}$ [Eq.~\eqref{eq:sigma-dE-final}] and is quantified in Fig.~\ref{fig:compare-estimate-exact}.
}
\label{fig:CDF-UDF}
\end{figure*}

To demonstrate the effectiveness of our approach, we analyze the ACDF and unnormalized distribution function (UDF) for small parameter shifts in both $h$ and $U$. Here, the UDF is defined as the analytic differentiation of ACDF.
Figure~\ref{fig:CDF-UDF} illustrates concrete examples of ACDF and UDF for energy differences $\Delta E_h$ and $\Delta E_U$, comparing our proposed method (solid lines) with SPE, both evaluated in the infinite-sample limit.
The two methods are compared at a common target precision, which is the criterion used throughout this work (Table~\ref{tab:comparison-proposal-SPE}, Fig.~\ref{fig:truncation_order_Nsample_full}, Table~\ref{tab:hubbard_resource_estimation_Pphys3}), rather than at a common resource budget.
For $\Delta E_h = E_0(h+\delta h, U) - E_0(h, U)$ with $h=1$, $\delta h=0.02$, and fixed $U=4$, Fig.~\ref{fig:CDF-UDF}(a) shows the ACDF.
Our method, configured with $\tau_{\rm ours}=4$ based on Eq.~\eqref{eq:eq:difference-tau-constraint-Hderivative}, clearly indicates the energy difference as a distinct step. Its corresponding UDF in Figure~\ref{fig:CDF-UDF}(c) exhibits a sharp, well-defined peak.
In contrast, when applying SPE to estimate $E_0(h, U)$ (orange dashed lines) and $E_0(h+\delta h, U)$ (light blue dotted lines) with $\tau_{\rm SPE}=0.058$, which is determined by Eq.~\eqref{eq:SPE-tau-constraint}, its two curves are indistinguishable on the scale of the main panel, because what it must resolve is a splitting of $\tau_{\rm SPE}|\Delta E_h|=7.4\times10^{-3}$ on an $\mathcal{O}(1)$ energy; the inset shows them magnified, where the splitting is recovered. Reaching the same precision requires $d=23\,142$ against $d=340$ for our method, while the maximum evolution times are essentially equal, $2724$ against $2693$.

Similarly, for $\Delta E_U = E_0(h, U+\delta U) - E_0(h, U)$ with $U=4$, $\delta U=0.02$, and fixed $h=1$, Figure~\ref{fig:CDF-UDF}(b) presents the ACDF. 
Our method, using $\tau_{\rm ours}=32$, again clearly shows the energy difference. The corresponding UDF in Figure~\ref{fig:CDF-UDF}(d) also displays a sharp peak.
For SPE the splitting is an order of magnitude smaller, $\tau_{\rm SPE}|\Delta E_U|=9.7\times10^{-4}$, and the reduction in truncation order correspondingly larger, $d=45$ against $d=23\,142$.
These results demonstrate that our proposed method accurately and distinctly resolves subtle energy differences.

\subsubsection{Scaling Advantages: Truncation Order and Sample Complexity}

\begin{figure*}[tbp]
\includegraphics[keepaspectratio, scale=0.65]{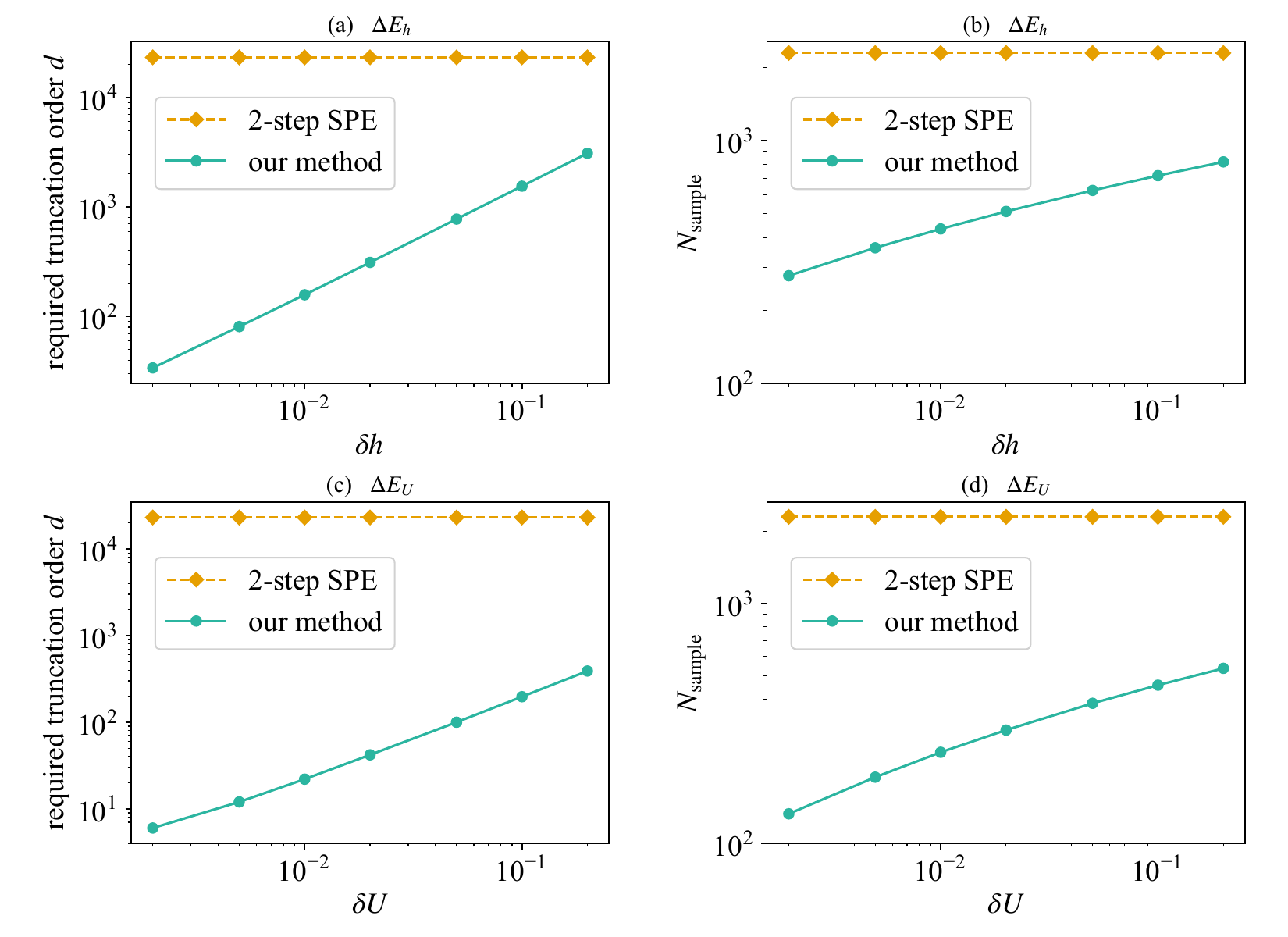}
\caption{
    Required truncation order and sampling complexity as a function of the parameter shift $\delta\lambda$, at $U=4$ and $\epsilon=10^{-3}$. Circles connected by solid lines are our method, diamonds connected by dashed lines the two-step SPE reference. The truncation order of our method is set by $\|\partial H/\partial\lambda\|\,\delta\lambda$ and therefore grows linearly with the shift, whereas that of SPE is set by $\|H\|$ and does not depend on it, so the two scalings of Table~\ref{tab:comparison-proposal-SPE} are displayed directly against each other.
(a) $d$ and (b) $N_{\rm sample}$ for $\Delta E_h$, with $\|\partial H/\partial h\|_1=12$.
(c) $d$ and (d) $N_{\rm sample}$ for $\Delta E_U$, with $\|\partial H/\partial U\|_1=3/2$.
At $\delta\lambda=0.02$, the shift used in Figs.~\ref{fig:CDF-UDF} and \ref{fig:compare-estimate-exact}, the time step here is taken at the bound of Eq.~\eqref{eq:eq:difference-tau-constraint-Hderivative}, whereas those figures use the rounded values $\tau=4$ and $\tau=32$; the truncation orders therefore differ by under ten percent.
For the two-step SPE, $N_{\rm sample}$ is summed over its two runs.
}
\label{fig:truncation_order_Nsample_full}
\end{figure*}

The superior performance of our proposed method is rooted in its ability to reduce the required truncation order $d$ and sampling complexity $N_{\rm sample}$ compared to SPE.
Figure~\ref{fig:truncation_order_Nsample_full} quantifies these as a function of the parameter shift $\delta\lambda$, which is the natural control parameter: by Eq.~\eqref{eq:diff-d-lower-bound} the truncation order of our method is set by $\|\partial H/\partial\lambda\|\,\delta\lambda$ and grows linearly with the shift, whereas that of SPE is set by $\|H\|$ [Eq.~\eqref{eq:SPE-d-lower-bound}] and does not depend on it.
The time steps $\tau=4$ and $\tau=32$ used for $\Delta E_h$ and $\Delta E_U$ are rounded values at about $92\%$ of the bound of Eq.~\eqref{eq:eq:difference-tau-constraint-Hderivative}, which gives $4.36$ and $34.9$ respectively; the same fraction is used for both shifts so that they are on an equal footing.
Over $\delta\lambda\in[0.002,0.2]$ our method reduces $d$ by factors of $7.5$ to about $3900$, and $N_{\rm sample}$ by factors of $2.8$ to about $17$.
The ratio is $\|H\|/(\|\partial H/\partial\lambda\|\,\delta\lambda)$ up to a constant of order unity, and therefore depends on the Hamiltonian representation through both norms; the values quoted here correspond to the cases in which the chemical potential is absorbed and the identity term removed (Sec.~\ref{subsec:simulation-setup}).
The advantage is largest in the small-shift regime, which is that of gradient estimation (Sec.~\ref{subsubsec:budget-gradient}), and is bounded from above by the finite-difference bias, which forbids taking $\delta\lambda$ arbitrarily large.

\subsection{Comparison with Exact Results and Overall Performance}
\label{subsec:results-direct}

\begin{figure*}[htbp]
\includegraphics[keepaspectratio, scale=0.58]{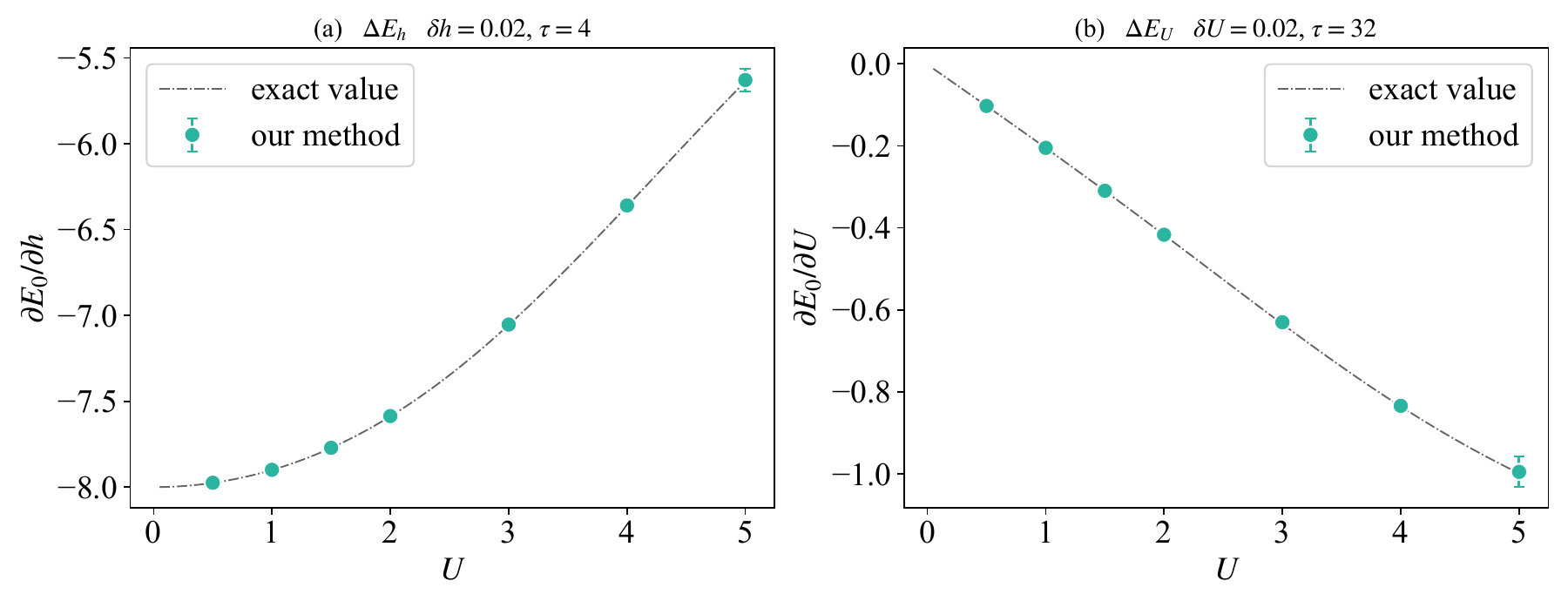}
\caption{
Comparison of the estimated ground-state energy derivatives with exact diagonalization.
(a) the derivative $\partial E_0/\partial h$, obtained from $\Delta E_h$ with $\delta h=0.02$, and (b) the derivative $\partial E_0/\partial U$, obtained from $\Delta E_U$ with $\delta U=0.02$. Error bars are the standard deviation measured over $2\times10^3$ independent realizations of the importance sampling and the shot noise, and do not include the systematic contributions of the Trotter step, the reconstruction grid, and neighboring rises, which together stay below $1.1\times10^{-4}$ throughout. The energy difference is read off the global maximum of the reconstructed UDF over $x\in[-\pi/2,\pi/2]$, with no additional restriction of the search range. The agreement with the exact finite differences from exact diagonalization, indicated by the dash-dotted lines, underscores the accuracy and reliability of our method in characterizing physical properties; both curves approach their free-fermion limits, $-8$ and $0$, as $U\to0$.
}
\label{fig:compare-estimate-exact}
\end{figure*}

Having established the efficiency gains of our proposed method, we now validate its accuracy and reliability by comparing our algorithm's estimates with exact diagonalization results across a range of $U$ values.
Figure~\ref{fig:compare-estimate-exact}~(a) displays the derivative $\partial E_0/\partial h$, obtained from the estimated difference $\Delta E_h$ with $\delta h=0.02$, as a function of $U$.
The dash-dotted curve represents the exact result, while the data points with error bars indicate our algorithm's estimates.
Similarly, Fig.~\ref{fig:compare-estimate-exact}~(b) displays $\partial E_0/\partial U$, obtained from $\Delta E_U$ with $\delta U=0.02$.

In both cases, the exact finite differences consistently fall within the error bars of the estimated results, confirming the high accuracy and reliability of our proposed method. 
The agreement is quantitative at the small-$U$ end, where the kinetic energy approaches $-8$, which is the sum of the three lowest single-particle levels of the six-site ring over both spins, and $\partial E_0/\partial U$ approaches zero, the value at which the double occupancy takes its uncorrelated value $N_{\rm site}/4=3/2$.
Both quantities vary monotonically with $U$, reflecting the suppression of double occupancy and of hopping by the on-site repulsion~\cite{LiebWu1968}.

The upper end of the range is set by the initial state rather than by the algorithm: beyond $U\simeq7.3$ the state prepared with the Givens-rotation network loses its dominant weight on the ground state, and the largest rise of the one-dimensional ACDF no longer belongs to it. That regime is not inaccessible, and it is what the joint estimation of Sec.~\ref{subsec:joint-estimation} is for, as demonstrated at $U=10$ in the following subsection.

\subsection{Results: Joint Estimation of Energy Levels and Differences}

\begin{figure*}[htbp]
\leftline{{\large (a)} \hspace{0.45 \hsize} {\large (b)}}
\begin{minipage}[t]{0.49\hsize}
\vspace{0pt}
\includegraphics[trim=65 0 10 35, clip, keepaspectratio, scale=0.65]{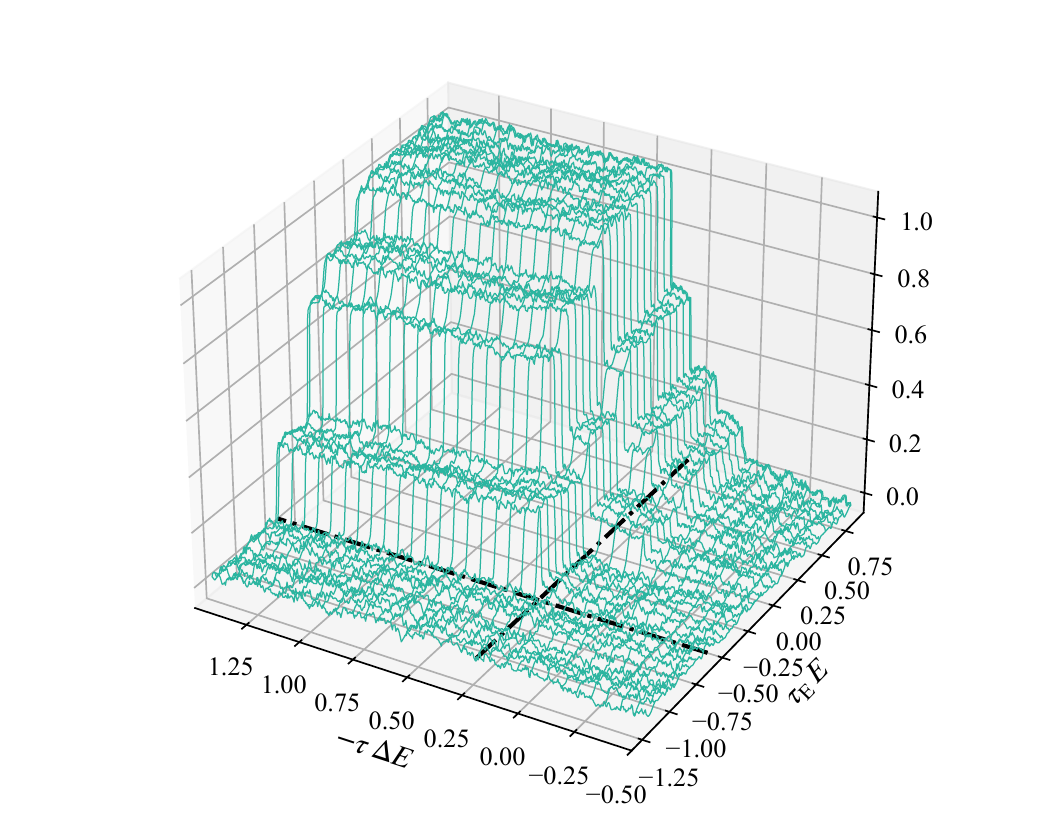}
\end{minipage}
\begin{minipage}[t]{0.49\hsize}
\vspace{0pt}
\includegraphics[keepaspectratio, scale=0.52]{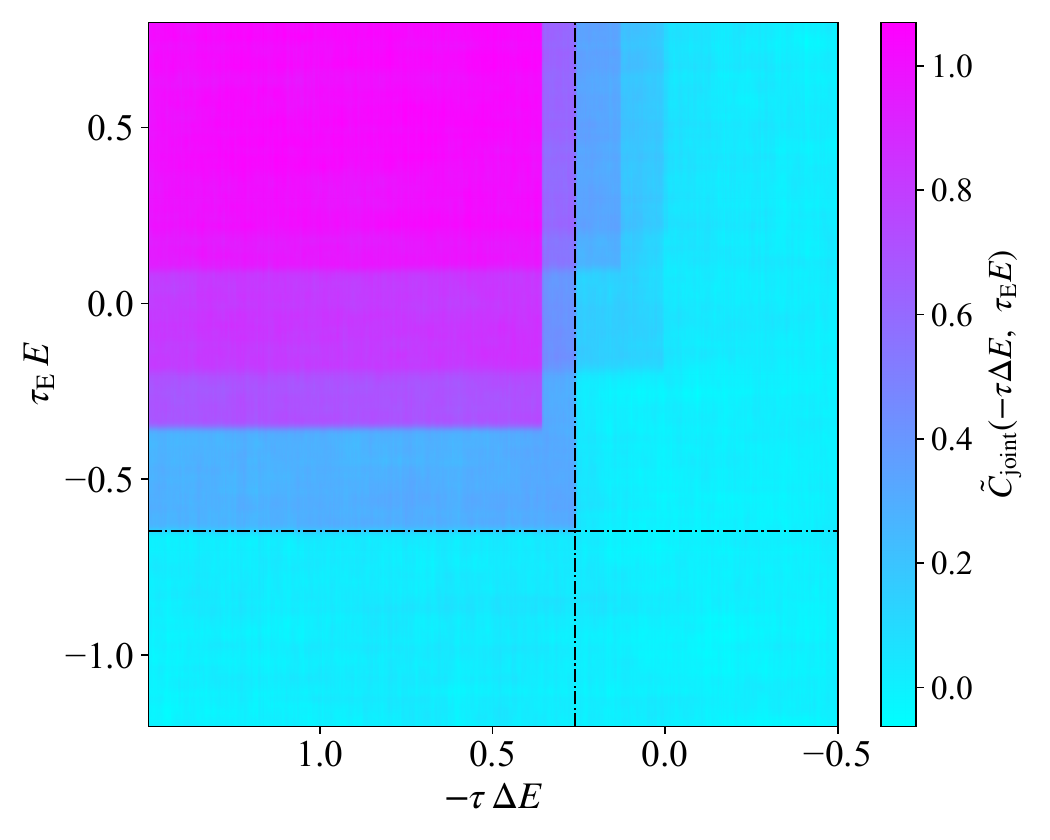}
\end{minipage}
\caption{2D-ACDF for $E_0$ and $\Delta E_h$ for $U=10$ and $\delta h = 0.02$.
The parameters are $\tau=4$, $\tau_{\rm E}=\pi/(3\|H\|_1)=0.0388$, $\epsilon=10^{-3}$ and $\epsilon_{\rm E}=0.5$; the surface is reconstructed from importance samples over the pair $(j,j')$, each index drawn from its own distribution $|\hat{F}_j|$ and $|\hat{F}_{j'}^{\rm (E)}|$ as assumed in Eq.~\eqref{eq:N-sample-joint}, so that every sample corresponds to a single circuit configuration. The reconstruction includes the resulting index-sampling noise but not the shot noise of the Hadamard tests.
The horizontal axis shows $-\tau\Delta E$, i.e.\ the coordinate $x_1$ of Eq.~\eqref{eq:2D-ACDF-final}.
 (a) The sharp rises in the three-dimensional surface plot, equivalent to (b) sharp changes in the color of the heatmap representation, indicate the estimated values for $(E_k, \Delta E_k)$. 
 Each {\it step} or {\it rise} in (a) corresponds to an energy eigenstate with significant overlap with the initial state $\ket{\psi_0}$. The height of each step reflects the magnitude of this overlap, while its position, i.e., coordinates, indicates the corresponding energy eigenvalue and its difference. 
 The estimated values for the ground-state energy and its difference well agree with the exact results, indicated by the black dash-dotted lines.
 The coupling $U=10$ lies deliberately outside the range scanned in Fig.~\ref{fig:compare-estimate-exact}: there the ground state carries the largest weight and the one-dimensional ACDF suffices, whereas here it carries $0.26$ against $0.42$ on an excited level, so the difference spectrum alone cannot identify which rise belongs to the ground state.
}
\label{fig:joint-CDF}
\end{figure*}

For the joint estimation, the classical post-processing involves constructing the 2D-ACDF over the energy and the energy-difference axes from the simulated expectation values.  
The ground-state level is identified as the first (i.e., left-most) jump along the energy axis $E(\lambda)$, which corresponds to the ground state because $E_0(\lambda)$ is the smallest eigenvalue, and the position of that same jump along the energy-difference axis then gives the ground-state energy difference $\Delta E_0$.

We take $U=10$ precisely because the one-dimensional estimation of Sec.~\ref{subsec:results-direct} no longer suffices there. Identification by weight requires the initial state to carry its largest weight on the target level, which for the state prepared with the Givens-rotation network holds up to $U\simeq7.3$. At $U=10$ the weights are $0.26$ on the ground state and $0.42$ on an excited level, so the largest rise of the one-dimensional ACDF lies at $\Delta E=-0.090$, corresponding to that excited level, well away from $\Delta E_0=-0.065$. This is the ambiguity the joint estimation removes: the ground state is unambiguous on the energy axis, being the left-most rise, and its difference coordinate is read off directly.

Figure~\ref{fig:joint-CDF} (a) shows a three-dimensional (3D) surface plot of the 2D-ACDF for $U=10$, where both the ground state energy $E_0$ and the energy difference $\Delta E_h = E_0(h+\delta h) - E_0(h)$ with $\delta h = 0.02$ are estimated.
The prominent steep rises in this function indicate the estimated $(E_0, \Delta E_h)$ pair. 
Each such {\it step} or {\it rise} on the 3D surface corresponds to an energy eigenstate of the system that has a significant overlap with our initial state. The number of visible steps reflects the number of such accessible eigenstates, and their heights are proportional to the magnitude of these overlaps. The coordinates at which these steps occur accurately pinpoint the corresponding energy eigenvalues and their differences.
In this plot, the rise at the position corresponding to the exact values of ground-state energy and its finite difference, indicated by black dash-dotted lines, is resolved above the sampling noise, demonstrating the agreement between our algorithm's estimations and the exact results.

For better visualization, Fig.~\ref{fig:joint-CDF} (b) presents the same 2D-ACDF as a heatmap. The region where the ACDF transitions from low to high values, represented by changes in the color, clearly indicates the estimated values. 
This highlights the method's versatility in associating each difference with its energy level, enabling unambiguous identification of specific energy differences and their corresponding energy eigenvalues.

\section{Resource Estimation}
\label{sec:resource-estimation}

This section discusses the resource estimation for implementing our proposed method, detailed in Sec.~\ref{sec:proposed-method}, on early-FTQC devices, building upon established frameworks for quantum simulation. 
In this section, as a prototypical example of early-FTQC devices, we specifically consider the recently proposed Space-Time efficient Analog Rotation quantum computing (STAR) architecture~\cite{Akahoshi2024, Toshio2025, Akahoshi2025, Toshio2026, Kanasugi2026}.
The STAR architecture is particularly promising for early-FTQC due to its ability to efficiently execute Trotter-decomposed time-evolution simulations with relatively limited quantum resources: the effective error rate of an analog rotation is smaller for smaller angles, and a product formula consists mainly of small-angle rotations. Our merged circuit exploits the same property (Appendix~\ref{app:merged-rotation}). 
These characteristics make it an ideal platform for evaluating the practicality of our algorithm within realistic constraints.

\subsection{General Procedure for Resource Estimation}
\label{subsec:resource-STAR}

The resource estimation process begins by defining the target Hamiltonian and its perturbed counterpart, along with key parameters such as the physical error rate of qubits and the required precision for the energy difference. This information is then used to determine the necessary quantum resources.

For a general Hamiltonian $H= \sum_{k=1}^L a_kP_k$, where $a_k$ are real coefficients and $P_k$ are Pauli string operators, its Pauli $L_1$-norm $\Lambda = \sum_{k=1}^L |a_k|$ (see Sec.~\ref{subsec:advantages-over-existing}) serves as a central quantity in determining the cost of time evolution. 
Our method requires implementing the time evolutions under both $H(\lambda)$ and $H(\lambda+\delta\lambda)$ within the Hadamard test circuit; as described below, the two are realized by a single merged sequence of rotations rather than by two controlled unitaries in series.
These time evolution operators are generally implemented via Trotter decomposition. 
The truncation order $d$ and the sampling complexity $N_{\rm sample}$ entering the cost are those of Table~\ref{tab:comparison-proposal-SPE}, evaluated with the algorithmic error $\epsilon_{\rm alg}<\epsilon$ that remains after the budget allocation described below.

A key step in determining the total cost is to translate the logical operations into physical qubit resources and execution time. The number $N_{\text{sys}}$ of logical qubits for encoding the simulated system, determined by problem size and qubit mapping, e.g., Jordan-Wigner transformation, dictates the logical resource requirement. The number $N_{\text{phys}}$ of physical qubits is then calculated based on factors such as $N_{\text{sys}}$, the physical error rate $p_{\rm phys}$, and the choice of the Quantum Error Correction (QEC) code, e.g., rotated planar surface codes~\cite{Fowler2012, Horsman2012}. The code distance $d_{\rm code}$, directly characterizing $N_{\text{phys}}$, is determined such that the total logical Clifford error accumulated within a single circuit remains below a target threshold, typically $1\%$~\cite{Babbush2018linearT, Toshio2025, Toshio2026}, i.e., 
\begin{align}\label{eq:code-distance}
p_L(p_{\rm phys},d_{\rm code})^{-1} \;\ge\; 100 \times d_{\rm code}\, C_{\rm total}\, N_{\text{patch}},
\end{align}
where the logical error rate per code cycle $p_L(p_{\rm phys}, d_{\rm code})$ is approximated as~\cite{Fowler2019, Litinski2019} 
\begin{align}
p_L(p_{\rm phys},d_{\rm code}) = 0.1 \cdot (100p_{\rm phys})^{(d_{\rm code}+1)/2}
\end{align}
for surface codes. In Eq.~\eqref{eq:code-distance}, $N_{\text{patch}}$ is the number of code patches, $C_{\rm total}$ is the number of clocks required to execute all analog rotations of the circuit, so that $d_{\rm code}\,C_{\rm total}$ is its duration in code cycles. Since every sampled circuit must satisfy the condition and the accumulated logical error grows with the circuit length, it suffices to impose Eq.~\eqref{eq:code-distance} on the longest one, namely that with the largest Fourier mode $j=2d+1$. 
The two time evolutions are implemented by the gate-efficient second-order Trotterization, which fixes the number of Trotter steps in a circuit of evolution time $t$ to $t\sqrt{W/\epsilon_{\rm Trotter}}$, where $W$ is the associated Trotter-error norm~\cite{Kivlichan2020} and $\epsilon_{\rm Trotter}$ the allocated Trotter error.
Writing $C_{\rm step}=\sum_{r} C_{\rm smm}(\theta_r)$ for the clock count of a single Trotter step, where $r$ runs over the analog rotations of that step rather than over the Pauli terms (Appendix~\ref{app:merged-rotation}) and each analog rotation contributes a number of clocks $C_{\rm smm}(\theta_r)$ determined by its own rotation angle $\theta_r$, assuming the recently proposed STAR-magic-mutation (SMM) protocol~\cite{Toshio2026, Kanasugi2026}, that circuit consists of $(2d+1)\tau\sqrt{W/\epsilon_{\rm Trotter}}$ Trotter steps and therefore has
\begin{align} \label{eq:C-total}
C_{\rm total} = C_{\rm step}\,(2d+1)\,\tau\,\sqrt{\frac{W}{\epsilon_{\rm Trotter}}} .
\end{align}
The physical qubit count $N_{\text{phys}}$ for such a surface code architecture is then given by $N_{\text{phys}} = 2N_{\text{patch}} d_{\rm code}^2$.
We adopt the fast block layout~\cite{Litinski2019} of the STAR architecture~\cite{Toshio2026, Kanasugi2026}, for which $N_{\text{patch}} = 2N_{\rm sys} + \lceil \sqrt{8 N_{\rm sys}} \rceil + 11$, where 10 of the 11 additional patches are allocated for non-Clifford state preparation~\cite{Gidney2024}.

In the STAR architecture, each analog rotation gate is realized through a repeat-until-success (RUS) process, and in the fast-block layout a single RUS trial is executed within a single clock, i.e., $d_{\rm code}$ code cycles~\cite{Toshio2026} via lattice surgery~\cite{Horsman2012, Litinski2019}. The number of clocks consumed by a single rotation depends on its angle through the threshold at which the protocol switches to $T$-gate synthesis~\cite{Toshio2026}; we specify this dependence below.
The time cost of a single code cycle is dictated by stabilizer measurement time and QEC decoding time. On current superconducting qubit chips~\cite{Arute2019, Google2023}, a single round of syndrome extraction takes less than \SI{1}{\micro\second}. Likewise, some latest decoding algorithms can process syndrome data in under \SI{1}{\micro\second} per round on a single CPU core~\cite{Higgott2025} or on low-end hardware, such as FPGA or ASIC~\cite{Barber2025}, for a moderate code distance. Consequently, in line with prior works~\cite{Babbush2018lowdepth, Kivlichan2020, Yoshioka2024, Akahoshi2024, Toshio2025, Akahoshi2025, Toshio2026, Kanasugi2026}, a single code cycle is assumed to take \SI{1}{\micro\second}.

The two controlled time evolutions of Fig.~\ref{fig:circuit-proposed-simple} are not implemented separately: for each Pauli term $P_k$ they are merged into an \emph{uncontrolled} rotation of the system by the half-sum $\theta^{+}_k$ of the two evolution angles and a \emph{joint} rotation of the system and the ancilla by their half-difference $\theta^-_k$ (see Appendix~\ref{app:merged-rotation} for details). 
An equivalent construction is used for BPDE in Ref.~\cite{Sugisaki2022}. What a parameter shift adds is that $\theta^{-}_k=\mathcal{O}(\delta\lambda)$, smaller than $\theta^{+}_k$ by the relative size of the shift, so that the rotations carrying the signal are the smallest in the circuit; on the STAR architecture, they are then executed at the minimum clock count.
The merging is exact, applies to a product formula of any order, and halves both the number of analog rotations and the total rotation angle relative to an implementation of the two evolutions in series; the same reduction is not available to the 2-step SPE reference, in which each controlled rotation must be realized as two analog rotations and each eigenvalue requires a separate circuit.

The total execution time $T_{\text{total}}$ for a single quantum processing unit (QPU) is evaluated as
\begin{align} \label{eq:total-execution-time}
T_{\text{total}} = 
2 d_{\rm code} & C_{\rm step
} \nonumber\\ 
\times & \sum_{j\in\mathcal{S}} \qty(\frac{N_{\rm sample}|\hat{F}_j|}{\sum_{j'\in\mathcal{S}} |\hat{F}_{j'}|})\gamma^2_{j\tau}  |j|\tau \sqrt{\frac{W}{\epsilon_{\rm Trotter}}} \ [\SI{}{\micro\second}],
\end{align}
where the time-evolution implementation error $\epsilon_{\rm TE}$ introduced in Sec.~\ref{subsec:advantages-over-existing} specializes to the Trotter error $\epsilon_{\rm Trotter}$, and $\gamma^2_{j\tau}$ is the factor for sampling overhead due to error mitigation with the probabilistic error cancellation~\cite{Temme2017, Li2017, Endo2018, Obrien2021, Cai2023, Toshio2026}, given explicitly in Eq.~\eqref{eq:gamma} below.
The number of samples is fixed as $N_{\rm sample}=C\,(\mathcal{F}/\eta)^2$ with $C=32$, giving $\sigma_{\Delta E} \lesssim \epsilon/4$ [Eqs.~\eqref{eq:sigma_C}--\eqref{eq:sigma-dE-final}], and each sampled circuit is executed twice (Sec.~\ref{subsec:simulation-setup}).

The parameter shift $\delta\lambda$ is taken as large as the finite-difference bias on $\Delta E$ permits, so that this bias is negligible on the scale of $\epsilon$ (the value adopted for the Fermi-Hubbard model is given in Sec.~\ref{subsec:estimation-condition}), and the precision budget is allocated between the algorithmic and Trotter errors as $ \epsilon_{\rm Trotter} + \epsilon_{\rm alg} < \epsilon$~\cite{Reiher2017, Kivlichan2020}, the remainder of $\epsilon$ being reserved for the finite-difference bias on $\Delta E$.
We treat $\epsilon_{\rm Trotter}$ as the Trotter error of a single evolution; by the cancellation discussed in Sec.~\ref{subsec:advantages-over-existing}, the actual error on $\Delta E$ is smaller by $\mathcal{O}(\delta\lambda)$, so this allocation is conservative.

For the SMM protocol, the mitigation cost accumulated over a single circuit at Fourier mode $j$ is
\begin{align}\label{eq:gamma}
\gamma^2_{j\tau} = e^{4 P_{j\tau}}, \hspace{3mm} 
P_{j\tau} = p_{\rm phys}\, |j|\tau \sqrt{\frac{W}{\epsilon_{\rm Trotter}}} 
\sum_{r} \alpha_{\rm RUS}(\theta_r)\,\theta_r
\end{align}
with $\alpha_{\rm RUS}(\theta_r)$ being the logical error factor of a single analog rotation of angle $\theta_r$. 
Both $\alpha_{\rm RUS}(\theta_L)$ and $C_{\rm smm}(\theta_L)$ depend on the threshold angle $\theta_{\rm th}$ at which the RUS process gives way to $T$-gate synthesis, as well as the logical rotation angle $\theta_L$.
We adopt the angle-resolved values reported in Ref.~\cite{Kanasugi2026}, evaluated with the interpolation routine of its public implementation, corresponding to the two prescriptions for choosing it: one that minimizes the error at each angle (``accuracy prioritized''), and one that keeps $\theta_{\rm th}/\theta_L$ above a fixed ratio (``speed prioritized'').
The two trade one quantity against the other: $\alpha_{\rm RUS}$ enters exponentially [Eq.~\eqref{eq:gamma}] while $C_{\rm smm}$ enters the runtime linearly, so neither dominates, the accuracy-prioritized setting being preferable when the mitigation overhead is the bottleneck. We evaluate both for every entry of Table~\ref{tab:hubbard_resource_estimation_Pphys3} and report the shorter time, indicating the selected prescription in the table. Data are available for $p_{\rm phys}=10^{-3}$ and $5\times10^{-4}$ in Ref.~\cite{Kanasugi2026}; for $p_{\rm phys}=10^{-4}$ we follow the same reference in reusing the performance calibrated at $5\times10^{-4}$, which is conservative since $\alpha_{\rm RUS}$ decreases steeply with the physical error rate. 
Those data are moreover obtained at a fixed code distance $d_{\rm code}=21$, whereas the distances selected below range from $13$ to $31$; we reuse the same values throughout. 
At $p_{\rm phys}=5\times10^{-4}$ the selected distances are $21$ and $23$, so the reuse is essentially exact there.
The data likewise span $\theta_L\in[10^{-7},10^{-1}]$ and are extrapolated as constants below that range.

Because $P_{j\tau}$ accumulates with the total rotation angle and the evolution time, the mitigation overhead grows exponentially with $\Lambda$ and $T_{\max}$, and hence with the system size and the inverse of the target precision. 
This ceiling is intrinsic to the partially fault-tolerant mitigation-based setting and applies equally to the 2-step SPE, which shares the same $\Lambda$ and a comparable $T_{\max}$ (Table~\ref{tab:app-circuit}).

We optimize the allocation between $\epsilon_{\rm Trotter}$ and $\epsilon_{\rm alg}$, the time step $\tau$, and the choice between the two prescriptions for $\theta_{\rm th}$~\cite{Toshio2026, Kanasugi2026} to minimize $T_{\text{total}}$ subject to Eq.~\eqref{eq:eq:difference-tau-constraint-Hderivative}.
The search is exhaustive over the two prescriptions and over a grid of $\tau$ spanning $0.06$ to $32$, with the error allocation minimized continuously for each combination.
The runtime of the single longest sampled circuit is also considered, which is given by
\begin{align} \label{eq:shot-execution-time}
T_{\text{shot}} &= d_{\rm code}\, C_{\rm step}\, (2d+1)\tau\sqrt{\frac{W}{\epsilon_{\rm Trotter}}} \nonumber\\
&= d_{\rm code}\, C_{\rm total} \ [\SI{}{\micro\second}].
\end{align}
In what follows, the execution times reported in our resource estimates (see Table~\ref{tab:hubbard_resource_estimation_Pphys3}) are obtained by numerically evaluating Eq.~\eqref{eq:total-execution-time} with the system parameters and error budget specified below. 

The sampled circuits are mutually independent, being drawn non-adaptively from the fixed distribution $|\hat{F}_j|$, and can therefore be distributed over multiple QPUs operating without coherent interconnection~\cite{Kanasugi2026}. For a total physical-qubit budget $\mathcal{Q}_{\rm budget}$, the number of QPUs that fit is $k^\ast= \lfloor \mathcal{Q}_{\rm budget}/N_{\rm phys}\rfloor$, and the time to solution is reduced to $T_{\rm total}/k^\ast$. We adopt $\mathcal{Q}_{\rm budget}=5\times10^{5}$ as an assumed single-cryostat capacity, following Ref.~\cite{Kanasugi2026}: it lies above the $1$--$2\times10^{5}$ that present-day refrigerator designs accommodate and below the $10^{6}$ scale requiring interconnects between refrigerators~\cite{Mohseni2025}. This form of parallelism is limited for adaptively scheduled schemes such as IPE, BPE, or BPDE, in which circuit parameters are updated from earlier outcomes.

\subsection{Resource Estimation for 2D Fermi-Hubbard Model}\label{subsec:estimation-condition}

We apply the general resource estimation procedure to the Fermi-Hubbard model on a two-dimensional square lattice~\cite{Campbell2022}, a canonical system in condensed matter physics.

\subsubsection{Estimation Conditions and Hamiltonian Parameters}

The target Hamiltonian is given by Eq.~\eqref{eq:Fermi-Hubbard-Hamiltonian} on an $N_{\rm size}\times N_{\rm size}$ square lattice.
We apply the Jordan-Wigner transformation with a fixed one-dimensional ordering of the lattice sites, giving
\begin{align}\label{eq:2D-Hubbard-JW}
H (h, U) = -\frac{h}{2} \sum_{\langle i,j \rangle,\sigma} &\qty(X_{i,\sigma}Z^{\leftrightarrow}_{i,j,\sigma} X_{j,\sigma} + Y_{i,\sigma}Z^{\leftrightarrow}_{i,j,\sigma} Y_{j,\sigma}) \nonumber\\
+ \frac{U}{4} \sum_{i} &\,  Z_{i,\uparrow}Z_{i,\downarrow},
\end{align}
where the sum $\langle i,j\rangle$ runs over nearest-neighbor pairs of the square lattice, and the Jordan-Wigner string $Z^{\leftrightarrow}_{i,j,\sigma} = \prod_{k=i+1}^{j-1} Z_{k,\sigma}$ acts on the same-spin sites lying strictly between $i$ and $j$ in this ordering (taking $i<j$ without loss of generality), reducing to the identity for consecutive sites $j=i+1$.
This model is implemented with $N_{\text{sys}} = 2 N_{\text{size}}^2 +1$ logical qubits, and the corresponding $L_1$-norm is given by $\Lambda = (4h + U/4)N_{\text{size}}^2$ at half filling $\mu=U/2$. Under the assumption of $h=1, U=4$, this simplifies to $\Lambda = 5N_{\text{size}}^2$. 

Our resource estimation assumes an initial-state fidelity $\eta=|\braket{E_0(\lambda)}{\psi_0}|^2 = 0.9$. This value is a requirement placed on state preparation rather than one attained by the Givens-rotation circuit used in Sec.~\ref{sec:numerical-demonstration}, whose overlap with the interacting ground state decays with system size; reaching $\eta=0.9$ presupposes the adiabatic state preparation, which is analyzed in Ref.~\cite{Yoshioka2024} for the closely related $J_1$--$J_2$ Heisenberg model. In this reference, an infidelity $1-|\braket{\psi_{\rm GS}}{\psi}|\sim0.1$, corresponding to $\eta\simeq0.81$, is estimated to require an evolution time of at most a few tens in units of $J_1^{-1}$ for $N\sim100$ sites; since that time depends on the target infidelity only logarithmically, $\eta=0.9$ costs a further factor of $1.3$. 
The runtimes in Table~\ref{tab:hubbard_resource_estimation_Pphys3} exclude this preparation. Since every circuit execution starts from a freshly prepared state, the preparation cost grows with the number of executions $N_{\rm circ}$, which is several times larger for the 2-step SPE reference (Table~\ref{tab:app-sampling}); including it would therefore not reduce the advantage of our method.

We report results for $p_{\rm phys}=10^{-3}$, $5\times10^{-4}$, and $10^{-4}$. The first is the value typically assumed for present-day superconducting devices, the second is the lowest rate for which the SMM performance has been characterized in Ref.~\cite{Kanasugi2026}, and the third sits at the upper end of what is currently attainable: two-qubit gate errors below $10^{-4}$ have been reported for trapped ions~\cite{Hughes2025}, although the physical error rate relevant to surface-code operation includes measurement and idling errors as well, and this level remains beyond reach for the superconducting devices assumed by the STAR architecture.

We aim for an energy difference estimation with a required precision of $\epsilon = 10^{-2}$ (in units of the hopping $h=1$). The same numerical value is adopted in Refs.~\cite{Yoshioka2024, Toshio2025}, which estimate the ground-state energy of the two-dimensional Fermi-Hubbard model at $U/h=4$ on lattices of comparable size by phase estimation, there as the target precision for a single energy eigenvalue rather than an energy difference; we use it here as the target accuracy for the difference $\Delta E$ itself, which is the quantity directly relevant to the gradient and force estimation that motivates our method~\cite{OBrien2022}.
For the 2-step SPE reference we likewise assign each individual energy estimate the precision $\epsilon_{\rm E}=\epsilon$, rather than the $\epsilon_{\rm E}\simeq\epsilon/2$ that resolving the difference to $\epsilon$ would require [Sec.~\ref{subsec:limitations-existing-methods}], which is conservative in favor of SPE.

We employ the lowest-order finite-difference approximation and adopt the $U$-direction shift, taking $\delta U$ as large as the finite-difference bias permits, as prescribed in Sec.~\ref{subsubsec:budget-gradient}. 
Evaluating the bound of Eq.~\eqref{eq:fd_deltalambda_var_bound} requires the gap $\Delta_{\rm gap}$ and the variance $\mathrm{Var}_{E_0}(\partial_U H)$.
Strictly, $\Delta_{\rm gap}$ is the gap to the lowest excitation coupled by $\partial_U H$, a neutral excitation that may lie below the charge gap in the antiferromagnetically ordered two-dimensional model. We nevertheless take the charge-gap scale as a representative value, using the one-dimensional Lieb--Wu value $\Delta_c\approx 1.29$ at $U=4,\,h=1$~\cite{LiebWu1968, EsslerBook2005}, so that the resulting $\delta U$ is an estimate rather than a rigorous bound. A smaller gap would require a smaller $\delta U$, which at fixed $\epsilon$ mainly lowers the attainable gradient precision $\epsilon_{\rm g}=\epsilon/\delta U$ and leaves the runtimes essentially unchanged.
Since $\partial_U^2 H = 0$ and $\partial_U H = \tfrac14\sum_i \hat Z_{i\uparrow}\hat Z_{i\downarrow}$ differs from $\sum_i n_{i\uparrow}n_{i\downarrow}$ by a constant within the particle-number sector, its variance is that of a sum of $N_{\rm size}^2$ projectors, each at most $1/4$, and is estimated, using the short-range nature of the ground-state correlations, as $\mathrm{Var}_{E_0}(\partial_U H)\le N_{\rm size}^2/4$.
The bias on $\Delta E$, obtained by multiplying Eq.~\eqref{eq:fd_deltalambda_var_bound} by $\delta U$, is therefore $\mathrm{Var}_{E_0}(\partial_U H)\,\delta U^{2}/\Delta_{\rm gap}$.
Allocating $10\%$ of the precision budget to it fixes the shift at $\delta U = 1.20$, $0.90$ and $0.72\times10^{-2}$ for the $6\times6$, $8\times8$ and $10\times10$ lattices, and leaves $\epsilon_{\rm alg}+\epsilon_{\rm Trotter}\le 9.0\times10^{-3}$.
For the Trotter-error norm $W$, we adopt the values in Table~6 of Ref.~\cite{Kivlichan2020} (fermionic swap network, $U=4$) as representative values for our implementation.

\subsubsection{Resource Estimation Results}

\begin{table*}[htbp]
\renewcommand{\arraystretch}{1.4}
\caption{\label{tab:hubbard_resource_estimation_Pphys3}%
Resource estimation for the energy-difference estimation on the two-dimensional Fermi-Hubbard model at $h=1$, $U=4$, half filling, with target precision $\epsilon=10^{-2}$ on $\Delta E$. The parameter shift is fixed by allocating $10\%$ of $\epsilon$ to the finite-difference bias, giving $\delta U = 1.20$, $0.90$ and $0.72\times10^{-2}$ for the $6\times6$, $8\times8$ and $10\times10$ lattices. 
``Ours'' denotes the direct difference estimation of Sec.~\ref{sec:proposed-method} and ``2-SPE'' the two-step SPE reference. The prescription column indicates which of the two rules for choosing $\theta_{\rm th}$ of Ref.~\cite{Kanasugi2026} yields the shorter time (A: accuracy prioritized, S: speed prioritized); the time step $\tau$ (for our method), the truncation order $d$, and the allocation of the precision budget are optimized jointly, with $\epsilon_{\rm alg}+\epsilon_{\rm Trotter}\le 9.0\times10^{-3}$.
The optimizer assigns $82$--$95\%$ of this budget to $\epsilon_{\rm alg}$; the allocation for each entry is listed in Table~\ref{tab:app-circuit} of Appendix~\ref{app:logical}.
The initial-state fidelity is $\eta=0.9$; at the two lower error rates, varying $\eta$ over $0.8$--$0.95$ changes the execution times of our method by factors of $0.8$--$1.8$. 
$T_{\rm shot}$ is the runtime of the longest sampled circuit [Eq.~\eqref{eq:shot-execution-time}]; $T_{\rm total}$ is the time on a single QPU, obtained by running all sampled circuits serially [Eq.~\eqref{eq:total-execution-time}]; $k^\ast$ is the number of QPUs that fit within a budget of $5\times10^{5}$ physical qubits, over which the circuits may be distributed without coherent interconnection, and $T_{\rm total}/k^\ast$ is the resulting time to solution.
Each sampled circuit is executed twice, once for the real and once for the imaginary part (Sec.~\ref{subsec:simulation-setup}), and $N_{\rm sample}=C(\mathcal{F}/\eta)^2$ with $C=32$.
The entries at $p_{\rm phys}=10^{-3}$ are retained to show how the mitigation overhead $\gamma^2=e^{4P}$ grows with $\Lambda T_{\max}p_{\rm phys}$; they are not practical estimates, and the discussion in the text rests on the two lower error rates.\\
$^{\dagger}$ $k^\ast = 0$: $N_{\rm phys}$ exceeds the assumed budget (by under $2\%$ for $8\times8$ and by $53$--$75\%$ for $10\times10$), so a single QPU does not fit and no time to solution is quoted.
}
\begin{ruledtabular}
\begin{tabular}{llcccccrrrr}
Lattice & Method & Prescr. & $\tau$ & $d$  &
$d_{\rm code}$ & $N_{\rm phys}$ & $k^\ast$ & \multicolumn{1}{c}{$T_{\rm shot}$} &
\multicolumn{1}{c}{$T_{\rm total}$} & \multicolumn{1}{c}{$T_{\rm total}/k^\ast$} \\
 & & & & & & & & [sec] & [days] & [days] \\
\hline
\multicolumn{11}{c}{$p_{\rm phys}=10^{-3}$}\\
\hline
$6\times6$   & Ours  & A & $0.25$              & $544$    & $27$ & $2.7\times10^{5}$ & $1$
& $5.2\times10^{4}$
& $2.09\times10^{6}$  & $2.09\times10^{6}$ \\
             & 2-SPE & A & $5.8\times10^{-3}$ & $23190$  & $29$ & $3.1\times10^{5}$ & $1$
             & $8.1\times10^{4}$
             & $1.10\times10^{7}$  & $1.10\times10^{7}$ \\
$8\times8$   & Ours  & S & $0.125$             & $1084$    & $29$ & $5.1\times10^{5}$ & ---$^{\dagger}$ & $1.2\times10^{5}$ & $2.97\times10^{11}$ & --- \\
             & 2-SPE & S & $3.3\times10^{-3}$ & $41250$  & $29$ & $5.1\times10^{5}$ & ---$^{\dagger}$& $1.8\times10^{5}$ & $2.05\times10^{12}$ & --- \\
$10\times10$ & Ours  & A & $0.06$            & $2252$   & $29$ & $7.6\times10^{5}$ & ---$^{\dagger}$& $2.2\times10^{5}$ & $1.22\times10^{18}$ & --- \\
             & 2-SPE & S & $2.1\times10^{-3}$ & $64488$  & $31$ & $8.7\times10^{5}$ & ---$^{\dagger}$& $3.6\times10^{5}$ & $1.51\times10^{19}$ & --- \\
\hline
\multicolumn{11}{c}{$p_{\rm phys}=5\times10^{-4}$}\\
\hline
$6\times6$   & Ours  & A & $6$                & $29$     & $23$ & $1.9\times10^{5}$ & $2$ & $1.3\times10^{5}$ & $123$              & $61.5$ \\
             & 2-SPE & S & $5.8\times10^{-3}$ & $23186$ & $21$ & $1.6\times10^{5}$ & $3$ & $7.4\times10^{4}$ & $603$               & $201$ \\
$8\times8$   & Ours  & A & $4$                & $41$    & $23$ & $3.2\times10^{5}$ & $1$ & $3.0\times10^{5}$ & $848$               & $848$ \\
             & 2-SPE & A & $3.3\times10^{-3}$ & $41251$  & $23$ & $3.2\times10^{5}$ & $1$ & $3.9\times10^{5}$ & $3859$              & $3859$ \\
$10\times10$ & Ours  & A & $1$                & $139$     & $23$ & $4.8\times10^{5}$ & $1$ & $5.8\times10^{5}$ & $6478$              & $6478$ \\
             & 2-SPE & A & $2.1\times10^{-3}$ & $64487$  & $23$ & $4.8\times10^{5}$ & $1$ & $6.3\times10^{5}$ & $2.60\times10^{4}$  & $2.60\times10^{4}$ \\
\hline
\multicolumn{11}{c}{$p_{\rm phys}=10^{-4}$}\\
\hline
$6\times6$   & Ours  & S & $8$                & $21$     & $13$ & $6.2\times10^{4}$ & $8$ & $2.2\times10^{4}$ & $13.8$              & $1.72$ \\
             & 2-SPE & S & $5.8\times10^{-3}$ & $26368$  & $13$ & $6.2\times10^{4}$ & $8$ & $2.8\times10^{4}$ & $72.1$              & $9.02$ \\
$8\times8$   & Ours  & S & $6$                & $26$     & $15$ & $1.4\times10^{5}$ & $3$ & $6.7\times10^{4}$ & $52.7$              & $17.6$ \\
             & 2-SPE & S & $3.3\times10^{-3}$ & $45139$  & $15$ & $1.4\times10^{5}$ & $3$ & $8.4\times10^{4}$ & $264$               & $88.0$ \\
$10\times10$ & Ours  & S & $4$                & $38$     & $15$ & $2.0\times10^{5}$ & $2$ & $1.4\times10^{5}$ & $148$               & $73.8$ \\
             & 2-SPE & S & $2.1\times10^{-3}$ & $68547$  & $15$ & $2.0\times10^{5}$ & $2$ & $1.8\times10^{5}$ & $692$               & $346$ \\
\end{tabular}
\end{ruledtabular}
\end{table*}

Table~\ref{tab:hubbard_resource_estimation_Pphys3} summarizes the estimated resources for the Fermi-Hubbard model on a square lattice of various sizes. Our approach of the energy level difference estimation (``Ours'') is compared against the difference calculation by 2-step SPE approach (``2-SPE'').
At $p_{\rm phys}=10^{-4}$, our method requires $13.8$, $52.7$, and $148$ days on a single QPU for the $6\times6$, $8\times8$, and $10\times10$ lattices, against $72.1$, $264$, and $692$ days for the 2-step SPE approach: a reduction by factors of $4.7$--$5.2$. 
Within the assumed qubit budget the available parallelism is $k^\ast=2$--$8$, so that these figures fall to $1.72$, $17.6$, and $73.8$ days, placing the $6\times6$ and $8\times8$ lattices within days and weeks respectively.
At $p_{\rm phys}=5\times10^{-4}$, the lowest rate for which the SMM performance has been characterized~\cite{Kanasugi2026}, the corresponding figures are $123$, $848$, and $6478$ days, with reductions by factors of $4.0$--$4.9$.
At $p_{\rm phys}=10^{-3}$, the mitigation overhead becomes the dominant cost and the required code distance pushes $N_{\rm phys}$ above the assumed budget for the $8\times8$ and $10\times10$ lattices, so those instances are out of reach at that error rate irrespective of runtime; our discussion of practicality below therefore rests on the two lower error rates, where every entry fits within the budget.

The advantage has three distinct sources. 
First, our method constrains only the small phase $\tau\|\partial H/\partial\lambda\|\delta\lambda \lesssim \pi/3$ [Eq.~\eqref{eq:eq:difference-tau-constraint-Hderivative}] rather than $\tau\|H\|\lesssim\pi/3$, so that the time step $\tau$ can be taken up to about three orders of magnitude larger and the truncation order is correspondingly smaller, which reduces the sampling complexity $N_{\rm sample}\propto\log^2 d$. 
Second, the two time evolutions are merged into a single sequence of rotations (Appendix~\ref{app:merged-rotation}), halving the number of analog rotations; the 2-step SPE reference admits no such merging and additionally requires two separate circuits. 
Third, in the small-shift regime, the joint rotations are the smallest in the circuit, $\theta^{-}_k\lesssim10^{-6}$, and are therefore executed at the minimum clock count.

It is instructive to compare with the cost of the corresponding classical calculation.
Reference~\cite{Yoshioka2024} reports DMRG runtimes, extrapolated from simulations that reach residual energy errors of 0.22--1.62, for the ground-state energy of the two-dimensional Fermi-Hubbard model on a cylinder at $U/h=4$ and a target accuracy of $0.01$ of $2.5$, $1.74\times10^{4}$, and $8.2\times10^{4}$ days on a single CPU for the $6\times6$, $8\times8$, and $10\times10$ lattices. Comparing single-processor figures on both sides, our method at $p_{\rm phys}=10^{-4}$ falls below the classical estimate by factors of $330$ and $556$ for the $8\times8$ and $10\times10$ lattices, while exceeding it for the smallest lattice; the crossover therefore lies between $6\times6$ and $8\times8$. 
Reference~\cite{Yoshioka2024} further notes that a GPU implementation is expected to shorten these times by a factor of ten; applying it still leaves our method a margin of $33$--$56$.
At $p_{\rm phys}=5\times10^{-4}$ the margins are $20$ and $13$, and the GPU factor would leave our method a margin by factors of about $2$ and $1.3$.

Two caveats must be attached to this comparison. First, the quoted times are those required to obtain the ground state to that accuracy; Ref.~\cite{Yoshioka2024} notes that estimating the energy alone is considerably cheaper, and that a GPU implementation and a multi-node parallelization are each expected to shorten them by about an order of magnitude. Second, and more importantly, the derivative used here as a benchmark is, by the Hellmann--Feynman theorem, a static ground-state expectation value of a local operator [Eq.~\eqref{eq:HF-double-occupancy}]; once a DMRG calculation has produced the ground state, such quantities follow at negligible additional cost, so this particular derivative is not itself a target for which a quantum computer is required. The comparison should therefore be read as locating the regime in which the two approaches reach comparable cost on a common model, rather than as a claim of advantage for this observable. The advantage the method is designed for lies with energy differences that a classical ground-state calculation does not deliver as a by-product, such as forces on an accurate correlated potential energy surface, response properties, and differences between excited-state surfaces, for which the relevant classical cost is that of repeating the whole calculation.

Extending this resource analysis to molecular Hamiltonians, where the force-operator $L_1$-norm is found empirically to grow only weakly with system size~\cite{OBrien2022}, is a particularly promising direction for geometry optimization and ab initio molecular dynamics, which we leave to future work.

\section{Discussion and Conclusion}
\label{sec:discussion}

In this work, we have introduced a quantum algorithm for the efficient and precise estimation of energy level differences in quantum systems, specifically tailored for execution on early-FTQC devices. Our method departs from conventional approaches that estimate individual energy levels and then compute their differences, instead directly targeting the difference through a refined statistical processing technique.
In the regime of small parameter shifts, this difference provides the energy gradient, making our method particularly suited to force and response-property calculations.

The core innovation lies in the read-out placed on the interferometer, and in establishing a refined classical post-processing. This is achieved by employing a modified Hadamard test with a composite unitary operator, which effectively probes the relative phases accumulated from two different time evolutions. Numerical simulations on the Fermi-Hubbard model showed that, at a common target precision, the direct estimation requires a truncation order smaller by up to nearly four orders of magnitude than a two-step application of SPE.
This is a direct consequence of our improved scaling for the truncation order of the discrete Fourier transform: unlike SPE-based approaches that scale with the operator norm of the full Hamiltonian, our method scales with the operator norm of the Hamiltonian derivative and the parameter shift. This renders the precise characterization of small energy shifts, vital for understanding subtle physical phenomena and chemical reactions, practically feasible on early-FTQC quantum hardware. The ability to jointly estimate both energy eigenvalues and their differences further enhances the diagnostic power of our method, offering a holistic view of the energy landscape.

The resource estimation, conducted within the framework of the STAR architecture, delineates the regime in which our algorithm becomes practical for the 2D Fermi-Hubbard model. On lattices up to $10\times10$ sites, the method reduces the total QPU time relative to the conventional 2-step SPE approach by factors of about $4$ to $5.2$, the reduction stemming from the relaxed time-step constraint and from the merging of the two controlled evolutions, whose joint rotations carry the entire signal at angles $\theta^-_k=\mathcal{O}(\delta\lambda)$ and are therefore executed at the minimum clock count. At a physical error rate of $10^{-4}$, the $6\times6$ and $8\times8$ lattices are reached in days and weeks once the sampled circuits are distributed over the number of QPUs that fit within the assumed qubit budget, reaching a cost comparable to, and on the larger lattices below, the corresponding DMRG estimate.
This parallelization is available to our non-adaptive scheme but is limited for adaptively scheduled alternatives, such as IPE, BPE or BPDE.

Looking ahead, several promising avenues for future development can further enhance the capabilities and applicability of our method.
One critical direction involves advancing the implementation of time-evolution operators and optimizing the preparation of initial states. This includes exploring techniques beyond low-order Trotterization~\cite{Berry2015, Campbell2019, Childs2019, Childs2021}, developing more sophisticated unitary compilation strategies~\cite{Low2017, Low2019, Gilyen2019}, and investigating advanced protocols for initial state preparation~\cite{Mcclean2016, Mitarai2018, Ge2019, Lin2020, Dong2022, Fomichev2024, Ollitrault2024}, potentially leveraging classical pre-computation~\cite{Dong2022, Fomichev2024, Ollitrault2024} or variational optimization~\cite{Mcclean2016, Mitarai2018}. Such improvements are vital for optimizing circuit depth, gate counts, and overall sampling complexity, especially when tackling complex molecular systems with high-norm Hamiltonians.
A particularly impactful application lies in molecular systems, crucial for tasks such as molecular geometry optimization~\cite{Pulay1969, Schlegel2011, Pulay2014, OBrien2019, Delgado2021, OBrien2022} and ab initio molecular dynamics (AIMD) simulation~\cite{Car1985, Marx2009}. These applications typically demand the precise estimation of forces derived from energy differences. For AIMD, a minimum of $10^6$ iterations is required, making the total run time dominated by this large number of iterations. This iterative bottleneck constitutes a challenge~\cite{Kar2023, Mo2024}, even for classical AIMD employing approximating methods like density functional theory~\cite{Hohenberg1964, Kohn1965}. 
To achieve practical computational speeds for such demanding applications on quantum computers, drastic reduction in the total iterative cost, beyond per-iteration efficiency gains, is essential, fully leveraging quantum computational advantage.
Finally, our method extends naturally beyond static energy differences to dynamical response properties and non-equilibrium dynamics~\cite{Roggero2019, Endo2020, Kosugi2020} induced by shifts in external parameters such as electric or magnetic fields~\cite{Helgaker1984}; while such responses have long been studied with classical methods, our approach offers a route to computing them in the strongly correlated regimes where classical approaches become unreliable. Continued benchmarking against state-of-the-art classical methods~\cite{Lee2023, Yoshioka2024} will also be crucial to precisely delineate the regimes where quantum advantage can be unequivocally demonstrated. The possibilities raised above are beyond the scope of this paper and are left for future work.

In conclusion, the methodology presented in this paper provides a practical path to satisfying the stringent precision requirements of advanced scientific applications within the constraints of current quantum hardware. By shifting the computational focus from energy eigenvalues to relative energy differences, this approach optimizes the utility of early-FTQC quantum devices for extracting high-fidelity physical insights. As we move toward the early-FTQC era, we anticipate that our strategies will serve as a foundational step toward the large-scale simulation of complex materials and chemical systems.

\section*{Acknowledgement}
The authors are profoundly indebted to Shintaro~Sato and Tennin~Yan for their invaluable support in fostering a conducive research environment.
N.M. is grateful to Riki~Toshio, Yutaro~Akahoshi, Koki~Chinzei, and Tran~Quachoan for helpful comments and discussions. 
N.M. acknowledges Masatoshi~Ishii for technical support in performing numerical simulations.

\appendix

\section{Overview of Bayesian Phase Estimation (BPE)}
\label{app:bpe_overview}

Bayesian Phase Estimation (BPE)~\cite{Svore2014, Wiebe2016, Paesani2017} is a quantum-classical hybrid algorithm designed to estimate the eigenvalues (phases) of a unitary operator $U$ acting on a quantum state $|\psi\rangle$, such that $U|\psi\rangle = e^{i\phi}|\psi\rangle$. Unlike traditional IPE algorithms, which reconstruct the binary digits of the phase $\phi$ one bit at a time, starting from the least significant bit, BPE directly infers the phase and provides an estimate of its uncertainty using Bayesian inference.

The idea behind BPE is to track the prior probability distribution over the possible values of the unknown phase $\phi$. After performing a quantum experiment and obtaining a measurement outcome, this prior distribution is updated to a posterior distribution using Bayes' theorem. This process is iterated, with the posterior from one step becoming the prior for the next, until the uncertainty in $\phi$ reaches a desired level.

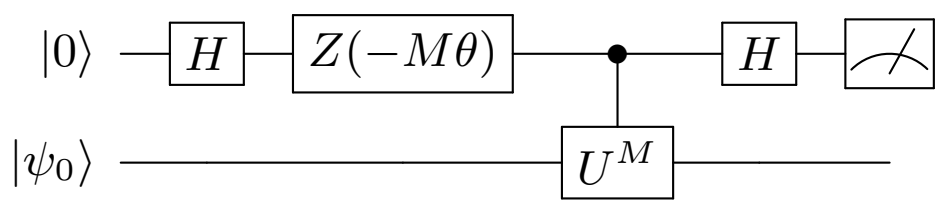
\begin{figure}[tbp]
\centering
\[
    	\Qcircuit @C=1em @R=.7em {\\
  	\lstick{\ket{0}} & \gate{H} & \gate{Z(-M\theta)} & \ctrl{1}  & \gate{H} & \meter \\
  	\lstick{\ket{\psi_0}} & \qw              & \qw   & \gate{U^M}            & \qw     & \qw
	}
    \]
    \caption{
    Quantum circuit for BPE. This circuit is used to estimate the phase $\phi$ of a unitary operator $U$ acting on an eigenstate $|\psi_0\rangle$. It consists of a Hadamard gate, a $Z(-M\theta)$ rotation, a controlled-$U^M$ operation, and a final Hadamard gate before measurement of the ancilla qubit.
    }
    \label{fig:circuit-BPE}
\end{figure}

The core components are summarized as follows:
\begin{enumerate}
    \item \textbf{Quantum Experiment}: The quantum circuit for BPE is shown in Fig.~\ref{fig:circuit-BPE}. An ancilla qubit, initialized in $|0\rangle$, is placed in superposition using a Hadamard gate,  rotated by a $Z(-M\theta)$ gate, exposed to a controlled-$U^M$ operation, and finally measured after another Hadamard gate. 
    Here, $|\psi_0\rangle$ is an eigenstate of $U$ corresponding to the phase $\phi$ to be estimated. The measurement outcome $E \in \{0, 1\}$ provides probabilistic information about $\phi$.

    \item \textbf{Likelihood Function}: Given the measurement outcome $E$ from an experiment with parameters $M$ and $\theta$, the likelihood function $P(E|\phi; M, \theta)$ quantifies the probability of observing $E$ if the true phase were $\phi$. For the quantum circuit in Fig.~\ref{fig:circuit-BPE}, if the measurement outcome if found to be $E=0$ or $E=1$, the likelihood function is given by
    \begin{align}
    P(0|\phi; M, \theta) &= \frac{1 + \cos\qty[M(\phi - \theta)]}{2}, \\
    P(1|\phi; M, \theta) &= \frac{1 - \cos\qty[M(\phi - \theta)]}{2}.
    \end{align}
    These functions describe the oscillatory nature of the measurement probabilities typical for interferometric phase estimation.

    \item \textbf{Bayesian Update}: If $P(\phi)$ is the prior distribution for $\phi$, and a measurement $E$ is observed from an experiment with $M$ and $\theta$, the posterior distribution $P(\phi|E; M, \theta)$ is calculated using Bayes' theorem:
    \begin{align}
    P(\phi|E; M, \theta) = \frac{P(E|\phi; M, \theta) P(\phi)}{\int P(E|\phi; M, \theta) P(\phi) \, d\phi}.
    \end{align}
    The denominator serves as a normalization factor. The resulting posterior distribution then becomes the prior distribution for the next iteration.

    \item \textbf{Classical Post-processing}: A Gaussian distribution $N(\mu, \sigma^2)$ can be used to approximate the prior distribution. The update of the mean $\mu$ and variance $\sigma^2$ (or an equivalent representation for a circular variable) is performed classically after each quantum experiment. Unlike the IPE algorithm, which estimates each bit of $\phi$ sequentially, BPE provides a continuous probability distribution over $\phi$, from which the most likely phase and its uncertainty can be extracted.

\end{enumerate}

A key feature of BPE is its adaptive nature. The parameters $M$, representing the number of unitary applications, and the reference phase $\theta$ for the next experiment are chosen based on the current posterior distribution, aiming to maximize the information gain. Heuristics like the ``Particle Guess Heuristic''~\cite{Wiebe2014} can be employed to select near-optimal parameters without complex pre-computation.

\section{Overview of Bayesian Phase Difference Estimation (BPDE)}
\label{app:bpde_overview}

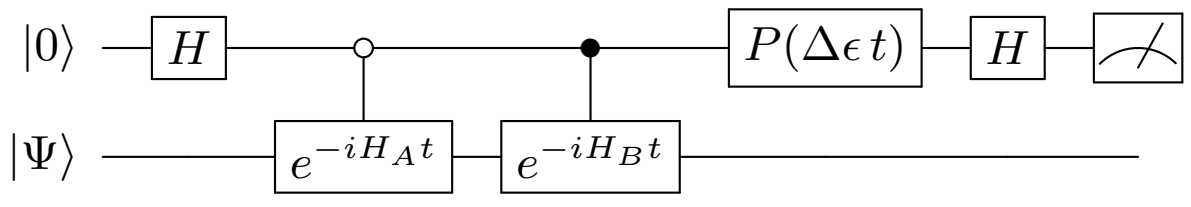
\begin{figure}[bp]
\centering
\[
    	\Qcircuit @C=1em @R=.7em {\\
  	\lstick{\ket{0}}   & \gate{H} & \ctrlo{1} & \ctrl{1} & \gate{P(\Delta\epsilon \, t)} & \gate{H} & \meter \\
  	\lstick{\ket{\Psi}} & \qw      & \gate{e^{-iH_A t}} & \gate{e^{-iH_B t}} & \qw                   & \qw     & \qw
	}
    \]
    \caption{
    Quantum circuit for BPDE. This circuit is designed to directly estimate the energy difference between two Hamiltonians, $H_A$ and $H_B$. It incorporates a Hadamard gate, controlled time-evolution operations under $H_A$ and $H_B$, a phase shift gate $P(\Delta\epsilon \, t)$ with parameter $\Delta\epsilon$, and a final Hadamard gate before measurement of the ancilla qubit.
    }
    \label{fig:circuit-BPDE}
\end{figure}

Bayesian Phase Difference Estimation (BPDE)~\cite{Sugisaki2021, Sugisaki2022} is a quantum-classical hybrid algorithm designed to directly estimate the energy gap between two electronic states~\cite{Sugisaki2021} or energy differences between two different molecular geometries~\cite{Sugisaki2022}. It extends the principles of BPE to focus specifically on the difference between energy eigenvalues, rather than individual eigenvalues. This approach has been explored for applications such as numerical energy gradient calculations and geometry optimization, positioning itself as an alternative to methods inferring differences from separate energy estimations.

BPDE shares the Bayesian inference framework with BPE, tracking a prior probability distribution over the unknown energy difference $\Delta E$. This prior distribution is iteratively updated to a posterior distribution based on measurement outcomes from quantum experiments. The key distinction lies in the design of the quantum circuit, which is specially tailored to encode information about the energy difference.

The core components of BPDE are given as follows:
\begin{enumerate}
    \item \textbf{Quantum Experiment Design}: BPDE utilizes a quantum circuit that incorporates time evolution under two different Hamiltonians, typically $H_A$ and $H_B$, corresponding to two different molecular geometries (e.g., at bond length $R$ and $R+\Delta r$ for geometry optimization) (see Fig.~\ref{fig:circuit-BPDE}). The circuit is designed such that the probability of a specific measurement outcome is maximized when the phase rotation angle $\Delta\epsilon$ corresponds to the energy difference.
    The initial state $|\Psi\rangle$ is chosen such that it approximates both $|\Psi^{(A)}\rangle$ and $|\Psi^{(B)}\rangle$, where $|\Psi^{(A)}\rangle$ ($|\Psi^{(B)}\rangle$) denotes the target state of $H_A$ ($H_B$).
    For the initial state, a series of time evolution operators is applied, implemented with the Trotter decomposition. 
    Here, the evolution time $t$ is set based on the standard deviation of the prior distribution $\sigma$, e.g., $t = 1.8/\sigma$.
    Then, $P(\Delta\epsilon t)$ represents a phase shift gate ($e^{i \Delta\epsilon t}$) whose parameter $\Delta\epsilon$ is optimized through Bayesian inference, as discussed below.

    \item \textbf{Likelihood Function}: For the above quantum circuit, the probability of the ancillary qubit state can be expressed in a form that reveals the energy difference. For the initial state $\ket{\Psi}$ that approximates both $\ket{\Psi^{(A)}} = \sum_j c_j^{(A)} \ket{\Psi_j^{(A)}}$ and $\ket{\Psi^{(B)}} = \sum_k b_k^{(B)} \ket{\Psi_k^{(B)}}$, the probability of obtaining the $|0\rangle$ state in the measurement of the ancillary qubit is, in the approximation of Ref.~\cite{Sugisaki2022}, given by
    \begin{align}
    \hspace{7mm}\text{Prob}(0) = \frac{1}{2}\Bigg(1 + &\sum_{j,k} |c_j^{(A)}|^2 |b_k^{(B)}|^2 \nonumber\\
    & \hspace{2mm} \times\cos\left[ (E_k^{(B)} - E_j^{(A)} - \Delta\epsilon)t \right] \Bigg),
    \end{align}
    where $E_j^{(A)}$ and $E_k^{(B)}$ are the energy eigenvalues corresponding to $\ket{\Psi_j^{(A)}}$ and $\ket{\Psi_k^{(B)}}$, respectively. By optimizing the phase shift gate $P(\Delta\epsilon t)$, the likelihood function can be maximized when $\Delta\epsilon$ matches the true energy difference.

    \item \textbf{Bayesian Update and Classical Post-processing}: Similar to BPE, a Gaussian function is typically used to represent the prior distribution for $\Delta E$. After each round of measurements, the mean $\mu$ and standard deviation $\sigma$ of this Gaussian are updated classically using Bayes' theorem. 
    The obtained likelihood function is fitted by a Gaussian function to calculate a posterior distribution.
    This iterative process continues until the uncertainty $\sigma$ falls below a predefined convergence threshold.

\end{enumerate}

Figure~\ref{fig:circuit-BPDE} depicts the conceptual (naive) implementation with two separate controlled time-evolution operators. As noted in Ref.~\cite{Sugisaki2022}, this doubling of the circuit depth offers no advantage over two separate energy estimations from the perspective of the total gate count; in practice, BPDE merges the two evolutions into a single time-evolution circuit in which each controlled-$R_z$ rotation angle is made to depend on the ancilla state, exploiting the shared Pauli-string structure of $H_A$ and $H_B$.
Reference~\cite{Sugisaki2022} gives two realizations: one using a pair of controlled $R_z$ gates with different angles, and one using two uncontrolled $R_z$ gates conjugated by CNOTs, whose angles are the half-sum and the half-difference of the two evolution angles. The latter is equivalent to the construction we adopt (Appendix~\ref{app:merged-rotation}) at the logical level.

\section{Fourier-series approximation of the Heaviside step function} \label{app:Fourier-Heaviside}

In this appendix, we summarize the Fourier-series approximation of the Heaviside function, $F_{\beta,d}(x)$, used in the main text, primarily following Ref.~\cite{Wan2022}. This approximation strategy involves sequentially applying approximations: first using a rescaled error function, then finally converting to a discrete Fourier series.
The function $F_{\beta,d}(x)$ is defined as
\begin{align}
&F_{\beta,d}(x) = \hat{F}_0 + \sum_{j=0}^{d} \hat{F}_{2j+1} \left[e^{i(2j+1)x} - e^{-i(2j+1)x}\right] \label{eq:Fbetad_appendix}
\end{align}
with coefficients
\begin{align}
\hat{F}_0 &:= \frac{1}{2},\nonumber\\
\hat{F}_{2j+1} &:= -i \sqrt{\frac{\beta}{2\pi}}e^{-\beta}\frac{I_j(\beta) + I_{j+1}(\beta)}{2j+1}\;\text{ for $0 \leq j \leq d-1$},\nonumber\\ \hat{F}_{2d+1} &:= -i\sqrt{\frac{\beta}{2\pi}}e^{-\beta}\frac{I_d(\beta)}{2d+1}, \label{eq:Fouriercoefficients_appendix}
\end{align}
where $I_n(\cdot)$ denotes the $n$-th modified Bessel function of the first kind.

\subsection{Chebyshev approximation}

The Heaviside function $\Theta(x)$ is first approximated using a rescaled error function, $\mathrm{erf}(\sqrt{2\beta}x)$, for large $\beta$. 
This error function is then approximated by a finite series of Chebyshev polynomials, resulting in $Q_{\beta,d}(x)$.
This $Q_{\beta,d}(x)$ is defined as
\begin{align} \label{Qbetad_appendix}
Q_{\beta,d}(x) := &\sqrt{\frac{2\beta}{\pi}} 2e^{-\beta}\Bigg[I_0(\beta)x  \nonumber\\
& + \sum_{j=1}^d I_j(\beta)(-1)^j \left(\frac{T_{2j+1}(x)}{2j+1} - \frac{T_{2j-1}(x)}{2j-1}\right)\Bigg],
\end{align}
where $T_n (\cdot)$ denotes the $n$-th Chebyshev polynomial of the first kind. 
The approximation error of $Q_{\beta,d}(x)$ and its relation to $\mathrm{erf}(\sqrt{2\beta}x)$ are bounded, as established in Lemma~5 of Ref.~\cite{Wan2022}, given $d \in \mathbb{N}$ satisfies~\footnote{In this Appendix, we write $\varepsilon_i$ for the approximation errors of the Fourier construction, to distinguish them from the estimation precision $\epsilon$ of the main text.}
\begin{align} \label{d_appendix}
d \geq {\sqrt{t w_{\varepsilon_1}}}, 
\end{align}
for a suitable integer $t$~\footnote{In our numerics, $t$ is treated as a real number, which may change $d$ by at most one.} chosen as in Lemma~5 of Ref.~\cite{Wan2022}, where $w_{\varepsilon_1} := W\left(\frac{8}{\pi \varepsilon_1^2} \right)$ with the principal branch of the Lambert-W function $W(\cdot)$.
The function $ P_{\beta,d}(x) := \frac{1}{2}\left[Q_{\beta,d}(x) + 1\right]\, $ then approximates $\Theta(x)$, with its properties and error bounds described in Theorem~2 of Ref.~\cite{Wan2022}. Specifically, for appropriate $\nu \in (0,1)$ and $\varepsilon_1,\varepsilon_2,\varepsilon_3 \in \mathbb{R}_{>0}$, if $\beta$ satisfies $\beta \geq \frac{1}{4\nu^2}W\big(\frac{2}{\pi\varepsilon_3^2}\big)$ and $d$ satisfies Eq.~\eqref{d_appendix}, then the approximation error $\max_{x\in [-1,-\nu] \cup [\nu,1]} \left| \Theta(x)- P_{\beta,d}(x)\right| \leq \frac{1}{2}(\varepsilon_1 + \varepsilon_2 + \varepsilon_3)$ holds, and $P_{\beta,d}(x)$ is bounded within $\qty[ -\frac{1}{2}(\varepsilon_1 + \varepsilon_2) , 1 + \frac{1}{2}(\varepsilon_1 + \varepsilon_2)]$.

\subsection{Fourier approximation}

The final step converts the Chebyshev polynomial approximation $P_{\beta,d}(x)$ into a Fourier series, $F_{\beta,d}(x) := P_{\beta,d}(\sin x)$.
The conversion leverages Lemma~6 of Ref.~\cite{Wan2022}, which details how a Chebyshev expansion of a function $f(x)$ transforms into a Fourier series for $f(\sin x)$, providing the direct relationship between the Chebyshev and the Fourier coefficients.
The explicit form of $P_{\beta,d}(\sin x)$ from Eq.~\eqref{Qbetad_appendix}, which is reorganized for clarity, allows for the direct derivation of the Fourier-series coefficients $\{\hat{F}_j\}$ presented in Eq.~\eqref{eq:Fouriercoefficients_appendix}. 

The key properties of this Fourier series approximation are summarized in Theorem~3 of Ref.~\cite{Wan2022}. For any $\delta \in (0,\pi/2)$ and $\varepsilon_1,\varepsilon_2,\varepsilon_3 \in \mathbb{R}_{>0}$, if $\beta$ is a real number satisfying
\begin{align} \label{eq:beta_final_theorem}
\beta \geq \max\left\{ \frac{1}{4\sin^2 \delta} W\left(\frac{2}{\pi\varepsilon_3^2}\right), 1\right\}
\end{align}
and $d$ is an integer satisfying Eq.~\eqref{d_appendix}, then $F_{\beta,d}(x)$ has the following properties:
\begin{enumerate}
    \item Error bound: The maximum error is bounded by $\frac{1}{2}(\varepsilon_1 + \varepsilon_2 + \varepsilon_3)$ on $x \in [-\pi + \delta, -\delta] \cup [\delta, \pi - \delta]$.
    \item Value range: $F_{\beta,d}(x)$ is bounded within $\qty[ -\frac{1}{2}(\varepsilon_1 + \varepsilon_2) , 1 + \frac{1}{2}(\varepsilon_1 + \varepsilon_2)]$  for all $x \in \mathbb{R}$.
    \item Sum of Fourier coefficients: The sum of absolute values of the Fourier coefficients is bounded as $ \sum\limits_{j = 0}^d |\hat{F}_{2j+1}| \leq \frac{1}{2}H_{d + {1}/{2}} + \ln 2$, where $H_n := \sum_{j=1}^n j^{-1}$ denotes the $n$-th harmonic number.
\end{enumerate}

\section{Merging the two controlled time evolutions}
\label{app:merged-rotation}

\begin{figure}[bp]
\centering
\[
\text{(a)}\hspace{10mm}
\Qcircuit @C=0.9em @R=1.1em {
  \lstick{\mathrm{anc}} & \qw                        & \ctrl{1} & \qw                        & \ctrl{1} & \qw \\
  \lstick{q}            & \gate{R_z(2\theta^{+}_k)}  & \targ    & \gate{R_z(2\theta^{-}_k)}  & \targ    & \qw
}
\]

\[
\text{(b)}\hspace{11.5mm}
\Qcircuit @C=0.9em @R=1.1em {
  \lstick{\mathrm{anc}} & \qw                              & \multigate{1}{\; e^{-i\theta^{-}_k P_k \otimes Z_{\rm anc}} \;} & \qw \\
  \lstick{\mathrm{sys}} & \gate{e^{-i\theta^{+}_k P_k}}    & \ghost{\; e^{-i\theta^{-}_k P_k \otimes Z_{\rm anc}} \;}       & \qw
}
\]
\caption{
Two realizations of the merged pair of Eq.~\eqref{eq:merged-rotation} for one Pauli term $P_k$ within a Trotter step.
(a) Gate-based realization used in Ref.~\cite{Sugisaki2022}: after a Clifford circuit that maps $P_k$ to a single $Z$ on one system qubit $q$ (omitted here, together with its inverse), the pair becomes two $R_z$ rotations and two CNOTs, with $R_z(\phi)=e^{-i\phi Z/2}$; the CNOTs conjugate the second rotation so that its sign is conditioned on the ancilla.
(b) Realization on the STAR architecture: a Pauli rotation of any weight is a single analog rotation implemented by lattice surgery, so no Clifford conjugation is needed and the joint factor is applied directly to the system and the ancilla together. The two panels are logically equivalent; what differs is that (b) needs two analog rotations per Pauli term, whereas the cost of (a) is dominated by the Clifford ladder and is independent of $\theta^{\pm}_k$.
}
\label{fig:merged-rotation-circuit}
\end{figure}
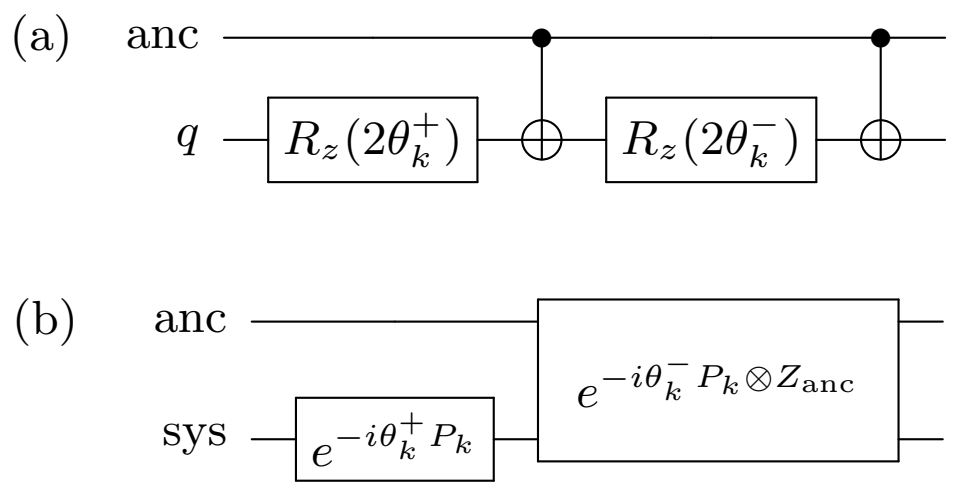

The circuit of Fig.~\ref{fig:circuit-proposed-simple} applies $e^{-ij\tau H(\lambda)}$ conditioned on the ancilla being in $\ket{1}$ and $e^{-ij\tau H(\lambda+\delta\lambda)}$ conditioned on $\ket{0}$. Implemented literally, this requires two controlled evolutions in series. Here we describe the merged implementation used in the resource estimates of Sec.~\ref{subsec:resource-STAR} and its consequences for the STAR architecture.

Expand both Hamiltonians in the same Pauli basis, $H(\lambda)=\sum_k a_k P_k$ and $H(\lambda+\delta\lambda)=\sum_k a_k' P_k$; a parameter shift changes the coefficients but not, in the cases considered here, the set of Pauli strings. For a Trotter step of width $\Delta t$, the circuit must apply $e^{-i a_k \Delta t P_k}$ on the $\ket{1}$ branch and $e^{-i a_k' \Delta t P_k}$ on the $\ket{0}$ branch. This pair is realized by
\begin{align} \label{eq:merged-rotation}
e^{-i \theta^{+}_k P_k}\; e^{-i \theta^{-}_k P_k \otimes Z_{\rm anc}}, \qquad
\theta^{\pm}_k = \frac{a_k' \pm a_k}{2}\,\Delta t ,
\end{align}
where the first factor acts on the system alone and the second acts jointly on the system and the ancilla. Since $Z_{\rm anc}$ takes the values $+1$ on $\ket{0}$ and $-1$ on $\ket{1}$, the exponents add to $-i a_k'\Delta t P_k$ on the former branch and to $-i a_k \Delta t P_k$ on the latter, as required. The two factors share the same Pauli string and hence commute exactly, so the implementation based on Eq.~\eqref{eq:merged-rotation} is exact: neither an approximation nor a reordering of noncommuting terms is involved. Because the substitution is local to each rotation, the ordering of the Pauli terms within a Trotter step is untouched and the construction applies to a product formula of any order. It also guarantees that the two evolutions are generated by the same decomposition and step size, which is the hypothesis underlying the cancellation of time-evolution implementation errors in Eq.~\eqref{eq:trotter-cancellation}.

Equation~\eqref{eq:merged-rotation} is equivalent to the construction of Fig.~3(c) of Ref.~\cite{Sugisaki2022}, used there for BPDE and summarized in Appendix~\ref{app:bpde_overview}, where the joint rotation is realized by conjugating an $R_z$ gate with two CNOTs [Fig.~\ref{fig:merged-rotation-circuit} (a)] and the resulting gate count is reported to be only slightly larger than that of a single-point energy calculation. On the STAR architecture no such conjugation is needed, since a Pauli rotation of any weight is a single analog rotation implemented by lattice surgery [Fig.~\ref{fig:merged-rotation-circuit} (b)]. The merged circuit therefore contains two analog rotations per Pauli-term rotation of the product formula, the same as a single controlled evolution, whereas two controlled evolutions in series would require four; the total rotation angle is correspondingly $\sum_k|\theta^{+}_k| = \Lambda\,\Delta t$ per step rather than twice that.

What distinguishes the present setting from that of Ref.~\cite{Sugisaki2022} is the size of the two angles rather than the construction. There the two Hamiltonians correspond to distinct molecular geometries and $\theta^{+}_k$ and $\theta^{-}_k$ are treated on the same footing, the cost of a rotation being independent of its angle in a $T$-gate-based implementation. For a parameter shift, $|\theta^{-}_k/\theta^{+}_k|=|(a_k'-a_k)/(a_k'+a_k)| =\mathcal{O}(\delta\lambda\,|\partial_\lambda a_k|/|a_k|)$, so that in the small-shift regime the joint rotations are the smallest in the circuit; on an architecture, such as the STAR discussed in Sec.~\ref{sec:resource-estimation}, whose analog rotations cost less at smaller angles, this smallness becomes a resource. They are executed at the minimum clock count, they contribute negligibly to $P_{j\tau}$, and they vanish altogether for any Pauli term whose coefficient does not depend on the shifted parameter. For the $U$-direction shift of Sec.~\ref{subsec:estimation-condition} the hopping coefficients are common to both Hamiltonians, so joint rotations are required only for the $N_{\rm size}^2$ interaction terms and the circuit contains $20N_{\rm size}^2$ analog rotations per Trotter step, against $36N_{\rm size}^2$ for the 2-step SPE reference, in which each controlled rotation must be realized as two analog rotations of half the angle and each eigenvalue requires a separate circuit.

\begin{table*}[htbp]
\renewcommand{\arraystretch}{1.4}
\caption{\label{tab:app-system}%
System-level parameters. Definitions are given in Appendix~\ref{app:logical}.
Energies are in units of the hopping $h=1$ and times in units of $1/h$.}
\begin{ruledtabular}
\begin{tabular}{lcccccccc}
Lattice & $N_L$ & $L$ & $\Lambda$ & $\|\partial_U H\|$ & $W$ & $\delta U$ & $\epsilon_g$ & $\tau_{\max}$ \\
\hline
$6\times6$   & $73$  & $324$ & $180$ & $9$  & $4.2\times10^{3}$ & $1.2\times10^{-2}$ & $8.35\times10^{-1}$ & $9.72$ \\
$8\times8$   & $129$ & $576$ & $320$ & $16$ & $7.5\times10^{3}$ & $9.0\times10^{-3}$ & $1.11$ & $7.29$ \\
$10\times10$ & $201$ & $900$ & $500$ & $25$ & $1.2\times10^{4}$ & $7.2\times10^{-3}$ & $1.39$ & $5.83$ \\
\end{tabular}
\end{ruledtabular}
\end{table*}

\vspace{5mm}

\begin{table*}[htbp]
\renewcommand{\arraystretch}{1.4}
\caption{\label{tab:app-circuit}%
Error budget and per-circuit logical parameters, for the operating points of Table~\ref{tab:hubbard_resource_estimation_Pphys3}.}
\begin{ruledtabular}
\begin{tabular}{llccccccc}
Lattice & Method & $\epsilon_{\rm alg}$ & $\epsilon_{\rm Trotter}$ & $\tau_{\rm Trot}$ & $\bar\theta_L$ & $N_{\rm rot}$ & $T_{\max}$ & $G_{\max}$ \\
\hline
\multicolumn{9}{c}{$p_{\rm phys}=10^{-3}$}\\
\hline
$6\times6$   & Ours  & $8.5\times10^{-3}$ & $4.7\times10^{-4}$ & $3.3\times10^{-4}$ & $8.3\times10^{-5}$ & $720$  & $272$ & $5.87\times10^{8}$ \\
& 2-SPE & $8.5\times10^{-3}$ & $4.5\times10^{-4}$ & $3.3\times10^{-4}$ & $4.6\times10^{-5}$ & $1296$ & $270$ & $1.07\times10^{9}$ \\
$8\times8$   & Ours  & $8.5\times10^{-3}$ & $4.6\times10^{-4}$ & $2.5\times10^{-4}$ & $6.2\times10^{-5}$ & $1280$ & $271$ & $1.40\times10^{9}$ \\
& 2-SPE & $8.5\times10^{-3}$ & $4.5\times10^{-4}$ & $2.5\times10^{-4}$ & $3.4\times10^{-5}$ & $2304$ & $270$ & $2.54\times10^{9}$ \\
$10\times10$ & Ours  & $8.5\times10^{-3}$ & $4.6\times10^{-4}$ & $1.9\times10^{-4}$ & $4.9\times10^{-5}$ & $2000$ & $270$ & $2.77\times10^{9}$ \\
& 2-SPE & $8.5\times10^{-3}$ & $4.5\times10^{-4}$ & $1.9\times10^{-4}$ & $2.7\times10^{-5}$ & $3600$ & $270$ & $5.02\times10^{9}$ \\
\hline
\multicolumn{9}{c}{$p_{\rm phys}=5\times10^{-4}$}\\
\hline
$6\times6$   & Ours  & $7.4\times10^{-3}$ & $1.6\times10^{-3}$ & $6.2\times10^{-4}$ & $1.5\times10^{-4}$ & $720$  & $354$ & $4.12\times10^{8}$ \\
& 2-SPE & $8.5\times10^{-3}$ & $4.5\times10^{-4}$ & $3.3\times10^{-4}$ & $4.5\times10^{-5}$ & $1296$ & $270$ & $1.07\times10^{9}$ \\
$8\times8$   & Ours  & $7.7\times10^{-3}$ & $1.3\times10^{-3}$ & $4.2\times10^{-4}$ & $1.1\times10^{-4}$ & $1280$ & $332$ & $1.00\times10^{9}$ \\
& 2-SPE & $8.5\times10^{-3}$ & $4.5\times10^{-4}$ & $2.5\times10^{-4}$ & $3.4\times10^{-5}$ & $2304$ & $270$ & $2.54\times10^{9}$ \\
$10\times10$ & Ours  & $8.5\times10^{-3}$ & $4.5\times10^{-4}$ & $1.9\times10^{-4}$ & $4.8\times10^{-5}$ & $2000$ & $279$ & $2.88\times10^{9}$ \\
& 2-SPE & $8.5\times10^{-3}$ & $4.5\times10^{-4}$ & $1.9\times10^{-4}$ & $2.7\times10^{-5}$ & $3600$ & $270$ & $5.02\times10^{9}$ \\
\hline
\multicolumn{9}{c}{$p_{\rm phys}=10^{-4}$}\\
\hline
$6\times6$   & Ours  & $8.2\times10^{-3}$ & $8.2\times10^{-4}$ & $4.4\times10^{-4}$ & $1.1\times10^{-4}$ & $720$  & $344$ & $5.60\times10^{8}$ \\
& 2-SPE & $7.5\times10^{-3}$ & $1.5\times10^{-3}$ & $5.9\times10^{-4}$ & $8.2\times10^{-5}$ & $1296$ & $307$ & $6.70\times10^{8}$ \\
$8\times8$   & Ours  & $8.4\times10^{-3}$ & $6.0\times10^{-4}$ & $2.8\times10^{-4}$ & $7.1\times10^{-5}$ & $1280$ & $318$ & $1.44\times10^{9}$ \\
& 2-SPE & $7.8\times10^{-3}$ & $1.2\times10^{-3}$ & $4.0\times10^{-4}$ & $5.5\times10^{-5}$ & $2304$ & $295$ & $1.71\times10^{9}$ \\
$10\times10$ & Ours  & $8.5\times10^{-3}$ & $4.8\times10^{-4}$ & $2.0\times10^{-4}$ & $5.0\times10^{-5}$ & $2000$ & $308$ & $3.07\times10^{9}$ \\
& 2-SPE & $8.0\times10^{-3}$ & $9.6\times10^{-4}$ & $2.8\times10^{-4}$ & $3.9\times10^{-5}$ & $3600$ & $287$ & $3.66\times10^{9}$ \\
\end{tabular}
\end{ruledtabular}
\end{table*}

The ratio above spans two extremes. It vanishes for a Pauli term whose coefficient does not depend on the shifted parameter, as just noted, and reaches unity for a term absent from $H(\lambda)$ altogether, as can happen when the shift breaks a symmetry of $H(\lambda)$: for such a term $a_k=0$ and $\theta^{+}_k=\theta^{-}_k=a_k'\Delta t/2$, so the joint rotation is no longer the smaller of the pair. Equation~\eqref{eq:merged-rotation} nonetheless remains exact, it being enough to take $\{P_k\}$ to be the union of the two supports, and for coefficients depending smoothly on $\lambda$ one has $a_k'=\mathcal{O}(\delta\lambda)$, so both rotations are still executed at the minimum clock count and contribute negligibly to $P_{j\tau}$. The only effect is on the rotation count, which should then be taken as that of the lower-symmetry Hamiltonian. For the global couplings $h$ and $U$ of Sec.~\ref{subsec:estimation-condition} the two supports coincide and the case does not arise.

\section{Logical resource estimates} \label{app:logical}

Table~\ref{tab:app-system} lists the system-level parameters entering the resource estimates of Sec.~\ref{sec:resource-estimation}, and Tables~\ref{tab:app-circuit} and~\ref{tab:app-sampling} the corresponding logical circuit and sampling requirements.
These are independent of the error-correcting architecture and may be used to re-derive physical costs under different assumptions.

In Table~\ref{tab:app-system}, $N_L=2N_{\rm size}^2+1$ is the number of logical qubits, including one ancilla, $L$ the number of Pauli terms after the Jordan--Wigner transformation, $\Lambda=\sum_k|a_k|$ the $\ell_1$ norm, $\|\partial_U H\|$ the operator norm of the shift generator, $W$ the second-order Trotter error norm of Ref.~\cite{Kivlichan2020}, $\delta U$ the finite-difference shift, $\epsilon_g$ the resulting precision on the gradient, and $\tau_{\max}$ the bound of Eq.~\eqref{eq:eq:difference-tau-constraint-Hderivative}.
In Table~\ref{tab:app-circuit}, $\tau_{\rm Trot}=\sqrt{\epsilon_{\rm Trotter}/W}$ is the Trotter step, $\bar\theta_L=\Lambda\tau_{\rm Trot}/N_{\rm rot}$ the mean rotation angle, $N_{\rm rot}$ the number of analog rotations per Trotter step ($20N_{\rm size}^2$ for our method and $36N_{\rm size}^2$ for the 2-step SPE reference, Appendix~\ref{app:merged-rotation}), and $G_{\max}=N_{\rm rot}T_{\max}/\tau_{\rm Trot}$ the rotation count of the longest sampled circuit. In Table~\ref{tab:app-sampling}, $N_{\rm circ}$ is the total number of circuit executions, equal to $N_{\rm sample}$ (per run, not rounded to an integer) times two, for the real and imaginary parts, times the number of runs (two for the 2-step SPE reference); $T_{\rm evo}=N_{\rm circ}\langle|j|\rangle\tau$ is the total evolution time with $\langle|j|\rangle=\sum_{j\in \mathcal{S}}|\hat F_j||j|/\mathcal{F}$; and $G_{\rm total}=N_{\rm circ}N_{\rm rot}\langle|j|\rangle\tau/\tau_{\rm Trot}$.

Two qualifications apply. First, $\tau$ and the split of the precision budget between $\epsilon_{\rm alg}$ and $\epsilon_{\rm Trotter}$ are chosen so as to minimize $T_{\rm total}$ for the architecture of Sec.~\ref{subsec:resource-STAR}, so the entries depend on $p_{\rm phys}$; on a different architecture they would be reoptimized, with $\tau$ bounded by the $\tau_{\max}$ of Table~\ref{tab:app-system}. Second, $N_{\rm circ}$, $T_{\rm evo}$ and $G_{\rm total}$ exclude the error-mitigation sampling overhead; $T_{\rm evo}$ and $G_{\rm total}$ are larger by the factor $\bar\gamma^2$, the mean of $\gamma^2_{j\tau}$ weighted by $|\hat{F}_j| |j|$, given in the last column of Table~\ref{tab:app-sampling}, which is included in $T_{\rm total}$ of Table~\ref{tab:hubbard_resource_estimation_Pphys3}.

\begin{table*}[htbp]
\renewcommand{\arraystretch}{1.2}
\caption{\label{tab:app-sampling}%
Sampling and aggregate logical resources, for the operating points of Table~\ref{tab:hubbard_resource_estimation_Pphys3}. The initial-state fidelity is $\eta=0.9$ and $N_{\rm sample}=C(\mathcal{F}/\eta)^2$ with $C=32$.}
\begin{ruledtabular}
\begin{tabular}{llcccccc}
Lattice & Method & $\mathcal{F}$ & $N_{\rm sample}$ & $N_{\rm circ}$ & $T_{\rm evo}$ & $G_{\rm total}$ & $\bar\gamma^2$ \\
\hline
\multicolumn{8}{c}{$p_{\rm phys}=10^{-3}$}\\
\hline
$6\times6$   & Ours  & $2.95$ & $343$ & $685$  & $1.29\times10^{4}$ & $2.79\times10^{10}$ & $7.28\times10^{4}$ \\
& 2-SPE & $4.15$ & $679$ & $2716$ & $3.67\times10^{4}$ & $1.45\times10^{11}$ & $8.63\times10^{4}$ \\
$8\times8$   & Ours  & $3.17$ & $396$ & $791$  & $1.38\times10^{4}$ & $7.16\times10^{10}$ & $4.25\times10^{9}$ \\
& 2-SPE & $4.33$ & $741$ & $2964$ & $3.85\times10^{4}$ & $3.62\times10^{11}$ & $6.89\times10^{9}$ \\
$10\times10$ & Ours  & $3.40$ & $456$ & $913$  & $1.49\times10^{4}$ & $1.52\times10^{11}$ & $8.90\times10^{15}$ \\
& 2-SPE & $4.47$ & $791$ & $3164$ & $4.00\times10^{4}$ & $7.43\times10^{11}$ & $2.42\times10^{16}$ \\
\hline
\multicolumn{8}{c}{$p_{\rm phys}=5\times10^{-4}$}\\
\hline
$6\times6$   & Ours  & $1.99$ & $157$ & $313$  & $1.06\times10^{4}$ & $1.23\times10^{10}$ & $2.69$ \\
& 2-SPE & $4.15$ & $679$ & $2716$ & $3.67\times10^{4}$ & $1.45\times10^{11}$ & $5.19$ \\
$8\times8$   & Ours  & $2.11$ & $175$ & $351$  & $1.07\times10^{4}$ & $3.23\times10^{10}$ & $7.64$ \\
& 2-SPE & $4.33$ & $741$ & $2964$ & $3.85\times10^{4}$ & $3.62\times10^{11}$ & $6.02$ \\
$10\times10$ & Ours  & $2.51$ & $248$ & $496$  & $1.11\times10^{4}$ & $1.14\times10^{11}$ & $24.4$ \\
& 2-SPE & $4.47$ & $791$ & $3164$ & $4.00\times10^{4}$ & $7.43\times10^{11}$ & $24.0$ \\
\hline
\multicolumn{8}{c}{$p_{\rm phys}=10^{-4}$}\\
\hline
$6\times6$   & Ours  & $1.87$ & $139$ & $277$  & $9.21\times10^{3}$ & $1.50\times10^{10}$ & $2.00$ \\
& 2-SPE & $4.19$ & $693$ & $2770$ & $4.22\times10^{4}$ & $9.21\times10^{10}$ & $1.62$ \\
$8\times8$   & Ours  & $1.95$ & $151$ & $301$  & $9.19\times10^{3}$ & $4.17\times10^{10}$ & $2.34$ \\
& 2-SPE & $4.36$ & $751$ & $3004$ & $4.25\times10^{4}$ & $2.46\times10^{11}$ & $1.89$ \\
$10\times10$ & Ours  & $2.07$ & $170$ & $340$  & $9.53\times10^{3}$ & $9.52\times10^{10}$ & $2.84$ \\
& 2-SPE & $4.49$ & $798$ & $3191$ & $4.27\times10^{4}$ & $5.45\times10^{11}$ & $2.21$ \\
\end{tabular}
\end{ruledtabular}
\end{table*}

\bibliography{energy_diff_estimation_v1}

\end{document}